# Understanding the origin of the particularly small and anisotropic thermal expansion of MOF-74

*Tomas Kamencek[1,2], Benedikt Schrode[3], Roland Resel[1], Raffaele Ricco[4], and Egbert Zojer[1]\**

[1]Institute of Solid State Physics, Graz University of Technology, NAWI Graz, Petersgasse 16, 8010 Graz, Austria

[2]Institute of Physical and Theoretical Chemistry, Graz University of Technology, NAWI Graz, Stremayrgasse 9, 8010 Graz, Austria

[3]Anton Paar GmbH, Anton-Paar-Str. 20, 8054 Graz, Austria

[4]School of Engineering and Technology, Asian Institute of Technology, 58 Moo 9, Khlong Luang, Pathum Thani, 12120, Thailand

**ABSTRACT**: Metal-organic frameworks often display large positive or negative thermal expansion coefficients. MOF-74, a material envisioned for many applications does not display such a behavior. For this system, temperature-dependent x-ray diffraction reveals particularly small negative thermal expansion coefficients perpendicular and positive ones parallel to the hexagonally arranged pores. The observed trends are explained by combining state-of-the-art density-functional theory calculations with the Grüneisen theory of thermal expansion, which allows tracing back thermal expansion to contributions of individual phonons. On the macroscopic level, the small thermal expansion coefficients arise from two aspects: compensation effects caused by the large coupling between stress and strain perpendicular to the pores and the small magnitudes of the mean Grüneisen tensor elements, $\langle\gamma\rangle$, which provide information on how strains in the material influence its phonon frequencies. To understand the small mean Grüneisen tensor in MOF-74, the individual mode contributions are analyzed based on the corresponding atomic motions. This reveals that only the lowest frequency modes up to ~3 THz provide non-negligible contributions, such that $\langle\gamma\rangle$ drops sharply at higher temperatures. These considerations reveal how the details of the anharmonic properties of specific phonon bands determine the magnitude and sign of thermal expansion in a prototypical material like MOF-74.

# 1. Introduction

Metal-organic frameworks (MOFs[1–4]) are a booming class of porous materials with an enormous accessible surface area, often amounting to several thousand square meters per gram.[5–7] Therefore, MOFs have a high potential for applications such as gas storage,[8–10] catalysis,[11–13] and gas separation.[14,15] They are also used in various functional devices.[16–20] In many of those applications, the thermal expansion of MOFs is a highly relevant property.[21,22] For example, it determines the thermal mismatch (and the resulting thermal stresses)[23–26] for heteroepitaxial growth processes,[27,28] or the change of the pore volume with temperature. Therefore, considerable efforts have been dedicated to studying thermal expansion in MOFs by means of both simulations and experiments. Experimental studies have, for example, focused on isoreticular MOFs (IRMOFs),[29–32] DUTs,[30,33] HKUST-1,[30,34–36] MOFs containing paddle-wheel secondary building units,[22,30,37] UiOs,[30] FMOF-1,[38] MCF-34,[39] ZIF-8,[40] MOF-74,[41] and others.[42–51] These experimental investigations are complemented by simulations, e.g., for IRMOFs,[21,26,52–55] DUTs,[56] HKUST-1,[26] paddle-wheel MOFs,[30] MOF-74,[57] and others.[26,56,58,59] For the meaning of the acronyms, we refer the reader to the respective references, as the detailed chemical nature of the above systems is not of immediate relevance for the following discussion. What is rather relevant is that these systems often show rather sizable linear and volumetric thermal expansion coefficients (either positive: PTE, or negative: NTE). This is shown in **Figure 1**. Notably, these thermal expansion coefficients in MOFs are usually much larger than for many common non-porous materials.[53,60,61] As a result, computational and experimental studies of thermal expansion in MOFs are typically concerned with explaining the origins of the large PTE and NTE coefficients.

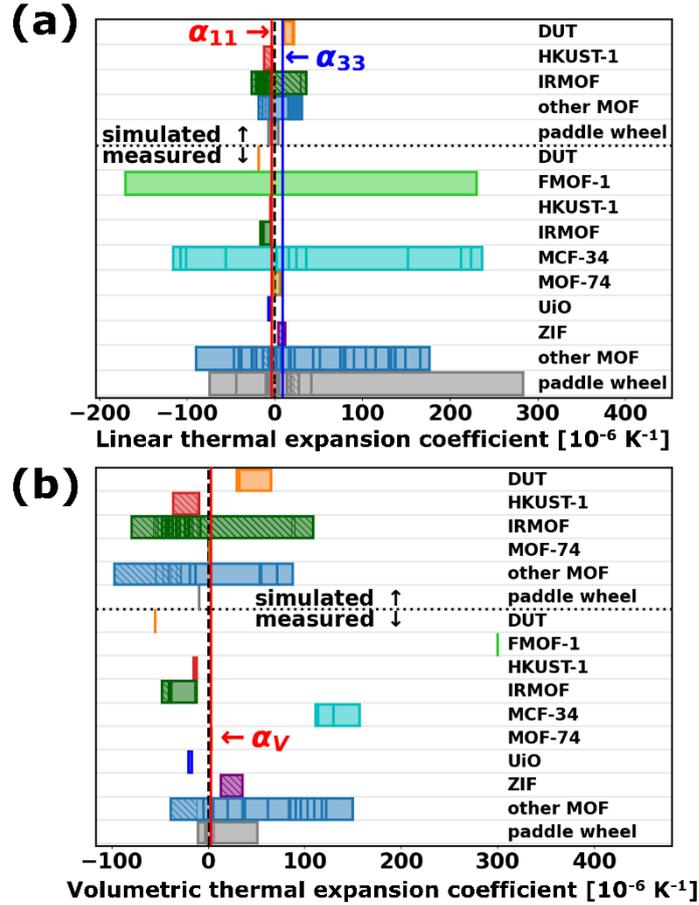

**Figure 1**. *(a) Linear and (b) volumetric thermal expansion coefficients for several MOFs reported in literature based on modelling (IRMOFs,[21,26,52–55] DUTs,[56] HKUST-1,[26] paddle-wheel MOFs,[30] MOF-74,[57] and others[26,56,58,59]) or experimental (IRMOFs,[29–32] DUTs,[30,33] HKUST-1,[30,34–36] paddle-wheel MOFs,[22,30,37] UiOs,[30] FMOF-1,[38] MCF-34,[39] ZIF-8,[40] MOF-74 [41], and others[42–51]) studies. The colossal thermal expansion coefficients of FMOF-1 in nitrogen atmosphere (-1.3·10$^{-2}$ and 1.2·10$^{-2}$ in the ab-plane and in the c-direction,[38] respectively) are not shown to enhance the visibility of the remaining MOF systems. The horizontal bars are drawn from the minimum to the maximum reported value of the respective MOF family, with vertical lines and hatched areas inside the colored bars representing individual values and ranges (e.g., upon varying linker lengths, adsorbates, topology, etc.) They have been taken from different papers and are reported for different crystallographic directions. The vertical red and blue lines in panel (a) denote the calculated linear thermal expansion coefficients (from the present work) for (activated) MOF-74(Zn) at 300 K in x(y)- ($\alpha_{11}$) and z-direction ($\alpha_{33}$) at 300 K, respectively. The red line in panel (b) indicates the low volumetric thermal expansion coefficient, $\alpha_V$, (at 300 K) calculated in this work.*

For archetypical MOFs such as IRMOFs and HKUST-1, several relevant mechanisms and factors of influence have been identified. These are related to optical[29,31,32,34,55,56] or acoustic[52] phonons, and potentially also to adsorbates,[54] where a correct treatment of anharmonicities in the potential energy surface has been shown to be crucial.[62] These considerations emphasize that studying thermal expansion is also an important first step towards understanding the anharmonic character of phonon modes in MOFs, which is relevant also for other properties like thermal conduction.[62–67] Particularly relevant quantities in this context are the so-called mode Grüneisen parameters, which describe the relative changes of the frequencies of the individual phonon modes with the unit-cell volume (or the applied strain). These mode Grüneisen parameters are directly connected to higher-order force constants beyond the harmonic approximation.[68–70] In this context, non-cubic MOFs are particularly interesting, as they allow studying the effects of anisotropic anharmonicities.

A practically relevant example of such an anisotropic system is zinc-based MOF-74,[71,72] which is in the focus of the present study. MOF-74 has been employed in different applications including colorimetric sensing,[73] adsorption chillers,[74] drug delivery,[75] and gas storage for, e.g., $CO_2$[76] and $H_2$.[9,77] When considering the thermal expansion of MOFs, MOF-74 shows comparably small linear expansion coefficients. Thus, MOF-74 can rather be defined as a low-thermal-expansion material (LTE). For reasons that will become apparent below, its volumetric thermal expansion coefficient is even lower at relevant temperatures, such that it lies in a similar range as certain ceramics often referred to as zero-thermal-expansion materials (ZTE).[78–82] This is consistent with existing "rough" theoretical estimates[57] (relying on a mixture of density-functional theory and force field calculations disregarding the anisotropy of the material). It is also in line with previous experimental findings for magnesium-based MOF-74,[41] which at room temperature suggest a positive thermal expansion coefficient of ~$6.7 \cdot 10^{-6}$ K$^{-1}$ along the pore axis and a negative thermal expansion coefficient of ~$-1.5 \cdot 10^{-6}$ K$^{-1}$ perpendicular to it. This raises the question, which mechanisms are responsible for the close to ZTE behavior of MOF-74. The structure of MOF-74 comprises hexagonal channels with 1D metal-oxide secondary building units (SBUs) located at the corners of the hexagons and organic linkers forming the walls in between (see **Figure 2**). The physical properties of MOF-74 distinctly differ for directions along its pores and perpendicular to them. This is not surprising, considering that for the former the characteristics of the inorganic SBUs dominate, while for the latter the organic linkers play a much more prominent role.[57,64,83] The fundamental reasons for the LTE/ZTE behavior of MOF-74, however, remain unclear. The present study aims at changing that by identifying the origins of the LTE/ZTE behavior of pristine zinc-based MOF-74 at an atomistic level, with a particular focus on the material's anisotropy. To this end, we combine diffraction studies with theoretical considerations based on state-of-the-art ab initio simulations within the framework of the Grüneisen theory of thermal expansion.[84] This approach (in contrast to, for example, molecular dynamics simulations) allows a phonon-resolved analysis of the situation such that one can assign to each phonon mode a quantitative contribution to the thermal expansion tensor. By analyzing the frequency ranges and wave vector-dependencies of the phonon bands with particularly relevant contributions, one is then able to identify trends and to systematically unravel the mechanisms behind the observed behavior.

The article is structured as follows: first, we briefly discuss the crystal structure of MOF-74 and provide the relevant synthetic, experimental, and theoretical details including a short summary of the Grüneisen theory of thermal expansion. Subsequently, the experimental results and their evaluation are discussed, followed by a detailed (phonon-resolved) analysis of the governing anharmonicities and contributions to the thermal expansion is presented. While the experiments yield observable thermal expansion values and represent a basis for the calculations, the simulations provide the desired mechanistic insights.

## 2. Methods
## 2.1. The Crystal Structure of MOF-74

The primitive crystallographic unit cell of MOF-74 with its three lattice vectors, $\vec{a_n}$ ($n \in \{1,2,3\}$), is shown in Figure 2. Although MOF-74 can be synthesized with a number of different metal ions,[85–91] with linkers consisting of more than one benzene ring,[72] and with modified docking groups,[92] whenever we refer to MOF-74 in the remainder of this article we mean pristine and activated MOF-74 based on $Zn^{2+}$ ions connected by 2,5-dihydroxy-terephthalate linkers. The structure of MOF-74 belongs to space group $R\bar{3}$ (international number 148) and has a rhombohedral (trigonal) Bravais lattice. Unlike many other MOFs, MOF-74 forms one-dimensionally extended pores arranged in a honeycomb-like lattice. This results in a distinct structural anisotropy. Accordingly, also the metal (oxide) nodes consist of (infinitely extended) quasi-one-dimensional arrangements of Zn and O atoms. Note that in the activated (pristine) state, the Zn atoms in MOF-74 are undercoordinated, but still exhibit an (approximately) octahedral coordination geometry. Due to its rhombohedral symmetry, the lattice has two independent lattice parameters, $a$ and $c$ (from which the primitive lattice vectors are constructed as explained in the caption of Figure 2): The distance between the metal atoms along the $z$-axis in the nodes equals $c$, while the pore diameter in the *xy*-plane (*i.e.,* the distance between two opposite corners of the hexagon) equals $4a/3$, as illustrated in Figure 2(a). Because of this distinct structural anisotropy, also the materials properties are expected to strongly depend on the crystallographic directions. This circumstance offers ideal conditions to separately study anharmonic effects in the *z*-direction and in directions within the *xy*-plane. These can then be associated with the nodes (dominating in *z*-direction) and the linkers of the MOF (relevant in the *xy*-plane).

In order to gain mechanistic insight and to qualitatively rationalize the signs of the mode Grüneisen tensor components in the Results section, we will analyze how certain geometric descriptors are impacted by the different phonons. This strategy is similar to the one used in Ref. [83], where such descriptors quantified the structural changes upon applying stress to the unit cell of MOF-74. The geometric descriptors include: the cross-sectional area of the nodes, $A_\Delta$, defined by the projection of the centers of the $Zn^{2+}$ ions onto the *xy*-plane, the inclination angle, $\phi$, between the edges of the nominal pore wall (i.e., the hypothetic regular hexagon connecting the centers of the nodes) and the projections of the linker backbones onto the *xy*-plane, the dihedral angle, $\tau$, between the $COO^-$ group of the linker and the benzene ring, and the polar linker tilt angle, $\theta$, between the node axes and the long molecular axes of the linkers. The mentioned descriptors are schematically depicted in Figure 2(c) and (d). For more details about their definition see Section S10 in the Supporting Information and Ref. [83].

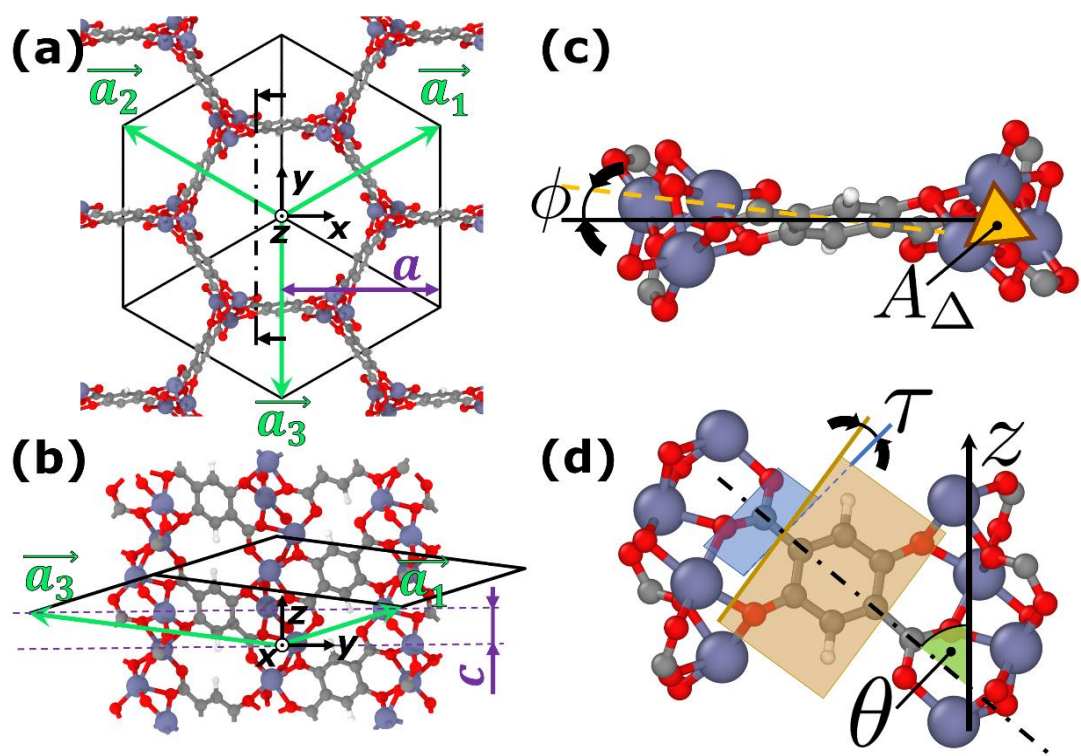

**Figure 2**. *(a) Top and (b) side view of the rhombohedral crystal structure of MOF-74 illustrating the two independent lattice parameters, a and c. The projections of the lattice vectors ($\vec{a_n}$ with n=1,2,3) are shown as colored arrows. The lattice parameter a is drawn in a way that it corresponds to the Cartesian x-component of lattice vector $\vec{a_1}$. The primitive Cartesian lattice vectors are constructed from a and c in the following way: $\vec{a_1} = [a, a/\sqrt{3}, c]$, $\vec{a_2} = [a, -a/\sqrt{3}, c]$, and $\vec{a_3} = [0, -2a/\sqrt{3}, c]$. The vertical dash-dotted line denotes the plane at which the structure is cut for the side view shown in panel (b), as seen in the direction of the black arrows. (b) Side view of the crystal structure illustrating the second independent lattice parameter c as the Cartesian z-component of all three lattice vectors. Atomic color coding: H…white, C…grey, O…red, Zn…purple. (c,d) Details of the structure of MOF-74 in which the most relevant geometric descriptors are shown: the cross-sectional area of the nodes, $A_\Delta$, defined by the projection of the centers of the $Zn^{2+}$ ions onto the xy-plane, the inclination angle, $\phi$, between the edges of the nominal pore wall (i.e., the hypothetic regular hexagon connecting the centers of the nodes; black horizontal line in panel (c)) and the projections of the linker backbones onto the xy-plane, the dihedral angle, $\tau$, between the $COO^-$ group of the linker and the phenylene ring, and the polar linker tilt angle, $\theta$, between the node axes (parallel to the z-axis) and the long molecular axes of the linkers (dash-dotted line).*

## 2.2. Experimental Methods

The studied MOF-74 powder samples were obtained following the solvothermal synthesis protocols of Ref. [93]. Temperature-dependent x-ray diffraction (XRD) was used to study the pristine MOF. Prior to the XRD measurements, the samples were activated directly in the in-situ temperature chamber (TTK 600 from Anton Paar) by heating them to 300°C in vacuum ($10^{-2}$ mbar) and keeping them at that temperature for 15 minutes. After the activation, the temperature was decreased stepwise, and x-ray diffractograms were recorded in Bragg-Brentano geometry using a PANalytical Empyrean diffractometer with a Cu radiation source, a primary Soller slit with a divergence angle of 0.02 rad, a divergence slit of 0.25°, a secondary 0.02 rad Soller slit, an anti-scatter slit with an angle of 0.25 rad, and an X'Celerator detector. To avoid a decrease in intensity, no additional monochromator was used, and to improve the thermal equilibration of the sample, a waiting time of up to 30 minutes was introduced between the time, when the internal sensor of the heating stage reached the set point temperature and when the XRD measurements were started. Furthermore, at each temperature the sample was aligned with respect to the x-ray beam to account for the thermal expansion of the setup.

The experimental results were analyzed using Bayesian probability theory to achieve meaningful results.[94] This was necessary due to the very minor change of the structural parameters with temperature combined with average peak widths of 0.19°. Additionally, a statistical analysis also allowed to take into account uncertainties in the determination of the sample temperature. Finally, employing a probabilistic approach also enabled the determination of estimators for the uncertainty of the calculated results. A mathematically rigorous description of the employed approach can be found in Section S2 in the Supporting Information.

**2.3. Theoretical Methods**

To generate the desired level of atomistic insight, ab initio calculations were carried out in the framework of density functional theory (DFT),[95,96] employing the VASP[97–100] code (version 5.4.4). Technically, these simulations were performed at 0 K (i.e., unlike in a molecular dynamics run, the atoms were kept fixed at their positions), but the temperature dependence of the phonon occupation and, thus, the thermal expansion tensor were accounted for, as described below. The exchange-correlation contributions were described via the PBEsol functional[101,102] in combination with the D3-BJ[103,104] a posteriori van der Waals correction. Additionally, tests employing the PBE functional[105,106] are contained in Sections S4 and S5 of the Supporting Information. Here it should not be concealed that they yield a different sign for the thermal expansion coefficient of the lattice parameter $c$ than when employing PBEsol. This is primarily a consequence of the fact that in MOF-74 one is dealing with overall very small effects. As discussed in Section S5 of the Supporting Information, the origin of the observation can be traced back to qualitatively different, small changes of certain geometric descriptors with strain or phonon displacements that we observe for the two functionals. In this context, it should, however, be stressed that PBEsol is supposed to yield more accurate structural properties[101,102,107] and thermal expansion coefficients[70,108,109] for solids (with smaller charge density gradients). Thus, the vast majority of studies calculating thermal expansion coefficients within DFT typically relies on the PBEsol functional. Moreover, only PBEsol yields results that are consistent with the most probable outcome of our experiments (see below) and with the neutron diffraction data of the Mg-analogue of MOF-74 in Ref. [41]. Finally, as again shown in Section S3 of the Supporting Information, only with PBEsol qualitatively consistent trends between the most sophisticated calculations performed within the quasi-harmonic approximation and in the framework of the Grüneisen Theory of thermal expansion are obtained.

Therefore, in the main manuscript we will exclusively focus on the results of the PBEsol calculations.

In the simulations, we first optimized the crystal structure of the primitive unit cell (including lattice parameters) to maximum residual forces below 1 meV/Å using a 3×3×3 *k*-mesh and a plane wave energy cutoff of 1000 eV for the optimization of the lattice constants and atomic positions. The elastic tensor and the compliance tensor were calculated according to the approach described in Ref. [83] and in Section S4 of the Supporting Information, keeping the particularly high cutoff energy of 1000 eV, following the convergence studies shown in Ref. [83]). Based on careful and extensive convergence tests discussed in Section S4 of the Supporting Information, we employed a 1×1×3 supercell of the conventional unit cell for the subsequent phonon calculations. This hexagonal supercell contains 9 primitive unit cells, as shown in Section S4.3 in the Supporting Information. As a consequence of the supercell size, its electronic structure had to be sampled only at the Γ-point to achieve convergence. An advantage of the conventional unit cell is that it allows to probe the interatomic force constants more efficiently in real space, since a corresponding 1×1×3 supercell has approximately the same unit-cell extents along all directions. For the phonon calculations, due to the enormous system size of the supercells, the plane wave energy cutoff was reduced to 800 eV (which required a reoptimization of the atomic positions with this cutoff, which was done for the primitive unit cell with a 3×3×3 *k*-mesh, see Section S4 in Supporting Information). In all simulations, an energy convergence criterion of $10^{-8}$ eV for the self-consistency cycle, and a Gaussian smearing of the electronic states with a width of 0.05 eV were employed.

For the chosen way of incorporating temperature (entropy) effects into the thermodynamic considerations, one first requires the phonon band structures of the material. For those phonon band structure calculations, we used the phonopy[110] package (version 2.9.1) and sampled the first Brillouin zone with a 20×20×20 mesh of phonon wave vectors, *q*, when calculating thermodynamic properties.

When following the evolution of the above-defined geometrical descriptors as a function of strain, we consider two distinct situations: (i) uniaxial strain parallel to the pore (i.e., in *z*-direction) and (ii) isotropic strain perpendicular to the pore (i.e., in the *xy*-plane). The latter is a more realistic scenario than applying uniaxial strain in either *x*- or *y*-direction because in the absence of external stresses, the thermal strain is isotropic in the *xy*-plane for a system with rhombohedral symmetry.

For the visualization of the crystal structure and of phonon modes, as well as for recording the corresponding animations, the Ovito package (version 3.3.1) was used.[111]

*2.3.1. Anisotropic Thermal Expansion*

One possible strategy for computing the thermal expansion tensor of a material would be to calculate the temperature-dependent (non-equilibrium) total Gibbs free enthalpy, $G$, (i.e., the contributions from electrons, phonons, and the elastic energy due to externally applied stress[112–114]) for various combinations of lattice constants. Then, at each temperature, one could identify the set of lattice constants, which minimizes the free enthalpy at each temperature to determine the thermal expansion coefficients. This minimization is typically achieved by fitting an analytical function (such as the well-known equations of state of Birch,[115] Murnaghan,[116] or Vinet et al.[117]) to the free enthalpy-vs.-lattice constants data and taking the minimum of the model function (which were systematically compared in Refs. [118,119]). This approach is commonly referred to as the quasi-harmonic approximation (QHA).[70,120,121] We have applied several flavors of this approach to the case of MOF-74, as discussed in Section S3 of the Supporting Information, including the procedure followed by George et al.[122] for several organic crystals and the approach by Holec et al.[123] who incorporated the anisotropy at a more sophisticated level. Overall, the results of the common QHA-based approaches sensitively depend on fitting and minimizing Gibbs free enthalpies with respect to finite strains. For thermal expansion coefficients as small as the ones encountered in MOF-74, the changes in the free enthalpy become, however, so small that fitting a model function and finding the associated minima (relying on well-established routines of the *numpy*,[124] *scipy*,[125] or scikit-learn[126] packages) becomes numerically stable only for applied strains much larger than the thermal strains suggested by our experiments. This means, for realistically small strains, the QHA fitting and minimization becomes highly inaccurate, while for unrealistically large strains, the results obtained within the QHA overestimate the thermal expansion especially for the lattice parameter $c$ (see the extensive discussion in Section S3 of the Supporting Information).

In the case of MOF-74, we, therefore, resorted to the Grüneisen theory of thermal expansion,[84] which in combination with the PBEsol functional yields results in rather good qualitative and also quantitative agreement with the experimental observations, as shown below.

The central quantity in the Grüneisen theory is the mode Grüneisen tensor, $\gamma_{ij}^{\lambda}$ (Cartesian components $i, j \in \{1=x, 2=y, 3=z\}$). It is defined as the (negative) first derivative of the phonon frequency, $\omega$, with respect to the components of the strain tensor, $\varepsilon_{ij}$, for each phonon mode, $\lambda$ (as indicated by super- and subscripts):[68,70,127]

$$\gamma_{ij}^{\lambda} = -\frac{1}{\omega_{\lambda}} \frac{\partial \omega_{\lambda}}{\partial \varepsilon_{ij}} \quad (1)$$

In practice, the partial derivatives in Equation (1) are determined based on the derivatives of the dynamical matrix (at 0 K), which are approximated by finite differences (see Section S3.7 in the Supporting Information). Hence, it is necessary to calculate the dynamical matrix for the equilibrium unit cell at 0 K as well as the dynamical matrices for suitably strained cells. In order to improve the numerical accuracy in the calculation of the finite differences, we implemented a more accurate fourth-order finite difference scheme, as detailed in Section S3.7 of the Supporting Information, where also the choice of $10^{-3}$ for the strain step size is briefly motivated. Based on the mode Grüneisen tensors, one can calculate the temperature-dependent mean Grüneisen tensor, $\langle \gamma_{ij} \rangle$, employing Equation (2). It can be interpreted as the average value of $\gamma_{ij}^{\lambda}$ weighted with the phonon mode contributions to the heat capacity (in the grand-canonical ensemble), $c_v^{\lambda}$, at a specific temperature, $T$:

$$\langle \gamma_{ij} \rangle(T) = \frac{\sum_{\lambda} c_v^{\lambda}(T) \gamma_{ij}^{\lambda}}{\sum_{\lambda} c_v^{\lambda}(T)} \quad (2)$$

Note that the temperature dependence of $\langle \gamma_{ij} \rangle$ arises solely from the temperature dependence of the mode contributions to the heat capacity, while, at the same time, the (0 K-) mode Grüneisen tensors are used.

Finally, the thermal expansion tensor, $\alpha_{ij}$, is directly related to $\langle \gamma_{ij} \rangle$, to the compliance tensor, $S_{ijkl}$, and to the phonon heat capacity per volume, $C_V/V$, according to Equation (3).[70]

$$\alpha_{ij} = \frac{C_V}{V} \sum_{kl} S_{ijkl} \langle \gamma_{kl} \rangle \quad (3)$$

A derivation of this equation is contained in Section S3.7 the Supporting Information. In contrast to the situation in cubic materials, the entire compliance tensor rather than the (inverse) bulk modulus enters Equation (3), such that the crystal anisotropy is considered when (thermal) stresses lead to (anisotropic) strains. In contrast to QHA-based approaches (see Section S3 in the Supporting Information), one cannot readily predict the temperature dependence of the compliance tensor with the Grüneisen theory. In fact, considering the variation of $S_{ijkl}$ with temperature would drastically increase the complexity of the calculations, making them prohibitively expensive for sophisticated ab initio methods. To avoid resorting to a lower level of theory, we, thus, used the compliance tensor obtained from precise DFT-calculations of the 0 K-structure. Consequently, the temperature dependence of the thermal expansion tensor is approximated to arise solely from the numerator of Equation (2) (as the denominator cancels with the heat capacity in Equation (3)). Neglecting the temperature dependence of the compliance tensor is a drawback of the Grüneisen theory of thermal expansion, but it makes it possible to rely on accurate ab initio calculations. Moreover, since MOF-74 is an LTE/ZTE material,[41,57] the assumption of a negligible temperature dependence of $S_{ijkl}$ appears well justified.

One of the distinct advantages of the Grüneisen theory of thermal expansion is its ability to provide information on the contributions of specific phonon modes, $\lambda$, to the thermal expansion tensor. These mode contributions, $\alpha_{ij}^\lambda$, can be defined in a straightforward manner by combining the thermal averaging of the Grüneisen tensor in Equation (2) with the general expression for $\alpha$ in Equation (3). This yields Equation (4) and (5), which define the mode contributions to $\alpha_{ij}$. They depend on the compliance tensor, the mode Grüneisen tensor, $\gamma_{ij}^\lambda$, and the mode contribution to the phonon heat capacity, $c_v^\lambda$ (with $N_q$ being the number of wave vectors used to sample reciprocal space):

$$\alpha_{ij}(T) = \frac{1}{N_q} \sum_\lambda \alpha_{ij}^\lambda(T) \qquad (4)$$

with

$$\alpha_{ij}^\lambda(T) = \frac{1}{V} \sum_{kl} S_{ijkl}\, c_v^\lambda(T) \gamma_{kl}^\lambda \qquad (5)$$

Due to the symmetry of MOF-74, which belongs to the rhombohedral crystal class, the equations above can be further simplified, as the associated mean Grüneisen tensor (with Cartesian components $ij$) has only two independent, non-zero elements, $\langle\gamma_{11}\rangle$ and $\langle\gamma_{33}\rangle$ with $\langle\gamma_{11}\rangle = \langle\gamma_{22}\rangle \neq \langle\gamma_{33}\rangle$ and $\langle\gamma_{ij}\rangle = 0$ for $i \neq j$[128] (see also Section S3 in the Supporting Information). The same applies to the thermal expansion tensor, $\alpha_{ij}$, with $\alpha_{11} = \alpha_{22} \neq \alpha_{33}$ and $\alpha_{ij} = 0$ for $i \neq j$. Moreover, when again exploiting the symmetries of MOF-74 and rewriting the elements of the compliance tensor in Voigt notation (as detailed for MOF-74, e.g., in Ref. [83]), Equation (3) can be simplified to

$$\alpha_{11} = \frac{C_V}{V}\left((S_{11} + S_{12})\langle\gamma_{11}\rangle + S_{13}\langle\gamma_{33}\rangle\right) \qquad (6)$$

and

$$\alpha_{33} = \frac{C_V}{V}\left(2S_{13}\langle\gamma_{11}\rangle + S_{33}\langle\gamma_{33}\rangle\right) \tag{7}$$

Given the diagonal form of $\boldsymbol{\alpha}$, the $\alpha_{11}$ element is equivalent to the thermal expansion coefficient of lattice parameter *a* (i.e., the lattice parameter perpendicular to the pore of MOF-74), while $\alpha_{33}$ corresponds to that of lattice parameter *c* (i.e., for the lattice parameter in pore direction). The mode contributions, $\alpha_{11}^\lambda$ and $\alpha_{33}^\lambda$, are defined in analogy to Equation (5) using the same symmetry-inequivalent elements as in Equation (6) and (7).

## 3. Results and Discussion
### 3.1. Experimental Results

While MOFs often show notable (negative) thermal expansion coefficients (see Figure 1) and comparably large shifts in the associated XRD peaks, for MOF-74 the peak positions display exceptionally small temperature-induced shifts, as illustrated in **Figure 3**(a). This severely complicates the determination of experimental thermal expansion coefficients, especially in view of the rather broad peaks compared, e.g., to metals.

Therefore, we evaluated the data using the Bayesian approach described in detail in Section S2 in the Supporting Information. This yielded the posterior probability distribution functions for the independent lattice constants, *a* and *c* (see Figure 3(b) and (c)) and for the associated thermal expansion coefficients, $\alpha_{11}$ and $\alpha_{33}$, shown in Figure 3(d) and (e). Equivalent plots for the temperature dependences of the probability densities of the unit-cell volume, *V*, and the associated volumetric thermal expansion coefficient, $\alpha_V$, can be found in Section S2 in the Supporting Information. Although the probabilistic uncertainty for the thermal expansion tensor elements is relatively large compared to the calculated expectation values, still certain trends can be observed: first and foremost, all data clearly show that the thermal expansion coefficients of MOF-74 are very small, confirming the notion that MOF-74 is an LTE/ZTE material, as indicated already in the introduction section (see Figure 1). Moreover, $\alpha_{11}$ tends to adopt slightly negative values, which is consistent with a weak drop of the lattice parameter *a* in Figure 3(b). The situation is more involved for lattice parameter *c* (see Figure 3(c)): it appears to grow between 123 K and 298 K and above ~400 K; correspondingly, $\alpha_{33}$ is positive for a significant fraction of the covered temperature region. There is, however, also a drop between 298 K and ~400 K, which is accompanied by particularly broad probability distributions at 273 K and at 298 K. The reason for that is not fully understood, but it might be an indication for certain structural rearrangements within the MOF resulting in an increased degree of disorder at those temperatures.

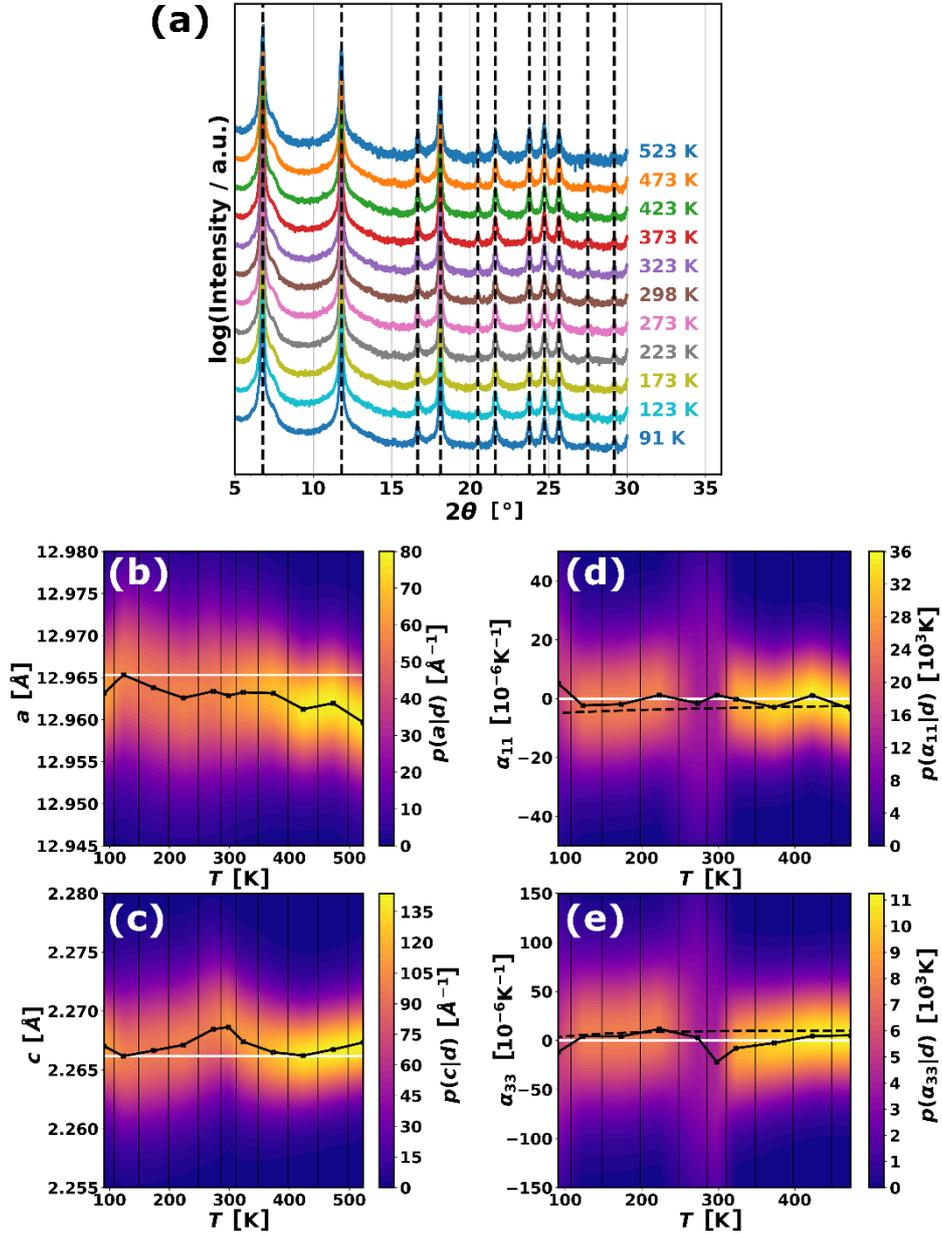

**Figure 3**. *(a) Powder XRD patterns of the activated MOF-74 sample for various temperatures. The vertical black dashed lines highlight the (fitted) peak positions at 300 K and serve as guide to the eye, emphasizing the particularly small peak shifts with temperature. (b,c) Posterior probability distributions, p(a|d) and p(c|d), (i.e., conditional probability densities given the entirety of experimental data referred to as "d") for the two independent lattice parameters, a and c, as a function of the setpoint temperature obtained from a Bayesian evaluation of the XRD patterns. The white horizontal lines denote the maximum measured value for a and the minimum measured value for c (in terms of their expectation values) for the given temperature*

*range and are guides to the eye. The thin black vertical lines indicate the boundaries of the discrete temperature stripes of which the probability density map is composed. Due to an offset in the lattice parameters between the values obtained from the DFT optimization (at 0 K) and the measured lattice parameters at the lowest temperature (91 K), which is larger than the temperature-induced changes (the simulations underestimate the unit-cell volume by ~2%), the theoretical values of a(T) and c(T) are not shown. (d,e) Posterior probability distributions, $p(\alpha_{ij}|d)$, for the two independent elements of the thermal expansion tensor, $\alpha_{11}$ and $\alpha_{33}$, as a function of temperature obtained from a Bayesian evaluation of the XRD patterns. These elements describe the temperature dependence of the lattice parameters a and c, respectively. The thick black lines connect the expectation value of the distributions at each temperature, while the dashed black lines show the temperature dependence of the theoretical results of the simulations presented in this work. The white horizontal lines denote the value zero and are guides to the eye. The thin black vertical lines indicate the boundaries of the discrete temperature stripes of which the probability density map is composed.*

Overall, the data suggest that $\alpha_{11}$ and $\alpha_{33}$ are not only different in magnitude, but also in sign. This interpretation is supported by qualitatively analogous results for magnesium-based MOF-74 measured by means of neutron powder diffraction.[41] We hypothesize that this difference in sign is a consequence of the one-dimensional nature of the pores (channels). As a result of the different changes, the volumetric thermal expansion coefficient, $\alpha_V \approx 2\alpha_{11} + \alpha_{33}$ (see Section S3 in the Supplementary Material), is approximately zero, albeit at an increased level of uncertainty compared to $\alpha_{11}$ and $\alpha_{33}$.

In passing we note that the temperature plotted on the horizontal axes in Figure 3(b-e) is the setpoint temperature measured at the temperature sensor inside the sample holder. In spite of the long equilibration times after reaching the setpoint temperature (see Methods section), there is a possibility that this temperature differs notably from the actual sample temperature due to the lack of convective heat transport for the measurements performed in vacuum (see activation of the MOF in the Methods section). Therefore, we additionally considered an uncertainty in the temperature in our Bayesian analysis (by means of Gaussian probability densities with a standard deviation of 20 K centered at the setpoint temperatures) and calculated an average linear expansion coefficient for the entire temperature range, $\bar{\alpha}$. This analysis, which is detailed in Section S2.1 in the Supporting Information, also yielded qualitatively unchanged trends of the linear ($\overline{\alpha_{11}} = (-2\pm1)\cdot10^{-6}\,\text{K}^{-1}$ and $\overline{\alpha_{33}} = (4\pm2)\cdot10^{-6}\,\text{K}^{-1}$) and volumetric ($\overline{\alpha_V} = (-1\pm4)\cdot10^{-6}\,\text{K}^{-1}$) thermal expansion coefficients.

To unravel the origin of the extraordinarily small (anisotropic) thermal expansion coefficients, we next discuss the results of the ab initio simulations, including a phonon mode-resolved analysis of the relevant contributions to the thermal expansion tensor.

### 3.2. Theoretical Considerations

*3.2.1. Grüneisen Theory and Macroscopic Effects*

The temperature dependence of the calculated thermal expansion tensor components is plotted in **Figure 4**(a), showing NTE perpendicular to the pore ($\alpha_{11} \leq 0$ for all temperatures) and for most temperatures PTE parallel to the pore ($\alpha_{33} > 0$ for temperatures > 70 K). The volumetric thermal expansion relevant, e.g., for the evolution of the pore volume, is given by $\alpha_V \approx \text{tr}(\alpha_{ij}) = 2\alpha_{11} + \alpha_{33}$. Due to the simultaneous NTE and PTE behavior of the material around room temperature its calculated value becomes extremely low (~$3 \cdot 10^{-6}$ K$^{-1}$ at 300 K), as shown in Figure 4(b). This is consistent with the experimental finding.

Beyond the partial cancellation of $\alpha_{11}$ and $\alpha_{33}$ for $\alpha_V$, also the individual tensor components are relatively small ($|\alpha_{11}| < 5 \cdot 10^{-6}$ K$^{-1}$ and $|\alpha_{33}| < 10.3 \cdot 10^{-6}$ K$^{-1}$), especially in view of the large (negative) thermal expansion coefficients frequently observed in MOFs and other porous materials (see Figure 1). This suggests that also for the individual tensor elements certain cancellation effects should occur, which keep the thermal expansion at such low values.

To identify these cancellation effects, the role of the compliance and the mean Grüneisen tensor in Equation (6) and (7) will be analyzed as a next step (see Methods section). Before discussing the more intricate influence of the mean Grüneisen tensor, it is useful to briefly comment on the role of the compliance tensor elements appearing in Equation (6) and (7). A more in-depth discussion of the elastic properties of MOF-74 can be found in Ref. [83], which also includes a discussion of their atomic origins. Here it is necessary to highlight that only four out of the seven[83] independent elements of $S$ play a role for thermal expansion within Grüneisen theory: the two elements $S_{11}$ and $S_{33}$ couple stress and strain that are both oriented in $x$-(or $y$-) and $z$-direction, respectively. The other two elements, $S_{12}$ and $S_{13}$, couple stress and strain in $x$- and $y$-direction (both perpendicular to the pore) with the corresponding quantities in $x$- (or $y$-) and in $z$-direction (i.e., perpendicular and parallel to the pore), respectively.

The compliances with the largest (absolute) magnitude are $S_{11}$ and $S_{12}$ ($S_{11}$ = 180 TPa$^{-1}$, $S_{12}$ = -155 TPa$^{-1}$). They, however, appear only as a sum in Equation (6) and, due to their different signs, the sum $S_{11}+S_{12}$ amounts to only 25 TPa$^{-1}$. This substantial cancellation is a result of the fact that when MOF-74 is exposed to, e.g., compressive uniaxial stress in $x$-direction, the resulting compression in $x$-direction almost equals the perpendicular expansion in $y$-direction.

Consequently, simultaneous stress in the *x*- and *y*-directions triggers expansions and compressions both in *x* and *y*, which largely cancel each other. In passing we note that this compensation effect not only impacts thermal expansion, but also results in a particularly small compressibility in the *xy*-plane (as discussed in more detail in Ref. [83]). Despite the above-described cancelation effect, the first term in Equation (6) still dominates $\alpha_{11}$, as the compliance $S_{13}$, which couples stress and strain parallel and perpendicular to the pore axis, is even smaller than $S_{11}+S_{12}$ ($S_{13}$ = -3 TPa$^{-1}$). Overall, as both compliance-related prefactors in Equation (6) are relatively small, it is not surprising that $\alpha_{11}$ also remains relatively small at all temperatures ($|\alpha_{11}|$ < 5·10$^{-6}$ K$^{-1}$).

Notably, similar cancellation effects do not appear for $\alpha_{33}$, since $S_{33}$ and $S_{13}$ enter Equation (7) separately. This lack of cancellation in the direction of the pores is again consistent with the mechanical properties of MOF-74, where one finds that the compressibility in *z*-direction is much larger than in the *xy*-plane.[83] The fact that $S_{33}$ (= 68 TPa$^{-1}$) is more than 11 times larger than $|2S_{13}|$ (=-6 TPa$^{-1}$), suggests that the $\langle\gamma_{33}\rangle$-related term in Equation (7) should dominate the evolution of $\alpha_{33}$.

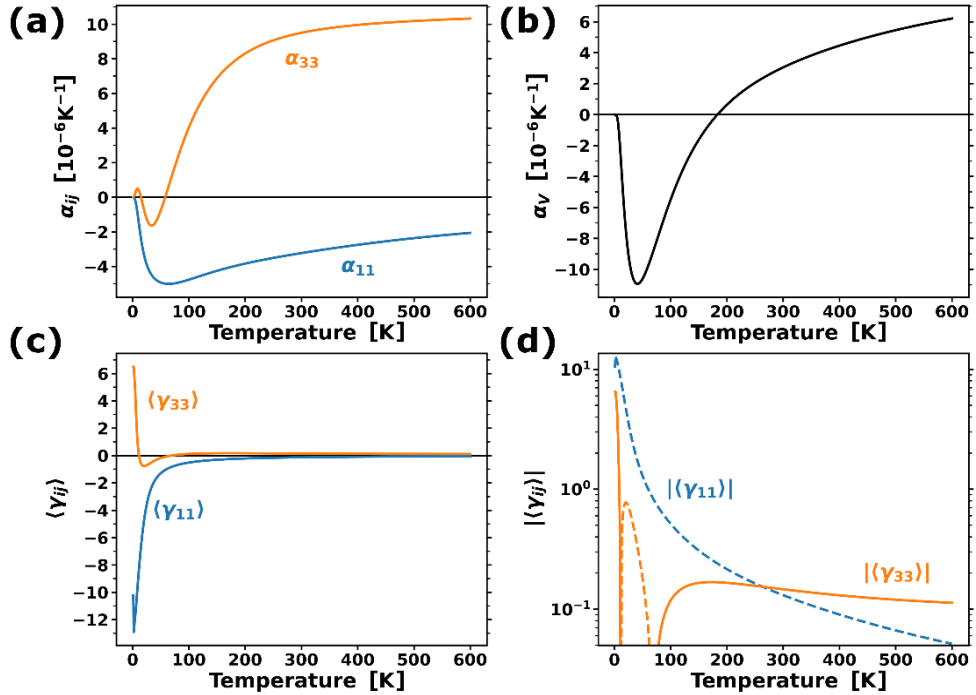

**Figure 4**. *Temperature dependence of (a) the two independent components of the thermal expansion tensor, $\alpha_{ij}$, calculated applying Grüneisen theory, (b) the resulting volumetric thermal expansion coefficient, $\alpha_V \approx 2\alpha_{11} + \alpha_{33}$, and (c) the two independent components of the mean Grüneisen tensor, $\langle\gamma_{ij}\rangle$, entering the expressions for the thermal expansion tensor. Panel (d) contains the absolute value of the two curves shown in panel (c) plotted on a logarithmic*

*scale. Dashed (solid) segments of the lines denote regions in which the mean Grüneisen tensor elements are negative (positive).*

Nevertheless, a complete picture concerning the sign and the magnitude of the thermal expansion tensor elements can only be obtained when also considering the temperature dependence of the elements of the mean Grüneisen tensor, $\langle \gamma_{ij} \rangle$, which is plotted in Figure 4(c). Their absolute values are plotted in Figure 4(d) on a logarithmic scale.

Notably, for temperatures above 100 K, both $|\langle \gamma_{11} \rangle|$ and $\langle \gamma_{33} \rangle$ are very small (below 0.51 and 0.19, respectively). The different signs of the components of the mean Grüneisen tensor in combination with the signs of the relevant compliance tensor elements discussed above lead to small negative values of $a_{11}$ and small positive values of $a_{33}$, consistent with the trends of weak negative and positive thermal expansion of the lattice parameters *a* and *c* deduced from the experiments.

For temperatures below 100 K, the components of the mean Grüneisen tensor display a pronounced temperature dependence: $\langle \gamma_{11} \rangle$ keeps its negative sign, but its absolute value increases significantly at low temperatures. As a result, also the dominating term in Equation (6) increases, resulting in a larger negative thermal expansion coefficient $a_{11}$. The reason for the less pronounced increase of $a_{11}$ compared to $\langle \gamma_{11} \rangle$ at low temperatures is the diminishing heat capacity $C_V$. The evolution of $\langle \gamma_{33} \rangle$ is fundamentally different: it starts out positive, becomes negative in a narrow temperature window (12 K ≤ *T* ≤ 70 K), and finally returns to the positive regime for high temperatures. This evolution is again directly correlated with the temperature dependence of $a_{33}$, which switches sign in a similar temperature range. The reason for $a_{33}$ still showing comparably high values at higher temperatures (> 100 K) is the increasing heat capacity in combination with $\langle \gamma_{33} \rangle$ dropping only rather weakly at higher temperatures.

We note in passing that the calculated heat capacities at constant pressure (stress) and at constant volume (strain) are essentially the same (see Section S3.1 in the Supporting Information). As the difference between the heat capacities depends on products of the thermal expansion tensor elements, this can be understood from the small thermal expansion coefficients due to the cancellation effects for the compliance tensor elements and the generally relatively small mean Grüneisen tensor elements.

The above considerations show that the thermal expansion behavior, on the one hand, is determined by the material's elastic properties (i.e., its compliance tensor), while, on the other hand, also the mean Grüneisen tensor plays a crucial role. As mentioned above, a microscopic explanation for the elastic properties of MOF-74(Zn) based on stress-induced atomistic displacements is already provided in Ref. [83]. A detailed explanation for the characteristics of the Grüneisen tensor has, however, not been provided yet. It requires an understanding of how its components depend on the individual phonon modes of the material. This understanding will be provided in the following section.

*3.2.2. Microscopic Origin: Phonon-Resolved Properties*

The phonon-resolved contributions to the thermal expansion, $\alpha_{ij}^\lambda$, and the mode Grüneisen tensor components are shown in **Figure 5**(a-d). Note that, in order to account for the rapid decrease of the $\gamma_{ij}^\lambda$ and $\alpha_{ij}^\lambda$ with phonon frequency, the frequency scale is plotted logarithmically (equivalent plots with linear frequency scales can be found in Section S8 in the Supporting Information). Figure 5(a) and (b) show that the mode Grüneisen tensor elements become very small above ~10 THz, and the most relevant modes are found below ~3 THz. When considering their contributions to the thermal expansion tensor, $\alpha_{ij}^\lambda$, the mode contribution to the heat capacity, $c_v^\lambda$, acts as another low-pass filter. Therefore, $\alpha_{ij}^\lambda$ (in contrast to $\gamma_{ij}^\lambda$) is temperature-dependent. Consequently, for low temperatures the frequency up to which phonon modes contribute to the thermal expansion is further reduced. This is explicitly shown for 10 K by the light blue and yellow data points in Figure 5(c) and (d).

Conversely, for temperatures > 300 K, the mode Grüneisen tensor elements show a similar frequency dependence as the $\alpha_{ij}^\lambda$. I.e., at room temperature $c_v^\lambda$ only excludes modes, which do not show relevant mode Grüneisen tensor elements. As a consequence, at even higher temperatures, the values of $\alpha_{ij}^\lambda$ hardly change anymore. This is shown in Figure 5(c) and (d) for a temperature of 500 K (dark blue and dark red data points).

To avoid the false impression that a significant portion of the overall phonon spectrum contributes significantly to thermal expansion, Figure 5(e) and (f) compare the density of phonon states (DOS) per logarithmic frequency interval, DOS$_\varphi$ (for mathematical details see Section S6 in the Supporting Information) plotted on a logarithmic frequency scale (analogous to panels (a)-(d)) with a conventional density of states plotted over a linear frequency scale. The vertical dashed lines in both plots refers to the frequency of 3 THz above which the components of the mode Grüneisen tensor become very small. This clearly illustrates that the modes with appreciable values of $\gamma_{ij}^\lambda$ and, thus, $\alpha_{ij}^\lambda$ cover only a small fraction of the entire phonon spectrum of MOF-74.

The observation that most phonon states are at higher frequencies, where the mode Grüneisen tensor elements have small values, directly impacts the temperature dependence of the mean

Grüneisen tensor elements: when the temperature rises, an increasing portion of the modes with particularly small $\gamma_{ij}^{\lambda}$ values becomes occupied, such that the averaging process described in Equation (2) generates very small values of $\langle \gamma_{ij} \rangle$ as shown in Figure 4(c) and (d). This is another key reason for the particularly small thermal expansion coefficients of MOF-74 at room temperature.

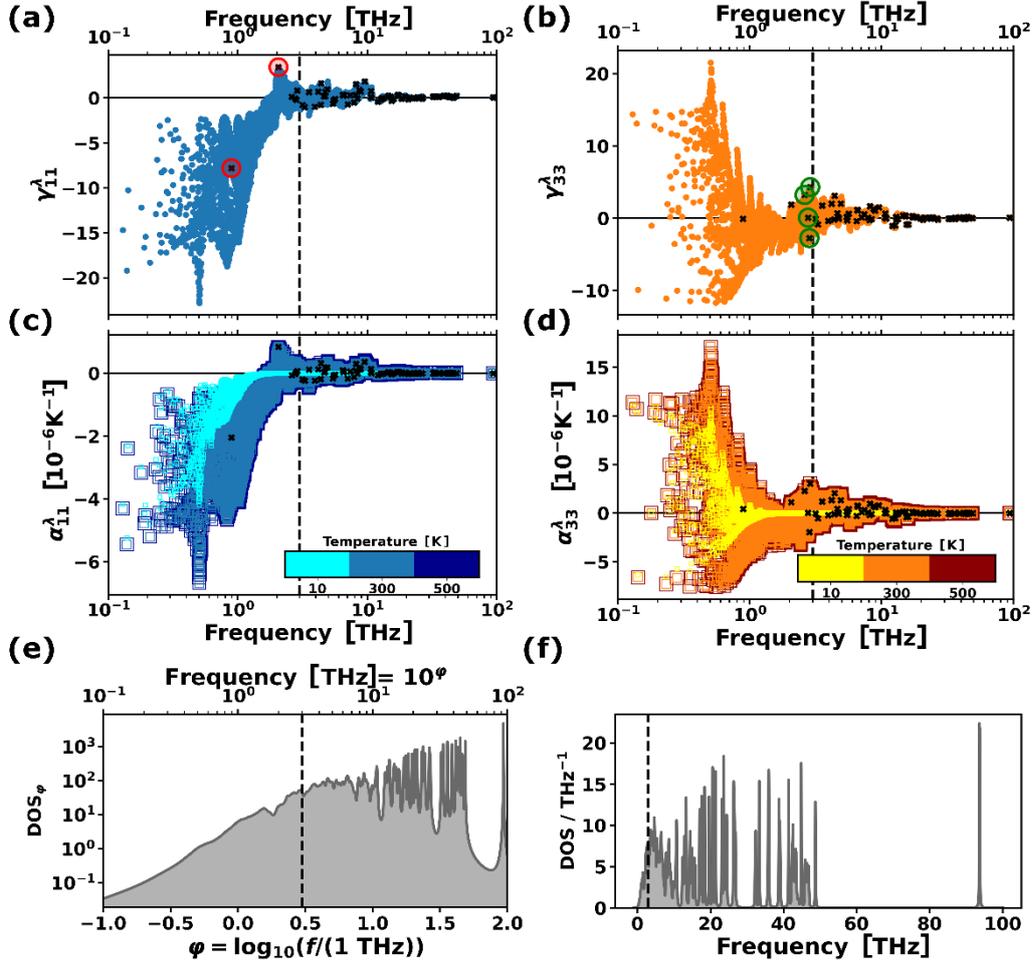

**Figure 5**. *(a,b) 11- and 33-components of the mode Grüneisen tensor and (c,d) mode contributions to the 11- and 33-components of the thermal expansion tensor at 10 K, 300 K, and 500 K of MOF-74 as a function of the phonon frequency. The colored data points correspond to the values for modes sampling the entire first Brillouin zone on a 20×20×20 mesh of wave vectors. In contrast, the black crosses denote the values for phonons at the Γ-point. The Γ-modes highlighted by semi-transparent circles in panels (a) and (b) are the optical modes, whose displacement patterns are displayed in **Figure 6** and **Figure 7**, respectively. The symbols in panels (c) and (d) are plotted with increasing size for higher temperatures to facilitate the visual recognition. Panel (e) shows the density of states per logarithmic frequency interval, $DOS_\varphi$, plotted on a logarithmic frequency scale, while panel (f) displays the*

*conventional density of states (DOS) and features a linear frequency scale. The detailed differences between the two quantities are explained in Section S6 in the Supporting Information. The vertical dashed lines in all panels are drawn at a frequency of 3 THz and serve as guide to the eye.*

Considering that the mode dependences of $\alpha_{ij}^{\lambda}$ and $\gamma_{ij}^{\lambda}$ are essentially identical for temperatures > 200 K, the following more in-depth discussion will be based on the data for $\gamma_{11}^{\lambda}$ and $\gamma_{33}^{\lambda}$ shown in Figure 5(a) and (b). They contain the following key messages: (i) the 11-components of the mode Grüneisen tensors are primarily negative (i.e., compressive strain in *x*- or *y*- direction decreases the frequency of those phonons, which is referred to as "phonon softening") (ii) conversely, the 33-components show negative and positive contributions, with a dominance of the latter (i.e., compressive strain *z*-direction primarily increases the frequency of the phonons). (iii) The magnitudes of the 11-components are typically somewhat larger than those of the 33-components (note the different scales in the figures). (iv) Finally, the contributions of the Γ-phonons (denoted as the black crosses in Figure 5(a-d)) typically deviate from those due to phonons with non-vanishing wave vector (especially ≤ ~2 THz). This emphasizes that calculations of $\alpha_{ij}$ based only on (more easily accessible) Γ-phonons are prone to missing the key anharmonicities responsible for the thermal expansion. In the following sections, we will separately discuss the most relevant optical and acoustic modes.

*3.2.3. Role of the Optical Phonons*

Despite the fact that also off-Γ phonons play a crucial role, for assessing the nature of the most relevant optical phonon bands with large mode Grüneisen parameters, it is useful to analyze the corresponding eigenmode displacements at the Γ-point. This is because – in the absence of avoided crossings – bands typically keep their qualitative character. Moreover, the Grüneisen tensor elements for the optical modes do not change as significantly throughout reciprocal space as those of the acoustic modes. For the analyzed optical Γ-modes, which are highlighted by semitransparent circles in Figure 5(a) and (b), the mode Grüneisen parameters are typically large than those of higher-lying Γ-phonons. In the following discussion, it also must be kept in mind that, overall, the mode Grüneisen parameters at Γ in MOF-74 are smaller in comparison to those in other MOFs,[52,55,67,129] since one is dealing with an LTE/ZTE material. Thus, the encountered correlations are expected to be sometimes less obvious than for a material with large positive or negative thermal expansion coefficients.

The relevant optical modes either typically involve pronounced motions of the metal nodes relative to the docking groups of the linkers, or they correspond to modes in which the linkers move approximately as rigid units ("rigid unit modes": RUMs[130–134]). The latter have actually been associated with negative thermal expansion effects observed in other MOFs.[34,54,56] Considering the nature of these modes, it is useful to not only follow the displacement patterns per se. Additionally, one ought to account for the evolutions of the descriptors defined in Section 2 (Figure 2(c) and (d)) in case of (i) applied strains and for (ii) the motion induced by specific modes. Within the harmonic approximation, the displacements of individual atoms show a strictly sinusoidal time dependence. This is, however, typically not the case for the more complex descriptors. Consequently, the excitation of specific phonons will induce net positive or net negative changes of the time averages of the descriptors relative to their equilibrium values (for more details see Section S10 in the Supporting Information).

The first two optical modes at 0.89 THz ($\gamma_{11}^{\lambda} \approx -7.80$) and 2.05 THz ($\gamma_{11}^{\lambda} \approx +3.45$) show comparably large values of $|\gamma_{11}^{\lambda}|$. They primarily correspond to rotations of the nodes around their axes (see **Figure 6**(a) and (b) and animations in the Supporting Information). The higher-frequency mode (Figure 6(b)) is characterized by rotations of all nodes in the same direction, while they rotate in alternating directions for the lower mode (Figure 6(a)). Although both modes display a relatively large level of anharmonicity (i.e., large Grüneisen tensor elements), only the lower one shows "phonon softening", as its frequency decreases upon compression in $x$-(or $y$-)direction. I.e., the associated 11-element of the mode Grüneisen tensor has a negative sign.

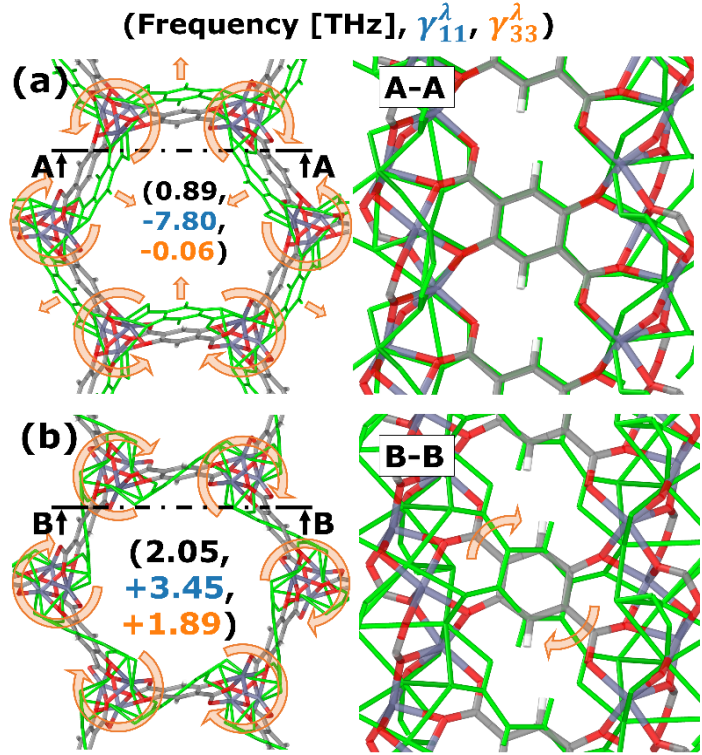

**Figure 6**. *Displacement patterns of the two lowest optical modes with large $|\gamma_{11}^\lambda|$-values for MOF-74 plotted for the Γ-point phonons. Panel (a) shows the situation for band 4 (at 0.89 THz) and panel (b) for band 5 (at 2.05 THz). The Γ-frequency (black text color) as well as the 11- and 33-components of the associated Grüneisen tensors (blue and orange text color) are shown for each mode in parentheses. The undisplaced geometry is shown in colors representing the respective atoms (C…gray, H…white, O…red, Zn…purple), while the structures displaced along the eigenmodes are shown in green. The relevant motions are stressed by arrows. The right panels show the cross-sections indicated by the black dash-dotted lines in the left columns (i.e., they illustrate the motions in the pore walls).*

The different signs of the 11-mode Grüneisen parameters of these modes can be understood from comparing the atomic motions in Figure 6(a) and Figure 6(b) and by computing the evolutions of the geometric descriptors of MOF-74 as a function of the displacements along the normal mode coordinates. In the lower frequency mode, as a result of the alternating node rotation, the linkers periodically move inwards and outwards relative to the pore center, similar to a bending motion. This mode is very similar to the "trampoline mode" partly responsible for the negative thermal expansion in MOF-5[29,31,32,52,55] and HKUST-1.[34] The inward- and outward motion of the linkers result in an increase of their average lengths, which is energetically unfavorable due to their stiffness.[83] The average cross-sectional area of the nodes also increases, and, ideally, the diameter of the pores would decrease such that there is less need for linker stretching. However, a change in the unit-cell parameters is conceptually impossible for Γ-point modes and is, therefore, not visible in Figure 6(a). It is nevertheless implied that, due to the anharmonicity of the actual potential, exciting the mode at 0.89 THz favors pore shrinking and, thus negative thermal expansion. In other words, if compressive *x*-stress were applied to the MOF, the motion associated with this phonon mode would be easier to accomplish, which would reduce the stiffness of the mode and, hence, its frequency such that the associated Grüneisen tensor element is negative.

In contrast, the frequency of the second mode involving linker rotations (shown in Figure 6(b)) increases upon compressive *x*-(or *y*-)strain. The most prominent change of an average geometric descriptor for this mode is a decrease in the linker inclination, $\phi$. This parameter decreases also when applying compressive strain in the *xy*-plane (see Section S10 in the Supporting Information). An observation that we consistently make is that, when dominant parameters change in the same direction due to compressive strain and excitation of the respective phonon, the corresponding mode Grüneisen parameter is positive (see also below). We attribute that to the fact that in such a case a simultaneous compression of the MOF and an excitation of the phonon displace the system more significantly from the global equilibrium position. Due to the nature of the anharmonicities, this leads to an increase of the phonon frequency.

Interestingly, "inclination modes" of the linker moieties similar to the mode at 2.05 THz were found in MOF-5 to contribute to the opposite effect, namely to negative thermal expansion.[29,31,32,52,55] This can be understood from the fact that in MOF-5 one starts from perfectly straight linkers such that an "inclination mode" increases the average linker inclination and shortens the unit cell, in contrast to the effect found in MOF-74.

As the strain-dependence of the above-discussed modes is intimately connected to the pore diameter rather than to the lattice parameter *c* (along the nodes), the frequencies of the modes change only very little for strain in *z*-direction, which results in small associated Grüneisen tensor elements, $\gamma_{33}^{\lambda}$. Conversely, most of the remaining optical modes below 3 THz exhibit a more pronounced anharmonic character for *z*-strain. The mode at 2.60 THz displayed in **Figure 7**(a) shows a comparably large positive $\gamma_{33}^{\lambda}$ element (+3.23). The associated atomic motion again involves a variation in the linker inclination, $\phi$, and also in the linker tilt, $\theta$, albeit here the average angles increase. Additionally, one can observe an increase in the average node cross-sectional area, $A_\Delta$, which impacts the local Zn-O bonding geometry of the nodes such that the average dihedral angle, $\tau$, between the COO⁻ group and the aromatic ring decreases. Upon compression in *z*-direction, both $A_\Delta$ and $\tau$ increase considerably. Considering the flat torsional potential of the COO⁻ group (as shown for IRMOFs[67]), the increase of $A_\Delta$ apparently dominates such that, in line with the above arguments, $\gamma_{33}^\lambda$ is positive.

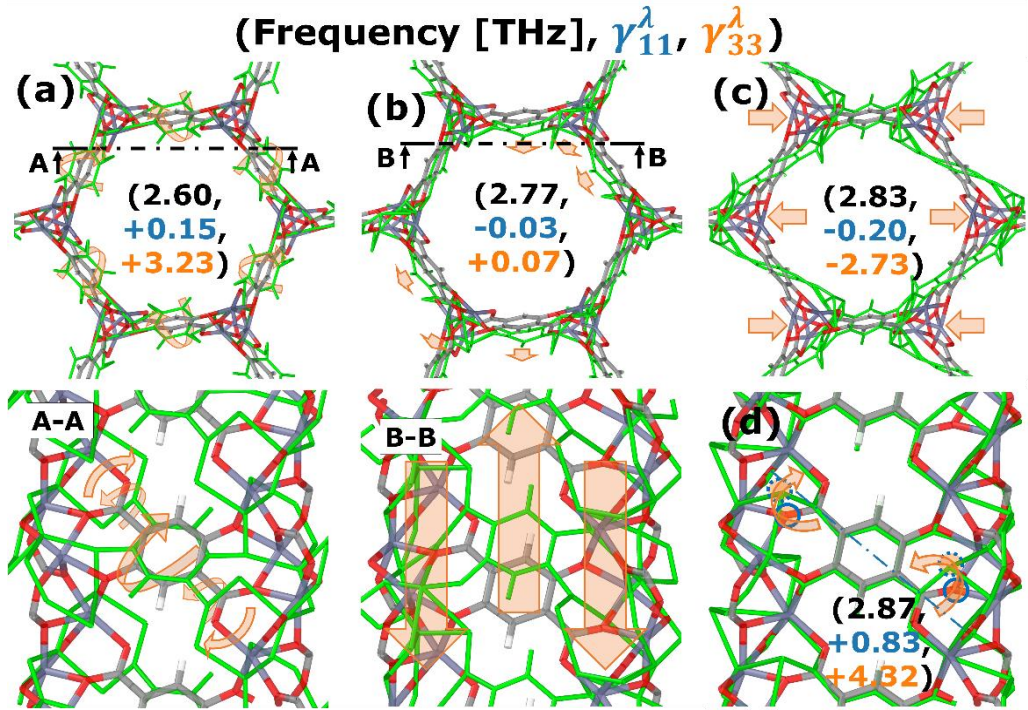

**Figure 7**. *Displacement patterns of low-frequency optical phonon modes with large (or particularly small) $|\gamma_{33}^\lambda|$-values for MOF-74 plotted for the Γ-point phonons. Panel (a) shows the situation for band 6 (at 2.60 THz), panel (b) for band 8 (at 2.77 THz), panel (c) for band 9 (at 2.83 THz), and panel (d) band 11 (at 2.87 THz). Except for the mode in panel (b), all displayed modes show comparably large values of $|\gamma_{33}^\lambda|$. The Γ-frequency (black text color) as well as the 11- and 33-components of the associated Grüneisen tensors (blue and orange text color) are shown for each mode in parentheses. The undisplaced geometry is shown in colors representing the respective atoms (C…gray, H…white, O…red, Zn…purple), while the structures displaced along the eigenmodes are shown in green. The relevant motions are stressed by arrows. The bottom panels of panels (a) and (b) show cross-sections indicated by the black dash-dotted lines in the top panels (i.e., they illustrate the motions in the pore walls). Notably, there are two bands, which are degenerate at the Γ-point both at 2.77 THz and at 2.83 THz (bands 7 and 8 as well as bands 9 and 10). Of those, only the Γ-phonons of band 8 and 9 are shown in panels (b) and (c). The other ones involve similar atomic motions (see animations in the Supporting Information). The blue circles in panel (d) highlight the (carboxylic) oxygen atom which exhibits the most notable torsional motion.*

The next higher optical, twofold degenerate mode at 2.77 THz shows comparably small Grüneisen tensor elements, both, for the 11- and the 33-component (see Figure 7(b)). This might be because they primarily trigger a deformation of the hexagonal shape of the pore, which is a very soft degree of freedom in MOF-74.[83] In *z*-direction, the MOF experiences no drastic structural changes for this vibration: the nodes and linkers are shifted alternatingly in +*z* and -*z* direction, which is not a deformation observed when applying *z*-strain.

The last two modes discussed here are again more relevant for the thermal expansion of MOF-74: the two-fold degenerate mode at 2.83 THz characterized by deformations of the pore shape (see Figure 7(c)) and the mode at 2.87 THz characterized by torsion of the COO⁻ groups of the linkers (see Figure 7(d)). They display negative (-2.73) and positive (+4.32) values of $\gamma_{33}^\lambda$, respectively. For the 2.83 THz mode with negative Grüneisen tensor elements, one can rationalize the strain-induced phonon softening considering that *z*-compression and the phonon mode act in opposite directions: during the (comparably soft) pore-deformation the average node area, $A_\Delta$, decreases and the linkers become flatter, i.e., $\theta$ increases. The opposite applies to the 2.87 THz mode: here, the most relevant vibration-induced changes in the average geometric descriptors are increases of $\tau$ and $A_\Delta$, which are both also triggered by compressive *z*-strain.

*3.2.4. Role of the Acoustic Phonons*

Despite the impact of the optical phonons, the largest contributions to the thermal expansion tensor arise from acoustic phonons with finite wave vectors, as previously indicated in Figure 5. The corresponding bands display a particularly large dispersion. Therefore, it is useful to analyze the associated mode Grüneisen constants in a wave-vector resolved fashion. The calculated low-frequency phonon band structure (≤ 3 THz) for MOF-74 is shown in **Figure 8** with the bands colored according to the elements of the mode Grüneisen tensors (a) and (b) and the respective mode contributions to the thermal expansion tensor (c) and (d). As the trends are similar for both quantities, we will again focus the discussion on the elements of the (mode) Grüneisen tensors.

Figure 8(a) and (b) confirm that the largest mode Grüneisen tensor elements are found for the acoustic bands below 1 THz. This is particularly true for phonons propagating in or close to the *xy*-plane of the structure, i.e., for bands towards the high-symmetry points F, P, and P₁. Interestingly, the two higher acoustic bands are characterized by negative values for $\gamma_{11}^\lambda$ and $\gamma_{33}^\lambda$ in the entire Brillouin zone. Conversely, the lowest acoustic band show negative $\gamma_{11}^\lambda$ tensor elements, while the $\gamma_{33}^\lambda$ elements are particularly large and positive. This causes the overall positive sign of $\gamma^\lambda_{33}$ (for the impact of the chosen functional on this trend mentioned already in the Methods section see Section S5 in the Supporting Information).

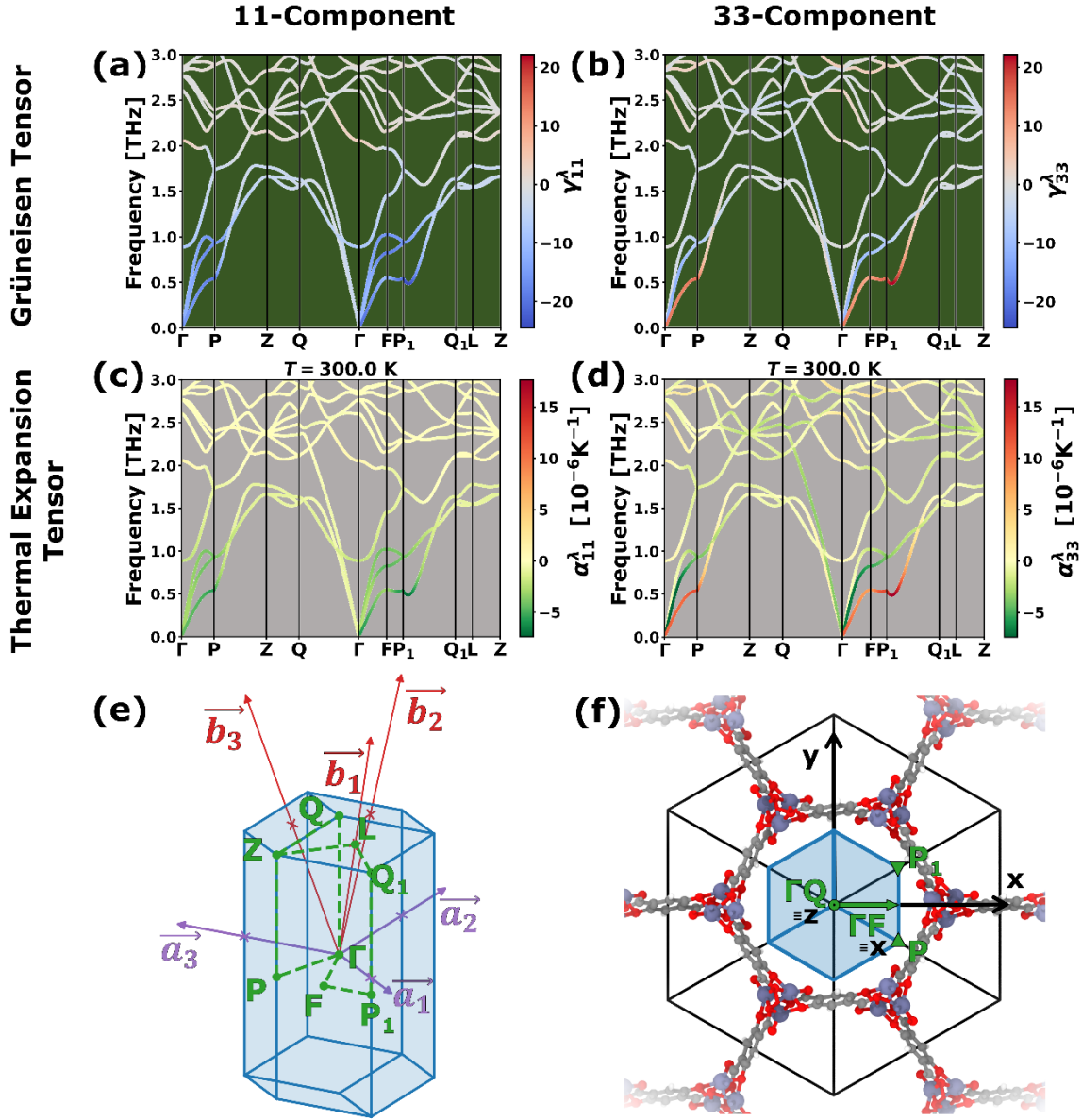

**Figure 8**. *PBEsol/D3-BJ-calculated phonon band structure of MOF-74 colored according to (a,b) the mode Grüneisen tensor components, $\gamma^\lambda_{11}$ and $\gamma^\lambda_{33}$, and (c,d) the mode contributions to the thermal expansion tensor components, $\alpha^\lambda_{11}$ and $\alpha^\lambda_{33}$, (see Equation (5)). (e) First Brillouin zone of MOF-74 including the lattice parameters, $\vec{a_n}$, the reciprocal lattice vectors, $\vec{b_n}$, and the high-symmetry points between which the band structures are drawn. (f) Directions of the high-symmetry paths ΓQ and ΓF in real space. They correspond to the z- and the x-axis, respectively and are, thus, parallel and perpendicular to the pore direction. The high-symmetry points P*

*and P₁ lie slightly above and below the xy-plane and are denoted by an upwards and a downwards pointing triangle.*

To understand the sign change between $\gamma_{11}^\lambda$ and $\gamma_{33}^\lambda$, it is again useful to consider the mode displacement pattern of the lowest acoustic mode at the high symmetry point F. This displacement pattern is shown in **Figure 9**(a) (animations can be found in the in the Supporting Information). It reveals that this acoustic mode has primarily transverse character, with the phonon polarization vectors showing nearly no components in *z*-direction (i.e., along the pores). For waves propagating in *x*-direction (along ΓF), the polarization of the said transverse acoustic phonons is oriented nearly fully in *y*-direction. At the F-point, neighboring nodes move in opposite directions, with correspondingly smaller phase shifts for smaller *q*. As a consequence, the average distance between the nodes increases, as shown in Figure 9(b) (selected snapshots of the motion associated with this acoustic phonon at F are shown in Figure 9(c-e)). This again results in the energetically very costly stretching of the linker molecules, which can be compensated by a unit-cell shrinkage, when this mode is increasingly excited with temperature. This causes a phonon softening and, thus, contributes to the negative thermal expansion in the *xy*-plane. Overall, this situation is reminiscent of similar observations reported for MOF-5.[52] Conversely, when MOF-74 is strained in *z*-direction, the nodes expand, but they also rotate such that the linkers are slightly elongated. The tendency to increase the linker length for *z*-strain and for the phonon displacement is again held responsible for the positive sign of the corresponding mode-Grüneisen tensor parameter $\gamma_{33}^\lambda$. This mode is additionally accompanied by a net lateral expansion of the nodes (an increase in $A_\Delta$) and a rotation of the nodes, $\eta$, like for compressive strain in *z*-direction.

Similar transverse acoustic modes with a "dual" character, i.e., a positive value of $\gamma_{33}^\lambda$ and a negative value of $\gamma_{11}^\lambda$ are observed, for example, for the ΓP and FP₁ directions (i.e., for wave vectors almost perpendicular to the pore). Interestingly, such modes do not occur for wave vectors parallel to the pore and here in particular in the ΓQ-direction. We attribute this to the observation that no modes with a strictly transverse character with displacements in the *xy*-plane occur for the respective *q*-direction, which results in only comparably small mode Grüneisen parameter components.

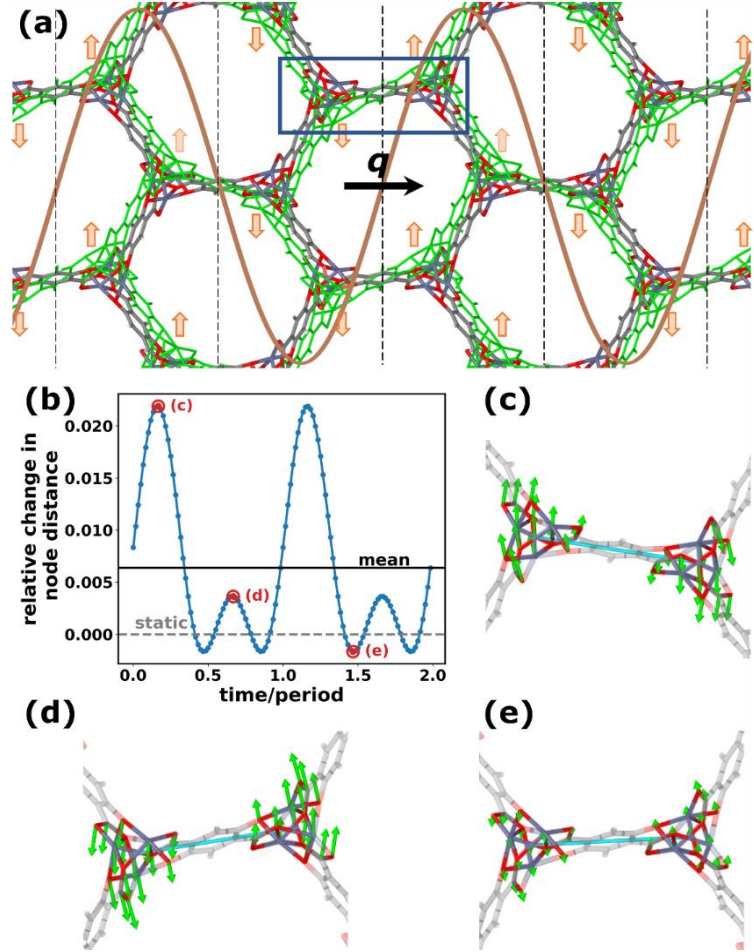

**Figure 9**. *(a) Displacement pattern of the lowest transverse acoustic phonon at the F-point (with the reduced wave vector [0.5, -0.5, 0]). The undisplaced geometry is shown in colors representing the respective atoms (C…gray, H…white, O…red, Zn…purple), while the structures displaced along the eigenmodes is shown in green. The wave vector is oriented in x-direction (see black arrow denoting the wave vector, **q**), while the polarization is primarily oriented in y-direction as denoted by the blue and orange arrows as well as the sinus-shaped waves implying the polarization amplitude for positive and negative displacements. The vertical dashed black lines indicate the nodal planes of the standing waves – i.e., the planes with vanishing displacements. The dark blue rectangle indicates the zoomed-in region shown in panels (c-e). (b) Time evolution of the change in distance between the axes of two neighboring nodes (i.e., the length of the blue line in panels (c-e)) during the atomic motion associated with the transverse acoustic phonon at F relative to the static (0-K optimized) value. The mean value of the node distance is drawn as a black horizontal line, while the static value is denoted by a grey dashed line. The snapshots of the motion in panels (c-e), in which the node distance is shown as a light blue line, are indicated by red circles in panel (b).*

## 4. Conclusions

The presented work combines temperature-dependent x-ray diffractometry with state-of-the-art dispersion-corrected density functional theory simulations (within the Grüneisen theory) to explain the atomistic origin of the low thermal expansion observed for pristine, zinc-based MOF-74. Comparing the thermal expansion coefficients in MOF-74 with the situation in other MOFs reported in literature, MOF-74 can be regarded as a low- or even zero-thermal-expansion material (LTE/ZTE). This is consistent with the results of the conducted diffraction experiments, revealing very minor temperature-induced shifts of the positions of the diffraction peaks. Therefore, a probabilistic approach is necessary to determine the two independent lattice constants of MOF-74, $a$ and $c$ (related to the pore diameter and the pore length per unit cell, respectively), and the associated components of the Cartesian thermal expansion tensor, $\alpha_{11}(=\alpha_{22})$ and $\alpha_{33}$.

The obtained probability densities suggest that lattice parameter $a$ decreases with temperature, while $c$ increases for most of the considered temperature range. The volumetric thermal expansion coefficient is close to zero (~$3 \cdot 10^{-6}$ K$^{-1}$) at 300 K due to the different signs of $\alpha_{11}$ and $\alpha_{33}$. The experimental results are in good qualitative agreement with theoretical results based on the Grüneisen theory of thermal expansion relying on ab initio data. In this context, it has to be kept in mind that it is virtually impossible to achieve a perfect quantitative agreement between theory and experiment for a quantity that is close to zero due to cancellation effects (see below). This also explains, why all our attempts to describe the situation via a free energy minimization within the quasi-harmonic approximation suffered from serious shortcomings.

As indicated in the previous paragraph, also the tensor elements, $\alpha_{11}$ and $\alpha_{33}$, are subject to cancellation effects: within Grüneisen theory, $\alpha_{11}$ and $\alpha_{33}$ arise from sums and products of elements of the compliance tensor, the heat capacity, and the mean Grüneisen tensors, with the latter incorporating the anharmonicities of all phonon modes in the crystal. For $\alpha_{11}$, the cancellation effect occurs due to close-to-equal compliance tensor elements with different sign ($S_{11} \approx -S_{12}$) resulting in a small value of $|\alpha_{11}| < 5 \cdot 10^{-6}$ K$^{-1}$ for all temperatures, regardless of the mean Grüneisen tensor. For the latter, a notable magnitude is observed only at low temperatures. The mean Grüneisen tensor also determines the negative sign of $\alpha_{11}$. For $\alpha_{33}$, the cancellation effects occur, when determining the mean Grüneisen tensor from the mode contributions due to a similar amount of negative and positive contributions especially below frequencies of ~2

THz. Only at higher frequencies, the positive contributions clearly dominate resulting in slightly larger positive values of $\alpha_{33}$ (with $|\alpha_{33}|$ remaining below $10.3 \cdot 10^{-6}$ K$^{-1}$).

Overall, the absolute magnitudes of the mode Grüneisen tensor elements, however, display a pronounced drop with frequency and become negligible for frequencies > 10 THz. Therefore, as the temperature increases, an increasing number of optical phonon modes with particularly small contributions to the thermal expansion become excited such that the mean Grüneisen tensor decreases, and the thermal expansion tensor is kept at a very low level.

Whether specific phonons trigger positive or negative thermal expansion is then analyzed by comparing the evolutions of carefully chosen geometric descriptors as a function of (i) the applied strains and (ii) normal-mode coordinates of the respective phonons. Interestingly, only few of the relevant contributions stem from phonons at the center of the first Brillouin zone, Γ. This means that the essence of the anharmonicities in MOF-74 is missed, when relying only on Γ-point phonons during the calculation of the thermal expansion tensor. Overall, we identify several optical phonon branches that have an impact on (negative) thermal expansion. The main contributions in MOF-74, however, arise from acoustic phonons propagating in directions almost perpendicular to the pore. In those directions, the main contributions arise from the lowest transverse acoustic bands, which simultaneously favor a reduction of the pore diameter and an expansion of the pore length.

These results show that the thermal expansion in MOFs can be determined by a subtle interplay of a wide variety of factors, which becomes particularly complex in situations dominated by cancellation effects.

**Supporting Information**

Supporting Information is available below.

**Acknowledgements**

The authors acknowledge Vienna Scientific Cluster (VSC3) as well as the Graz University of Technology for the use of the HPC resources provided by the ZIP. Moreover, we thank Timo Müller, Barbara Puhr, and Andrew Jones from the company Anton Paar who provided the infrastructure and their know-how to conduct the temperature-dependent XRD measurements. Finally, the authors acknowledge Paolo Falcaro for his appreciated input and for providing the resources for the sample preparation.

# Supporting Information

# Understanding the origin of the particularly small and anisotropic thermal expansion of MOF-74

*Tomas Kamencek, Benedikt Schrode, Roland Resel, Raffaele Ricco, and Egbert Zojer\**

## Table of Contents

## S1. Data availability

The simulation input and output required to reproduce the results presented in the main text are publicly available in the NOMAD database.

<span style="color:red">https://</span>

(A final DIO will be placed here after the acceptance of the manuscript, considering that additional simulations might become necessary.)

## S2. Bayesian Analysis of the experimental x-ray diffraction measurements

As explained in the main text, the particularly small thermal expansion coefficients of MOF-74 in combination with the given experimental conditions (using a lab device and studying the sample in powder form instead of sufficiently large single crystals) was a significant obstacle for relying on more common means of analyzing the data of the x-ray diffraction (XRD) experiments. For that reason, a probabilistic approach relaying on Bayesian probability theory was employed. It is based on the general concepts which can, *e.g.*, be found in Ref. [1]. The mathematical concepts employed to calculate the probability density functions of the thermal expansion coefficients will be detailed in the following.

In the Bayesian approach, one considers conditional probabilities and probability distributions for random variables $x$ given a certain information $I$. To denote these conditional probabilities, we will use the notation $p(x|I)$. In particular, one is interested in probability distributions given the experimental information. To this end, one calculates (*a posteriori*) probability densities ("posterior probability"), $p(x|d)$, of a random variable, $x$, given the data, $d$. This can be achieved with Bayes' theorem using the *a priori* probability density ("prior"), $p(x)$, and the likelihood to observe the data, *p(d|x)*, with properly taking care for the normalization *via* $p(d) (= \int dx\, p(x)p(d|x))$:

$$p(x|d) = \frac{p(x)p(d|x)}{p(d)} \tag{S1}$$

Note that, technically, all the appearing probability distribution functions (PDFs) above and in the following are different mathematic functions, despite the consistent use of *p(x)* denoting a probability density as a function of *x*. However, for the ease of notation, we presume that different arguments of the probability density functions implicitly denote different PDFs.

In our case, we are looking for the probability density of the two independent lattice parameters, *a* and *c*, properly considering the data:

$$p(a,c|d) = \frac{p(a,c)p(d|a,c)}{p(d)} \tag{S2}$$

In order to use the above equation, one must first find the peak positions in the XRD data. To this end, we fit (pseudo) Voigt profiles to the measured data and extract their peak positions, *(2θ$_{hkl}$)$_{exp}$*. These peak positions act as the data, *d*, in equation (S2).

As a next step, we make use of the analytical expression for a general reciprocal lattice vector, $\vec{G_{hkl}}$, as a function of the two lattice parameters, *a* and *c* (from which the three lattice vectors are constructed in the following way: $\vec{a_1} = \left[a, \frac{a}{\sqrt{3}}, c\right], \vec{a_2} = \left[-a, \frac{a}{\sqrt{3}}, c\right]$ and $\vec{a_3} = \left[0, \frac{2a}{\sqrt{3}}, c\right]$):

$$||\vec{G_{hkl}}|| = 2\pi \sqrt{\frac{(h-k)^2}{4a^2} + \frac{(h+k-2l)^2}{12a^2} + \frac{(h+k+l)^2}{9c^2}} \tag{S3}$$

Using Bragg's law, an analytical expression for the peak positions, 2θ, can be derived:

$$2\theta_{hkl}(a,c) = 2\arcsin\left(\frac{\lambda}{2}\sqrt{\frac{(h-k)^2}{4a^2} + \frac{(h+k-2l)^2}{12a^2} + \frac{(h+k+l)^2}{9c^2}}\right) \tag{S4}$$

This directly yields the likelihood, *p(d|a,c)*, under the assumption that the deviation of the measured values, *(2θ$_{hkl}$)$_{exp}$*, from the true values, *2θ$_{hkl}$*, are normally distributed:

$$p(d|a,c;\sigma) = \left(\frac{1}{\sqrt{2\pi}\sigma}\right)^{N_{hkl}} \prod_{hkl}^{N_{hkl}} \exp\left(-\left(\frac{(2\theta_{hkl}(a,c) - (2\theta_{hkl})_{exp})^2}{2\sigma^2}\right)\right) \tag{S5}$$

However, the use of Equation (S5) requires a new parameter, σ, which is *a priori* unknown. In order to avoid a sensitiveness of the calculation on this parameter, one can use the marginalization rule and Jeffrey's prior normalized in the interval $[\sigma_1, \sigma_2]$ $(p(\sigma) = 1/\ln\left(\frac{\sigma_2}{\sigma_1}\right)\sigma^{-1})$ [1] to arrive at another likelihood function:

$$p(d|a,c) = \int_{\sigma_1}^{\sigma_2} d\sigma \; p(\sigma)\, p(d|a,c;\sigma) \tag{S6}$$

$$= \frac{1}{2\ln\frac{\sigma_2}{\sigma_1}} (\pi Q)^{-\frac{N_{hkl}}{2}} \left[\Gamma_{inc}\left(\frac{N_{hkl}}{2}, \frac{Q}{2\sigma_1^2}\right) - \Gamma_{inc}\left(\frac{N_{hkl}}{2}, \frac{Q}{2\sigma_2^2}\right)\right]$$

with
$$Q = \sum_{hkl}^{N_{hkl}} \left((2\theta_{hkl}(a,c) - (2\theta_{hkl})_{exp})^2\right) \tag{S7}$$

and
$$\Gamma_{inc}(x,b) = \int_0^b dt \; t^{x-1} \exp(-t) \tag{S8}.$$

This slightly altered form of the likelihood function allows to incorporate the fact that we do not know, how large the fluctuations of $(2\theta_{hkl})_{exp}$ around $2\theta_{hkl}$ are in practice. Therefore, we consider all fluctuations that result in standard deviations between $\sigma_1$ and $\sigma_2$, which we chose as 0.005° and 0.05°, respectively.

Finally, to evaluate equation (S2), one has to choose a prior probability density, which incorporates the preknowledge that we have about the lattice parameters. To consider our preknowledge, we chose gaussian priors for the lattice parameters according to the following expression (assuming that both are uncorrelated):

$$p(a,c|\mu_a,\mu_c,\sigma_a,\sigma_c) = \frac{1}{2\pi\sigma_a\sigma_c} \exp\left(-\frac{(a-\mu_a)^2}{2\sigma_a^2}\right) \exp\left(-\frac{(c-\mu_c)^2}{2\sigma_c^2}\right) \tag{S9}$$

In this prior, again, additional parameters appear, namely, the expectation values, $\mu_{a/c}$, and standard deviations, $\sigma_{a/c}$, of the Gaussian distributions of lattice parameters $a$ and $c$. From the DFT optimization, we know that the lattice parameters should have some values around 12.95 Å and 2.3 Å, which we use as the expectation values in the prior. Additionally, we consider the fact that we are not too sure about the precise values by using a rather broad density function, *i.e.*, large standard deviations of 0.05 Å.

Using the likelihood of equation (S6) and the prior of equation (S9) (as well as the proper normalization in the denominator), one can evaluate equation (S2) for each temperature separately and obtain the (joint) probability function (given the data), $p(a,c|d)$. From this two-dimensional density, the corresponding probability densities for lattice parameter $a$ and $c$ are evaluated by one-dimensional integration:

$$p(a|d) = \int dc\, p(a,c|d) \qquad (S10)$$

$$p(c|d) = \int da\, p(a,c|d) \qquad (S11)$$

Once one has calculated the probability functions for the two lattice parameters, one can derive the posterior PDF for the unit cell volume, $V$ ($=2\sqrt{3}\, a^2 c$), by:

$$\begin{aligned} p(V|d) &= \int da \int dc\, p(V|a,c,d)\, p(a|d)\, p(c|d) \\ &= \int da \int dc\, \delta(V - 2\sqrt{3}\, a^2 c)\, p(a|d)\, p(c|d) \\ &= \int da\, \frac{1}{2\sqrt{3}a^2}\, p(a|d)\, p\left(c = \frac{V}{2\sqrt{3}a^2}\bigg|d\right) \end{aligned} \qquad (S12)$$

Here, the conditional PDF $p(V|a,c,d)$ in the equation above is replaced by a delta distribution, $\delta(V - 2\sqrt{3}\, a^2 c)$, because the volume is known precisely without any uncertainty if both $a$ and $c$ are given.

These *a posteriori* PDFs (*i.e.*, $p(a|d)$, $p(c|d)$, and $p(V|d)$) are separately calculated for each temperature, at which the XRD data were obtained. In order to accelerate the calculation of the PDFs, we updated the prior expectation values (in equation (S9)) with the expectation values of $p(a|d)$ and $p(c|d)$ of the previous temperature step.

The procedure described above yields the *a posteriori* PDFs for the lattice parameters $a$ and $c$ (see Figure 2 in the main text) as well as the unit-cell volume $V$ displayed in Figure S 1. Already from the PDF of lattice parameter $a$ (see Figure S 1(a)) one can draw the conclusion that the associated thermal expansion should be negative. This impression is also supported by the drawn evolution of the expectation value of the PDF ($\langle a \rangle = \int da\, a\, p(a|d)$) with temperature. The observation that the actual values slightly fluctuate and do not show a very smooth temperature dependence, this can most probably be ascribed to the small magnitude of the observed changes in the lattice parameter (sub-pm changes).

For lattice parameter $c$ (see Figure S 1(b)), the trend is less clear. Although one can observe clear region in which $c$ increases with temperature, the data at and around 300 K seem to yield larger $c$-values than at the remaining temperatures. These outliers can most probably again be ascribed to the exceptionally small changes in $c$ being even smaller in (absolute) magnitude than for $a$.

Finally, one can clearly see that the PDFs of both lattice parameters are sharper at higher temperatures, while they are slightly more smeared-out at lower temperatures. The same observation can also be made for the PDF of the unit-cell volume in Figure S 1(c). Here, *V*, appears to stay at an essentially constant value, yet with an increased uncertainty (as a result of the error propagation). Additionally, the outliers of *c* at around 300 K impact also the PDF of *V*, which shows expectation values close to 300 K shifted to volumes slightly larger than the trend based on the remaining expectation values would suggest.

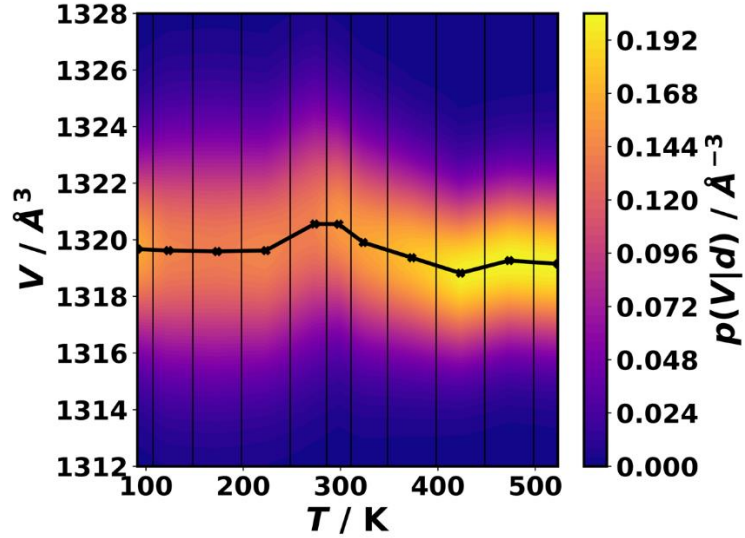

*Figure S 1: A posteriori probability density functions of the unit-cell volume V as a function of temperature. The probability density is encoded in the color scale of the plots. The black dots denote the calculated expectation values connected by the solid black lines.*

Within the presented Bayesian analysis, the posterior PDFs can be further used to calculate PDFs for the thermal expansion tensor elements, $\alpha_{11}$ and $\alpha_{33}$, describing the thermal expansion of *a* and *c*, respectively. In the following, we will briefly summarize the main ideas of the calculation for the thermal expansion coefficient of *a*, for which we will – for the ease of notation – omit the subscript 11, thus, referring only to $\alpha$. The same equation can directly be used for the thermal expansion coefficients of *c* and *V*.

In general, the posterior PDF for $\alpha$ at a given temperature, *T*, is given by the following integral:

$$\begin{aligned} p(\alpha|d;T) &= \int da \int da_0 \, p(\alpha|a, a_0, T, T_0) \, p(a|d;T) \, p(a_0|d;T_0) \\ &= \int da \int da_0 \, \delta\big(a - a_0 e^{\alpha(T-T_0)}\big) \, p(a|d;T) \, p(a_0|d;T_0) \\ &= \int da_0 \, p\big(a = a_0 e^{\alpha(T-T_0)}\big|d,T\big) \, p(a_0|d;T_0) \end{aligned} \quad \text{(S13)}$$

This equation is subsequently evaluated for all experimentally temperatures, $T_i$, where we always adapt $T=T_i$ and $T_0=T_{i-1}$. Note that we again replaced the conditional PDF $p(\alpha|a, a_0, T, T_0)$ with a delta distribution $\delta(a - a_0 e^{\alpha(T-T_0)})$ (as $\alpha$ is unambiguously determined if $a$ and $a_0$ at temperatures $T$ and $T_0$ are known). Because the temperature step, $(T-T_0)$, is always evaluated as the difference between two adjacent experimentally set temperatures, we assume $\alpha$ to be constant between $T$ and $T_0$. Without this assumption, the evaluation of the above integral would become much more complicated, as the delta distribution in this case would read $\delta\left(a - a_0 \exp\left(\int_{T_0}^{T} dT' \alpha(T')\right)\right)$ such that one would have to calculate PDFs of (ansatz) functions instead of mere real-valued numbers.

Using equation (S24) with the correct associated posterior PDFs, also the PDF for the thermal expansion coefficient of lattice parameter $c$ and the unit-cell volume $V$ can be calculated. The latter is shown in Figure S 2 together with the results from the Grüneisen theory of thermal expansion as shown in the main text (equivalent plots for $\alpha_{11}$ and $\alpha_{33}$ are already contained in the main text). Here, the increased width of the posterior PDF of $\alpha_V$ at around 300 K can be seen very clearly. Additionally, the uncertainty (due to error propagation) lies generally at a higher level than for the PDFs of the lattice parameters or the unit cell-volume.

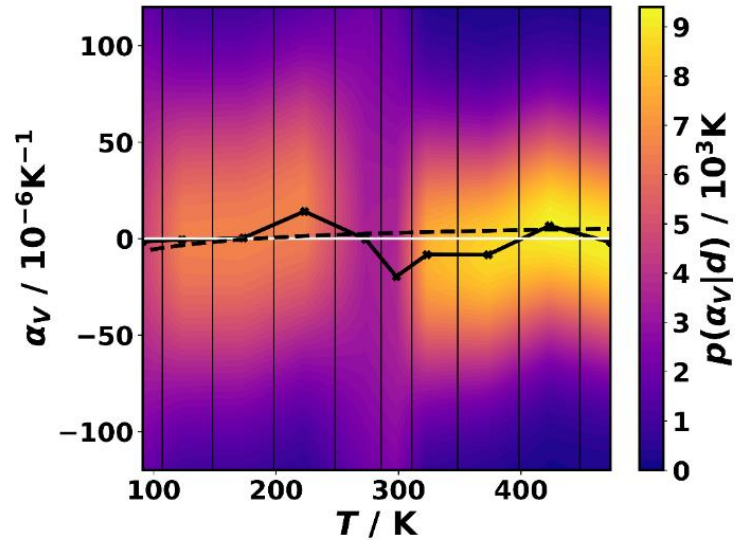

*Figure S 2: Calculated a posteriori probability distribution of the volumetric thermal expansion coefficient, $\alpha_V$, as a function of temperature. The probability density is encoded in the color scale of the plots. The black dots denote the calculated expectation values connected by the solid black lines. The horizontal white line is a guide to the eye and shows $\alpha_V = 0$. The black dashed line*

*corresponds to the volumetric thermal expansion coefficient calculated from the Grüneisen theory of thermal expansion as shown in the main text.*

### S2.1 Bayesian analysis considering the uncertainty in the sample temperature

Following the approach presented in Ref. [1], we also considered the case of the sample temperature being known only within a certain uncertainty. This uncertainty can be incorporated within Baysian probability theory by means of, *e.g.*, a Gaussian probability density to find the actual sample temperature, $T_s$, around the measured temperature, $T_i$, with a standard deviation, $\sigma_T$, which we chose as 20 K. In passing we note that such a treatment does not account for a possible systematic error in the determination of the sample temperature (like a too high or low measured temperature when heating (cooling) the sample), but considering such systematic errors goes beyond the scope of the present manuscript. In addition to the uncertainty in the temperature, also the lattice parameters, *a* and *c*, and the unit-cell volume, *V*, show an uncertainty, which we obtain from the posterior probability distributions of the respective quantities (see equations (S10), (S11), and (S12)) in terms of their variances, $\sigma_x^2$, at each measured temperature, $T_i$:

$$\sigma_{x,i}^2 = \int dx\, p(x|T_i, d)x^2 - \int dx\, p(x|T_i, d)x \quad \text{(S14)}$$

Here and in the following, the variable *x* is a placeholder for either *a*, *c*, or *V*. The used approach is presented for this general variable, *x*, and the associated thermal expansion coefficient, $\alpha_x$. Due to the uncertainty in temperature being relatively large in view of the small expected changes in *x*, we refrain from explicitly considering a temperature dependence of $\alpha_x$. Instead, we rely on the average thermal expansion coefficient, $\overline{\alpha_x}$, for temperatures between $T_0$ and $T$ according to the following functional dependence of *x* on the temperature:

$$x(T) = x_0 \exp(\overline{\alpha_x}(T - T_0)) \quad \text{(S15)}$$

In the equation above, $x_0$ is the value of *x* at $T_0$. For the sake of simplicity, $T_0$ has been chosen to be 0 K. Using the above model, one can obtain posterior distribution functions for the two parameters of the model, $x_0$ and $\overline{\alpha_x}$. To do so, one again uses Bayes' theorem relying on prior probability distribution functions for the two parameters (incorporating previous knowledge) and a likelihood distribution. Since both of the measured quantities (indexed with the data point index, $i$), $\{x_i\}$ and $\{T_i\}$, have a finite accuracy, the likelihood for such a problem has the following form: [1]

$$p(\{T_i\}, \{x_i\} \mid x_0, \overline{\alpha_x}, \{\sigma_{T,i}\}, \{\sigma_{x,i}\}) = \frac{1}{Z} \exp\left[-\frac{1}{2} \sum_i \Phi_i^*(T_i, x_i \mid x_0, \overline{\alpha_x}, \sigma_{T,i}, \sigma_{x,i})\right] \quad (S16)$$

In this equation, $Z$ is a normalization constant and the function $\Phi_i^*$ is the *minimal misfit* if the $i^{\text{th}}$ data point according to the following expression:

$$\Phi_i^*(T_i, x_i \mid x_0, \overline{\alpha_x}, \sigma_{T,i}, \sigma_{x,i}) = \frac{(x_i - x_i^*)^2}{\sigma_{x,i}^2} + \frac{(T_i - T_i^*)^2}{\sigma_{T,i}^2} \quad (S17)$$

with
$$T_i^* = \underset{T_i'}{\mathrm{argmin}} \left\{ \frac{(x_i - x_0 \exp(\overline{\alpha_x} T_i'))^2}{\sigma_{x,i}^2} + \frac{(T_i - T_i')^2}{\sigma_{T,i}^2} \right\} \quad (S18)$$

and
$$x_i^* = x(T_i^*) = x_0 \exp\{\overline{\alpha_x} T_i^*\} \quad (S19)$$

In other words, one has to find each temperature, $T_i^*$, that minimizes the misfit function for each data point ($x_i$, $T_i$). This is done by numerically solving equation (S18) (or, more precisely, by numerically finding the root of the derivative of the expression in braces with respect to $T_i'$) and inserting $T_i^*$ in the model for $x(T_i^*) = x_i^*$. Once the pair ($x_i^*$, $T_i^*$) is found, the minimal misfit, $\Phi_i^*$, can be calculated. Finally, the sum of the minimal misfits for all $i$ data points enters the exponent of the likelihood in equation (S16). Using the latter and the priors for $x_0$ and $\overline{\alpha_x}$, which we approximated as (broad) Gaussian distributions, one arrives at the joint posterior distribution for $x_0$ and $\overline{\alpha_x}$. The posterior distributions for the individual parameters are, then, obtained from an integration of the joint posterior over the respective other parameter (see above). The calculated expectation values and variances are listed in Table S 1, while Figure S 3 shows the respective posterior distribution functions and the comparison of the expected models to the measured data.

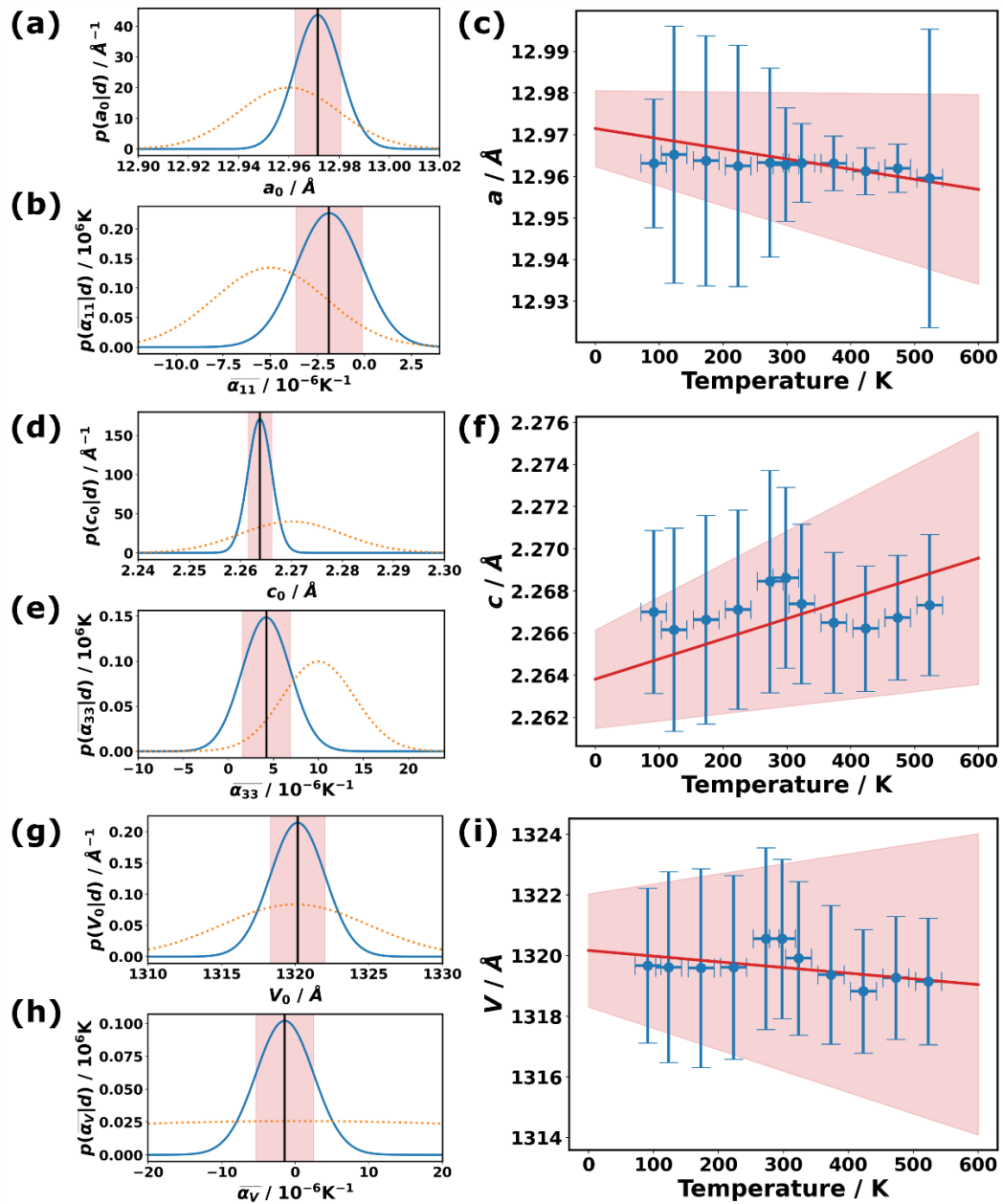

*Figure S 3: Comparison of the calculated posterior distributions for (a,d,g) the 0-K parameters $a_0$, $c_0$, and $V_0$, and (b,e,h) the associated average thermal expansion coefficients, $\overline{\alpha_{11}}$, $\overline{\alpha_{33}}$, and $\overline{\alpha_V}$, (blue solid lines) with their respective priors (orange dashed lines). The expectation values are denoted by the vertical black solid lines, while the red semitransparent areas indicate the intervals of ± the square root of the variances around the respective expectation values. (c,f,i) Measured values of a, c, and V, respectively, as a function of the setpoint temperature. The vertical errorbars correspond to the square root of the associated posterior variances, while the horizontal*

ones were set to a standard deviation of 20 K. The red line denotes the model function (see equation (S15)) evaluated for the expectation values of the parameters shown in the panels in the left column. The semitransparent red areas around that line denote the possible model functions within one standard deviations of the model parameters.

Table S 1: Evaluated model parameters $x_0$ and $\overline{\alpha_x}$ (see equation (S15)) for $x$ = a, c, and V. The parameters are given in the following format: expectation value ± (variance)$^{1/2}$

| $x$ | $x_0$ / Å$^{(3)}$ | $\overline{\alpha_x}$ / $10^{-6}$ K$^{-1}$ |
|---|---|---|
| a | 12.971 ± 0.009 | -2 ± 1 |
| c | 2.264 ± 0.002 | 4 ± 2 |
| V | 1320 ± 2 | -1 ± 4 |

## S3. Theoretical details for the *ab initio* calculation of thermal expansion tensors

### S3.1 General relations and properties of the thermal expansion tensor

Prior to recapitulating some general physical considerations on thermal expansion, a definition of the thermal expansion tensor and its relation to the volumetric thermal expansion coefficient are briefly given. Throughout this document, we will use the Einstein convention of implicitly carrying out summations over indices that occur multiple times on the same side of an equation. In general, the thermal expansion tensor of a solid is defined as the change in strain with temperature at constant stress, $\sigma$, (or pressure) [2–6]:

$$\alpha_{ij} = \left.\frac{\partial \varepsilon_{ij}}{\partial T}\right|_\sigma \tag{S20}$$

This means that, to first order, the lattice vectors, $\vec{a_i}$ (with $i$=1,2,3), of an arbitrary crystal can be written as a matrix product involving the lattice vectors at a reference temperature and the (small) temperature step, $\delta T$, with $\delta_{jk}$ being the Kronecker delta:

$$(\vec{a_i})_j = (\delta_{jk} + \varepsilon_{jk})\left(\vec{a_i^0}\right)_k = (\delta_{jk} + \alpha_{jk}\delta T)\left(\vec{a_i^0}\right)_k \tag{S21}$$

Essentially, $(\vec{a_i})_j$ and $\left(\vec{a_i^0}\right)_k$ can be written as matrices, $A_{ji}$ and $A^0{}_{ki}$, such that the determinant on both sides of the equation can be calculated. With $\det(A_{ji})=V$ and $\det(A^0{}_{ki})=V_0$, i.e., the strained and unstrained unit-cell volumes, one obtains:

$$V = \det(\mathbb{1} + \boldsymbol{\alpha}) V_0 \, \delta T = V_0 (1 + \alpha_V \, \delta T) \tag{S22}$$

The volumetric thermal expansion coefficient, $\alpha_V$, is, thus, equivalent to the determinant of $(\mathbb{1} + \boldsymbol{\alpha})$. Especially in the case of diagonal thermal expansion tensors (with small elements), the approximation $\det(\mathbb{1} + \boldsymbol{\alpha}) \approx tr(\boldsymbol{\alpha})$ holds. *I.e.*, for a rhombohedral material, as MOF-74, for which the thermal expansion tensor (and all rank-2 materials-property tensors) reduces to the following diagonal form [7], the volumetric thermal expansion tensor is given by equation (S24).

$$\boldsymbol{\alpha} = \begin{pmatrix} \alpha_{11} & 0 & 0 \\ 0 & \alpha_{11} & 0 \\ 0 & 0 & \alpha_{33} \end{pmatrix} \tag{S23}$$

$$\alpha_V = \det(\mathbb{1} + \boldsymbol{\alpha}) - 1 = (1 + \alpha_{11})^2 (1 + \alpha_{33}) - 1 \approx 2\alpha_{11} + \alpha_{33} \tag{S24}$$

Several more properties and consequences of the thermal expansion tensor can be understood and pointed out from statistical physics. Moreover, these thermodynamic considerations important for the practical approaches to calculate the thermal expansion coefficients. The thermodynamic ensemble which most closely resembles typical experimental conditions, is the Gibbs ensemble. The associated thermodynamic potential is the Gibbs free enthalpy, $G$, which is (in the absence of any electromagnetic fields) a function of the externally applied stress tensor, $\sigma_{ij}$, and the temperature, $T$ [2,3,6,8]. The following equation relates $G$ to the internal energy of the system, $U$, the (Helmholtz) free energy, $F$, and its entropy, $S$:

$$G(T, \boldsymbol{\sigma}) = F - V \varepsilon_{ij} \sigma_{ij} = U - TS - V \varepsilon_{ij} \sigma_{ij} \approx E_{elec} + F_{ph} - V \varepsilon_{ij} \sigma_{ij} \tag{S25}$$

The last approximate equality in the equation above splits the total free energy of the system, $F$, into a phonon contribution, $F_{ph}$, which includes the phonons' entropy, and an electronic energy, $E_{elec}$. The latter is, in fact, only the internal energy of the electrons and does not consider temperature effects, as electronic entropies in insulators are typically negligible [2,3,9]. Additionally, in density-functional theory, electronic entropies are not accessible.

Based on the (exact) expression for $G$ above, the total derivative of it, $dG$, can be written as shown in the following equation.

$$dG(T, \boldsymbol{\sigma}) = dU - SdT - V \varepsilon_{ij} d\sigma_{ij} \tag{S26}$$

$$S = -\left.\frac{\partial G}{\partial T}\right|_{\sigma} \quad ; \quad \varepsilon_{ij} = -\frac{1}{V} \left.\frac{\partial G}{\partial \sigma_{ij}}\right|_T \tag{S27}$$

Mathematically, the entropy and the strain tensor can, thus, be related to the partial derivatives of $G$ (leading to the Maxwell relations in the Gibbs ensemble). These can be exploited to show that the zero-temperature limit of the thermal expansion tensor equals zero due to the third law of thermodynamics:

$$\alpha_{ij} = \left.\frac{\partial \varepsilon_{ij}}{\partial T}\right|_\sigma = -\frac{1}{V}\frac{\partial^2 G}{\partial \sigma_{ij} \partial T} = \frac{1}{V}\left.\frac{\partial S}{\partial \sigma_{ij}}\right|_T \xrightarrow[T\to 0]{} 0 \tag{S28}$$

The limit at $T\to 0$ follows from the fact that the entropy must vanish at 0 K such that also the stress-derivative (at constant temperature) must vanish.

An interesting consequence of the thermal expansion of a crystal is the difference in its heat capacity at constant pressure, $C_p$, (or constant stress, $C_\sigma$) and its heat capacity at constant volume, $C_V$ (or constant strain, $C_\varepsilon$). The following derivation of their difference is based on the similar derivation for the isotropic (cubic) case found in Ref. [2].

In general, the heat capacity at a constant variable, $X$, is defined as the temperature-derivative of the internal energy with $X$ held constant, which can also be expressed *via* the temperature-derivative of the entropy:

$$C_X = \left.\frac{\partial U}{\partial T}\right|_X = T\left.\frac{\partial S}{\partial T}\right|_X \tag{S29}$$

In the Gibbs ensemble, the thermodynamic variables are the temperature and the externally applied stress tensor. For these variables, the total derivative of $S$ can be written as:

$$dS = \left.\frac{\partial S}{\partial T}\right|_\sigma dT + \left.\frac{\partial S}{\partial \sigma_{ij}}\right|_T d\sigma_{ij} \tag{S30}$$

Differentiating this equation with respect to $T$ at constant strain yields the following expression.

$$\left.\frac{\partial S}{\partial T}\right|_\varepsilon = \left.\frac{\partial S}{\partial T}\right|_\sigma + \left.\frac{\partial S}{\partial \sigma_{ij}}\right|_T \left.\frac{\partial \sigma_{ij}}{\partial T}\right|_\varepsilon \tag{S31}$$

The term on the left-hand side and the first term on the right-hand side can already be expressed as the desired heat capacities at constant strain (volume) and stress (pressure), respectively. Additionally, the stress-derivative of the entropy (at constant temperature) can be rewritten using equation (S28) such that the temperature-derivative of the stress tensor at constant strain remains the only unknown factor in the equation:

$$\frac{C_\varepsilon}{T} = \frac{C_\sigma}{T} + \left.\frac{\partial S}{\partial \sigma_{ij}}\right|_T \left.\frac{\partial \sigma_{ij}}{\partial T}\right|_\varepsilon = \frac{C_\sigma}{T} + V\, \alpha_{ij} \left.\frac{\partial \sigma_{ij}}{\partial T}\right|_\varepsilon \tag{S32}$$

This partial derivative can be easily evaluated with a common trick in statistical physics. As we want to evaluate the derivative at constant strain, we start with the total derivative of the strain tensor in the Gibbs ensemble.

$$d\varepsilon_{ij} = \left.\frac{\partial \varepsilon_{ij}}{\partial T}\right|_{\sigma} dT + \left.\frac{\partial \varepsilon_{ij}}{\partial \sigma_{kl}}\right|_{T} d\sigma_{kl} \tag{S33}$$

At constant strain $\varepsilon_{ij} = \text{const}\ (i.e., d\varepsilon_{ij} = 0)$, the equation above can be rearranged such that the desired temperature-derivative of the stress tensor at constant temperature can be calculated:

$$-\left.\frac{\partial \sigma_{kl}}{\partial \varepsilon_{ij}}\right|_{T} \left.\frac{\partial \varepsilon_{ij}}{\partial T}\right|_{\sigma} = \left.\frac{\partial \sigma_{kl}}{\partial T}\right|_{\varepsilon} \tag{S34}$$

The two partial derivatives on the left hand-side can be identified with the elastic tensor, $C_{klij}$, and the thermal expansion tensor

$$-C_{klij}\, \alpha_{ij} = \left.\frac{\partial \sigma_{kl}}{\partial T}\right|_{\varepsilon} \tag{S35}$$

Inserting this expression in equation (S32) yields equation (S36).

$$C_{\sigma} = C_{\varepsilon} + T\, V\, \alpha_{ij} C_{ijkl} \alpha_{kl} \tag{S36}$$

The difference between the heat capacity at constant pressure and the one at constant volume is, therefore, proportional to the product $\alpha_{ij} C_{ijkl} \alpha_{kl}$. I.e., the thermal expansion tensor appears in this product to second order such that for systems for which the thermal expansion is already very low (like for MOF-74), this product becomes very small, as emphasized in Figure S 4. The observation that both heat capacities become essentially identical, has a relevant consequence for MOF-74: the (phonon contribution to the) heat capacity at constant pressure (obtained from experiments) and at constant volume (typically obtained from phonon band structure calculations) are directly comparable due to the exceptionally small thermal expansion coefficients. In other words, one does not necessarily have to compute the much more challenging heat capacity at constant pressure if one wants to compare this property to experimental measurements, but already the much easier to obtain heat capacity at constant volume will be a good estimator.

We note, in passing, that the obtained general result in equation (S36) is essentially identical to the equivalent solution for isotropic (cubic) solids [2]. However, in that case, the volumetric thermal expansion (squared) and the bulk modulus, $B$, enter the difference between $C_p$ and $C_V$:

$$C_p = C_V + T\, V\, \alpha_V^2 B \tag{S37}$$

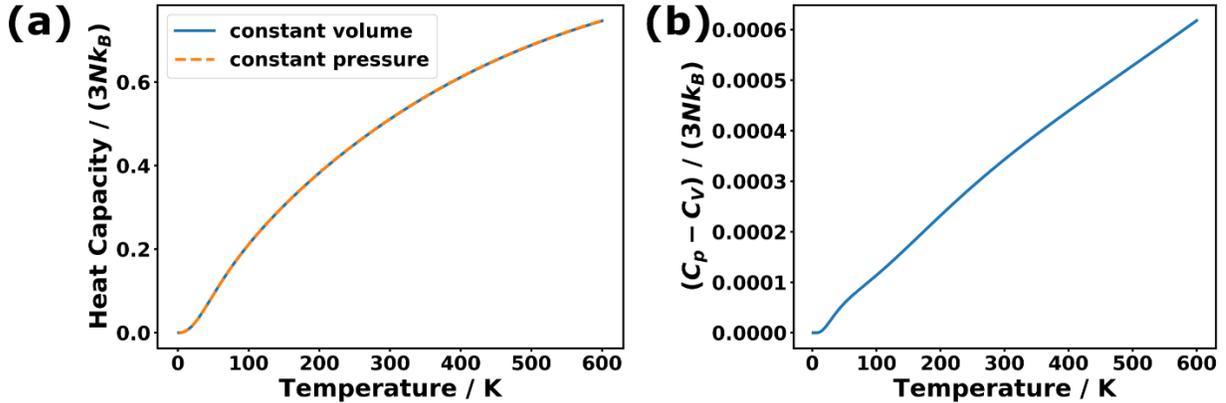

*Figure S 4: (a) Comparison of the heat capacity at constant pressure (stress), $C_p$, with the heat capacity at constant volume (strain), $C_V$, of MOF-74 as a function of temperature. $C_p$ was calculated from equation (S36) using the thermal expansion tensor calculated from the Grüneisen theory of thermal expansion as presented in the main text. (b) Difference of $C_p$ and $C_V$ as a function of temperature. In both panels, the heat capacities are normalized with $3Nk_B$, with $3N$ being the number of phonon bands in the crystal and $k_B$ being the Boltzmann constant.*

### S3.2  General concepts of the quasi-harmonic approximation

A general problem when trying to calculate the thermal expansion tensor of a solid from *ab initio* methods is the limited capability to sample the Gibbs ensemble (*i.e.*, the situation for a constant temperature and stress tensor applied to the systems). The quasi-harmonic approximation (QHA) is a workaround to overcome this limitation based on calculations entirely carried out at ~0 K, with the temperature dependence incorporated afterwards *via* the suitable thermodynamic expressions.

In the Gibbs ensemble, the strain tensor would always adapt to the externally set stress tensor and temperature (resulting in the thermal expansion of the system). As one does not carry out calculations in the Gibbs ensemble in the QHA, only "non-equilibrium" Gibbs free enthalpies, $G'$, can be calculated with a fixed rather than a free strain tensor. This means that the real Gibbs free enthalpy is, then, recovered by taking the minimum of the non-equilibrium Gibbs free enthalpies with respect to the strain tensor:

$$G(T, \boldsymbol{\sigma}) = \min_{\boldsymbol{\varepsilon}}\{G'(T, \boldsymbol{\sigma}, \boldsymbol{\varepsilon})\} = \min_{\boldsymbol{\varepsilon}}\{E_{elec}(\boldsymbol{\varepsilon}) + F_{ph}(\boldsymbol{\varepsilon}, T) - V\,\varepsilon_{ij}(T)\,\sigma_{ij}\} \qquad (S38)$$

This minimization is separately carried out at each temperature. The strain tensor that minimizes $G'$ at each temperature yields the temperature dependence of the (equilibrium) strain tensor and, thus the thermal expansion.

In practice, one, must, thus, carry out several calculations of the total electronic energy, $E_{elec}$, and the phonon free energy, $F_{ph}$, with a number of strained unit cells, from which the minimum Gibbs free enthalpy (with respect to the applied strain tensors) can be evaluated. This procedure captures, in part, the anharmonic nature of the phonons in the crystal, whose frequencies are implicitly dependent on the applied strain tensor [10].

Alternatively, one can formulate the equation above as a minimization with respect to the lattice parameters, which would, in the case of MOF-74, correspond to $a$ and $c$:

$$G(T, \boldsymbol{\sigma}) = \min_{a,c}\{G'(T, \boldsymbol{\sigma}, a, c)\} = \min_{a,c}\{E_{elec}(a,c) + F_{ph}(a,c,T) - V\,\varepsilon_{ij}(a,c,T)\,\sigma_{ij}\} \qquad (S39)$$

Regardless of the chosen space in which the minimization is carried out, one will typically have to fit an analytical function to the sampled $G'$ values in the relevant parameter space. Usually, the model functions used here are so-called equations of state (EoSs), such as the Birch-Murnaghan [11,12] EoS) which has its origin in a Taylor expansion of the free enthalpy with respect to finite strains) or the Rose-Vinet EoS [13]. We note, in passing, that in those equations of state, also the choice of the calculation of the strain is a relevant parameter: infinitesimal strain or finite strain (Eulerian or Lagrangian), with Eulerian finite strain being typically the best choice for large strains [14,15].

Unfortunately, those and other well-established equations of state are only available in an explicit form for cubic crystals, for which the minimization space of equation (S38) becomes one-dimensional because the only parameter is the unit-cell volume, significantly facilitating the minimization procedure. For crystals with non-cubic symmetry, the fitting and minimization procedure is not that straightforward. Thus, in the following, several QHA-based approaches are presented to calculate the thermal expansion tensor in MOF-74.

### S3.3 Quasi-harmonic approximation I: considering only the change in unit-cell volume

The simplest QHA-approach relies on well-established (one-dimensional) equations of state for optimizing $G'$. In a first step, one optimizes (at constant unit-cell volume) all relative magnitudes of the lattice parameters based only on electronic energies. In a second step, a one-dimensional Rose-Vinet EoS is used to find the volume that minimizes non-equilibrium Gibbs free enthalpies. Mathematically, this two-step procedure can be defined for MOF-74 as follows:

$$\zeta^*(V) = \underset{\zeta=c/a}{\mathrm{argmin}}\{E_{elec}(V,\zeta)\} \tag{S40}$$

$$G(T,\boldsymbol{\sigma}) = \min_V\{G'(T,\boldsymbol{\sigma},V,\zeta^*(V))\} = \min_V\{E_{elec}(V,\zeta^*(V) + F_{ph}(V,\zeta^*(V),T) \\ - V\,\varepsilon_{ij}(V,\zeta^*(V),T)\,\sigma_{ij}\} \tag{S41}$$

In practice, one chooses a set of unit-cell volumes of the system of interest and optimizes the crystallographic lattice parameters under the constraint that the volume must remain constant. With the *VASP* code, this can, *e.g.*, be done with the `ISIF=4` option during a geometry optimization. As a result, one has optimized ratios of lattice constants (in the case of MOF-74, $\zeta = c/a$) for various volumes. This procedure is computationally relatively easy to follow because no (expensive) phonon band structure calculations must be computed at this point, as only the electronic energy is considered in the minimization in equation (S40).

As a second step, one calculates phonon band structures for all the unit cells with (electronically) optimized $\zeta$ but different volumes to be able to fit established equations of state to the non-equilibrium Gibbs free enthalpies, $G'$, to find that volume which minimizes $G'$ at each temperature. This yields $V(T)$ and $\zeta(V(T))=\zeta(T)$, from which the thermal expansion tensor can be calculated. This procedure has, *e.g.*, been used in Ref. [16] to calculate the thermal expansion in various organic semiconductor crystals.

This procedure, which will be referred to as QHAiso in the following, has been applied to MOF-74 (with the PBEsol functional and the numeric settings described in Section S4). Figure S 5 shows the case for zero applied stress (pressure), in which the Gibbs free enthalpy equals the free energy: nine unit cells with (electronically) optimized *c/a*-ratio were used to calculate the total (non-equilibrium) free energy as a function of temperature and volume. The volumes that minimize those non-equilibrium free energies and, therefore, yield the free energy in thermodynamic equilibrium are marked as the green crosses in Figure S 5. One can see that these volumes shift to higher values for higher temperatures, *i.e.*, a positive (volumetric) thermal expansion is expected. Interestingly, already at 0 K, the optimal volume when considering the total free energy notably differs from the volume when only considering the electronic energy. This means that the phonon contribution to the total free energy at low temperatures (*i.e.*, the zero-point energies of the harmonic oscillators at 0 K) affects the predicted low-temperature volume.

Figure S 6 confirms that the QHAiso approach indeed yields a positive volumetric thermal expansion coefficient. In combination with the information from the electronic energy of how the *c/a*-ratio changes as a function of the unit cell volume, also the lattice parameters *a* and *c* and the associated thermal expansion tensor elements $\alpha_{11}$ and $\alpha_{33}$ can be calculated. Figure S 6 shows that in this approach, the thermal expansion of *a* does not meet the expectations from the XRD measurements as it shows a small but significant positive thermal expansion coefficient. Also $\alpha_{33}$ is much larger than expected based on experiments and the data of the Grüneisen theory presented in the main text. Therefore, there is no cancellation effect between $\alpha_{11}$ and $\alpha_{33}$, and $\alpha_V$ is distinctly positive in this approach.

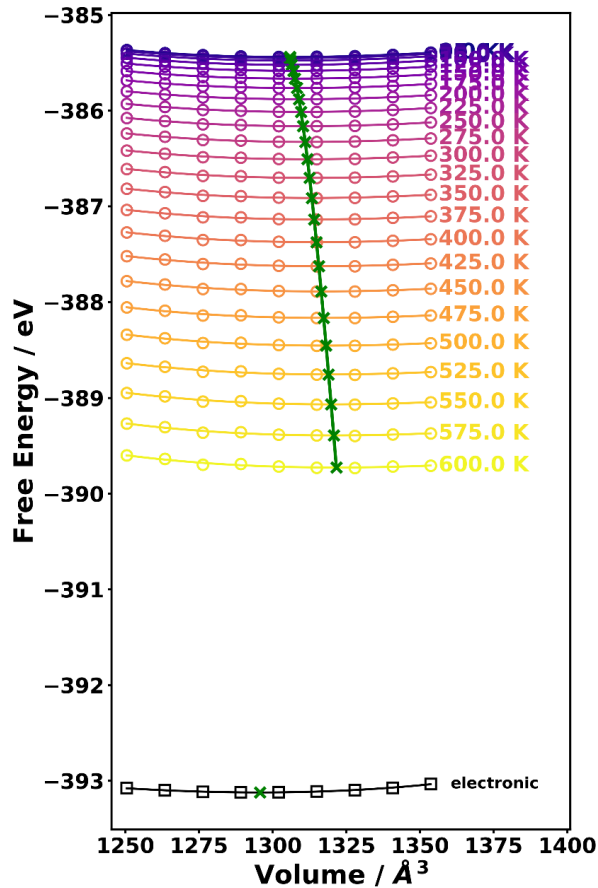

*Figure S 5: Total free enthalpy at zero applied stress (=free energy) of MOF-74 as a function of the unit-cell volume (abscissa) and the temperature (color scale and labels) obtained with the QHAiso approach. The open symbols denote the total (i.e., electronic and phonon) free energies calculated as a function of the unit-cell volume with (electronically) optimized c/a-ratio, while the solid lines correspond to the fitted Rose-Vinet EoS at each temperature. The green crosses mark the minimum free energy and the volume that minimizes the free energy at each temperature. The temperature dependence of those marked volumes yields the volumetric thermal expansion. Additionally, the electronic energies (together with a fitted Rose-Vinet EoS) are shown.*

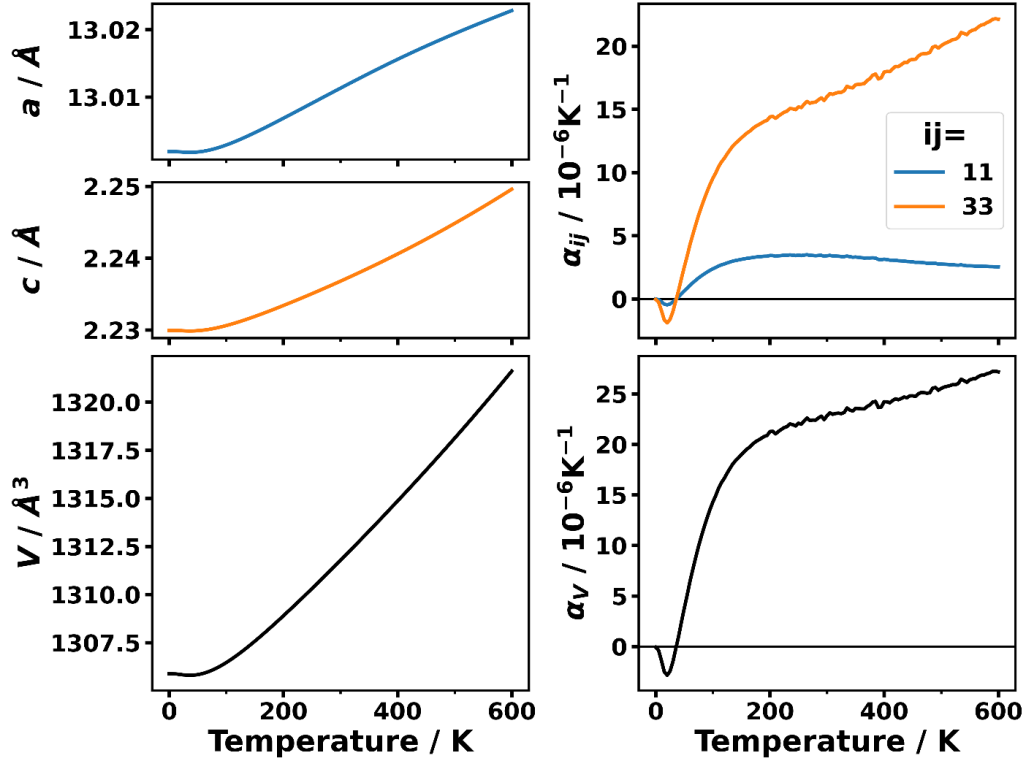

*Figure S 6: Temperature dependence of the two independent lattice parameters, a and c, and the unit-cell volume, V, of MOF-74 calculated with the QHA-procedure described in Section S3.3 (QHAiso). The right column shows the associated thermal expansion coefficients, $\alpha_{11}$, $\alpha_{33}$, and the volumetric thermal expansion coefficient $\alpha_V$.*

It seems as if the implicit assumption in the QHAiso approach, namely that the *c/a*-ratio is only an explicit function of the unit cell volume from a purely electronic calculation, but not of temperature, is not justified well.

### S3.4 Quasi-harmonic approximation II: considering the ratio of lattice parameters and subsequently the unit-cell volume

As an improvement of the QHAiso approach, one can explicitly consider the temperature dependence of $\zeta=c/a$. As shown in Ref. [17], this approach, which we will refer to QHA2fitiso, can improve the quality of *ab initio* calculations of the thermal expansion for certain non-cubic system.

In contrast to the QHAiso approach described in Section S3.3, the QHA2fitiso needs a second fit (hence, QHA"2fit"iso) of the free enthalpy considering the temperature dependence explicitly. This is possible if $\zeta^*$ is not taken as the *c/a*-ratio that minimizes the electronic energy only, but the total free enthalpy (or free energy for zero applied stress). Mathematically, the QHA2fitiso approach can be described as:

$$\zeta^*(V,T) = \underset{\zeta=c/a}{\mathrm{argmin}}\{G'(T,\boldsymbol{\sigma},V,\zeta)\} \tag{S42}$$

$$G(T,\boldsymbol{\sigma}) = \min_V\{G'(T,\boldsymbol{\sigma},V,\zeta^*(V,T))\} = \min_V\{E_{elec}(V,\zeta^*(V,T) + F_{ph}(V,\zeta^*(V,T),T) - V\,\varepsilon_{ij}(V,\zeta^*(V,T),T)\,\sigma_{ij}\} \tag{S43}$$

Unfortunately, this approach comes at increased computational cost: in order to find the optimal *c/a*-ratio, at each temperature and volume one must carry out more than one phonon band structure calculation. In fact, to fit a polynomial of $n^{th}$ order, one must calculate *(n+1)* phonon band structures for unit cells with varying *c/a*-ratio for each volume.

For the results obtained with the QHA2fitiso approach, we used second-order polynomials to find the temperature- and volume-dependent optimal *c/a*-ratio. *I.e.*, we calculated (at least) three complete phonon band structures for (at least) three different *c/a*-ratios at each of the nine volumes of the QHAiso approach.

At each of the volumes, for these three *c/a*-ratios the non-equilibrium free energy was calculated and fitted to a second order polynomial as a function of $\zeta=c/a$. From this fit, the optimum *c/a*-ratio as well as the minimal free energy, $F^*$, were found. The latter was subsequently used for the fits of Rose-Vinet equations of state to the $F^*$-*vs.*-*V*-data. Interestingly, the fitted $\zeta^*$ values show notable variations with temperature depending on the unit-cell volume. This is illustrated in Figure S 7. At smaller volumes, $\zeta^*$ decreases with temperature, at slightly higher ones, $\zeta^*$ is nearly temperature-independent, while it increases with temperature for larger volumes. At even larger volumes, $\zeta^*$ appears to become mostly temperature-independent again.

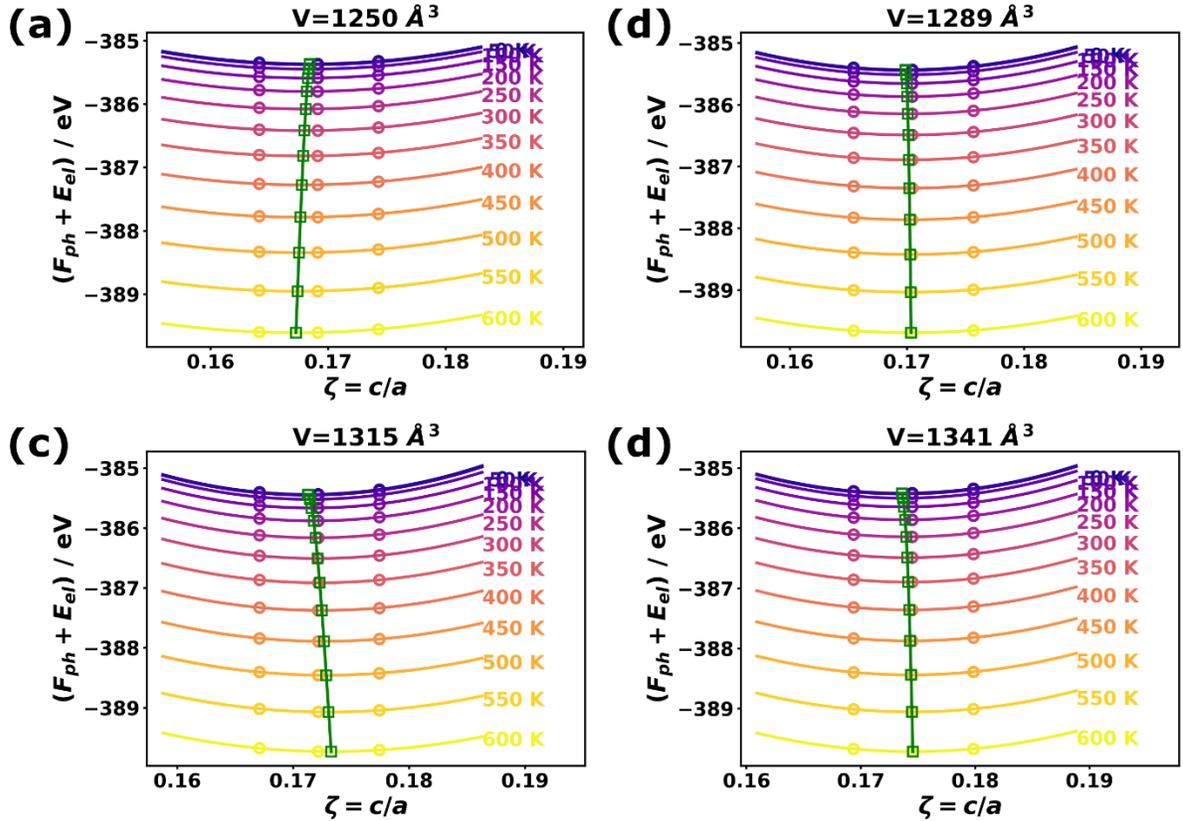

*Figure S 7: Total free enthalpy at zero applied stress (=free energy) of MOF-74 as a function of the c/a-ratio, ζ, for various temperatures (color scale) and volumes (panels (a)-(d)) obtained with the QHA2fitiso approach. The solid lines fitted to the three data points at each volume and temperature correspond to second order polynomials. The positions of the minima of these polynomials, ζ*, are highlighted by open green squares. The evolutions of the ζ* with temperature at each volume show that the optimum c/a-ratio (a) decreases with temperature for small volumes, (b) is nearly temperature-independent at intermediate volumes, (c) increases for large volumes, and (d) becomes hardly temperature-dependent at even larger volumes again.*

When the free energies minimized with respect to ζ are used for the fits with a (one-dimensional) Rose-Vinet EoS as function of the unit-cell volume, a similar positive volumetric thermal expansion is observed as shown in Figure S 8 as with the above-described approach shown in Figure S 5.

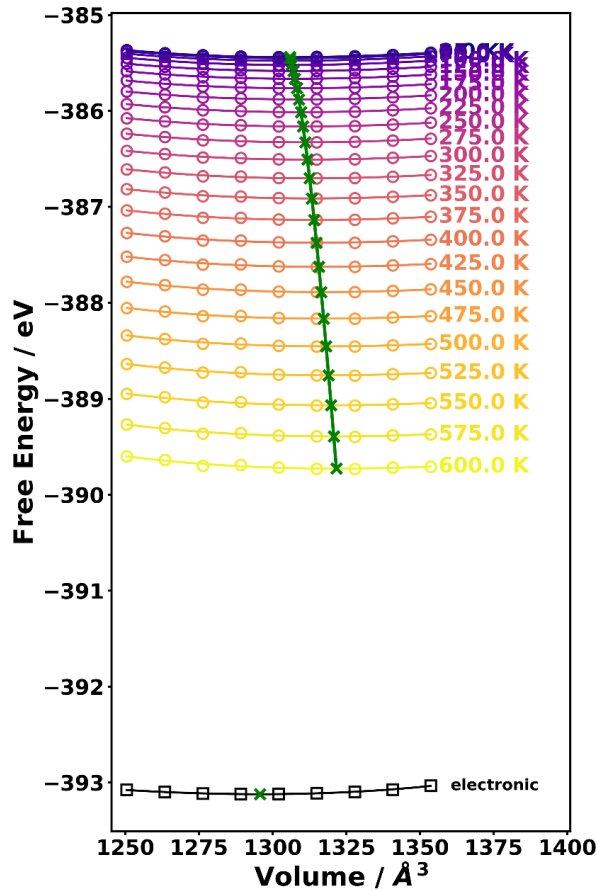

*Figure S 8: Total free enthalpy at zero applied stress (=free energy) of MOF-74 as a function of the unit-cell volume (abscissa) and the temperature (color scale and labels) obtained with the QHA2fitiso approach. The open symbols denote the total (i.e., electronic and phonon) free energies, which were found from minimizing the total free energies at constant volume with respect to the c/a-ratio, as a function of the unit-cell volume. The solid lines correspond to the fitted Rose-Vinet EoS at each temperature. The green crosses mark the minimum free energy and the volume that minimizes the free energy at each temperature. The temperature dependence of those marked volumes yields the volumetric thermal expansion. Additionally, the electronic energies (together with a fitted Rose-Vinet EoS) are shown.*

When analyzing the thermal expansion tensor obtained with the QHA2fitiso approach (see Figure S 9), a significant qualitative improvement concerning the anisotropy can be found. Now, $\alpha_{11}$ is negative and lies in the same order of magnitude as the $\alpha_{11}$ element obtained with the Grüneisen theory of thermal expansion. Also the thermal expansion of lattice parameter $c$, $\alpha_{33}$, shows the correct (positive) sign, although its magnitude is still too large compared to the results from Grüneisen theory by about a factor of 3. Although cancellation effects between $\alpha_{11}$ and $\alpha_{33}$, in principle, occur, the latter is overestimated too severely to reduce the volumetric thermal expansion to a level comparable to the experiments or the Grüneisen theory.

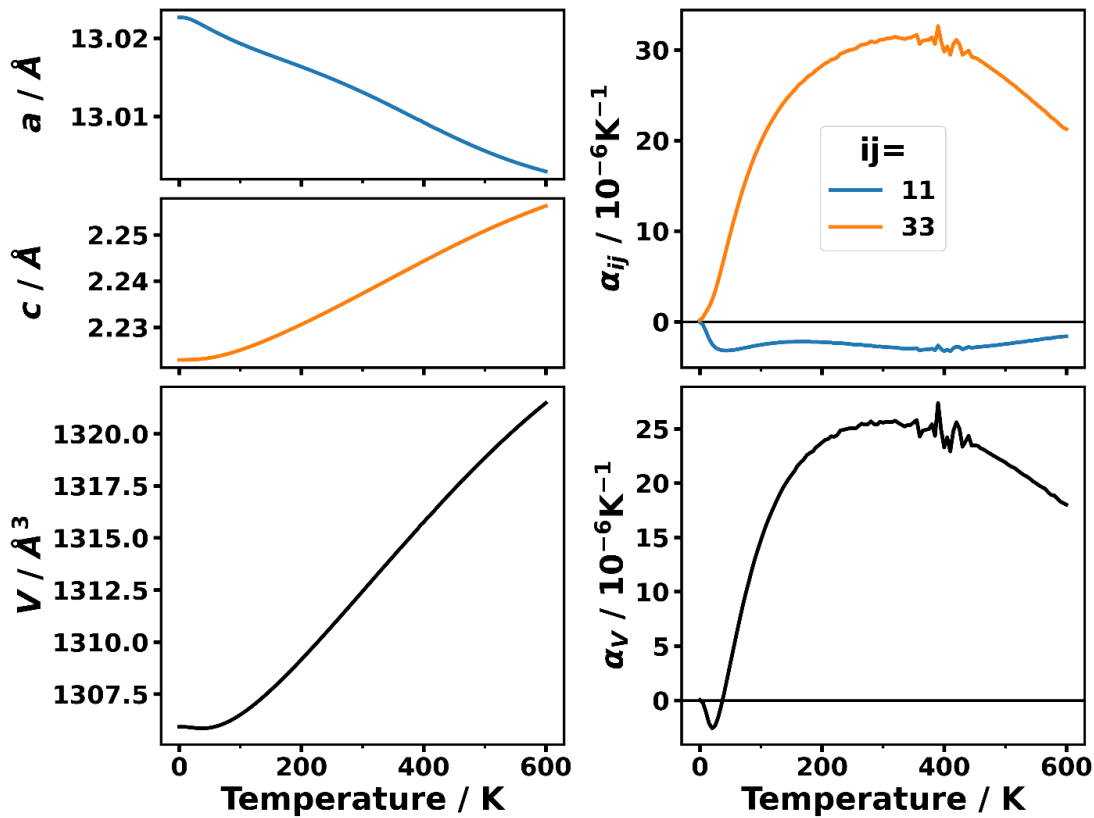

*Figure S 9: Temperature dependence of the two independent lattice parameters, a and c, and the unit-cell volume, V, of MOF-74 calculated with the QHA-procedure described in Section S3.4 (QHA2fitiso). The right column shows the associated thermal expansion coefficients, $\alpha_{11}$, $\alpha_{33}$, and the volumetric thermal expansion coefficient $\alpha_V$.*

### S3.5 Quasi-harmonic approximation III: considering all independent lattice parameters simultaneously

Although the two-step fitting procedure of the QHA2fitiso approach was able to achieve a certain improvement, still the step-by-step fitting of (i) the *c/a*-ratio and (ii) the unit-cell volume might result in different results than when directly looking for the combination of lattice parameters, *(a,c)*, that minimizes the free enthalpy at each temperature. To this end, following the suggestions of Ref. [4], several attempts were made to fit the (non-equilibrium) free enthalpies directly in the two-dimensional space of lattice parameters in analogy to equation (S38). This procedure will be referred to as the QHAaniso approach in the following. To stabilize such a computationally more demanding fit of (non-equilibrium) free energies (at constant stress and temperature) as a function of the lattice parameters, the full phonon band structures for 41 *(a,c)*-combinations were calculated. The considered *(a,c)*-combinations are shown in Figure S 10(a).

In this QHAaniso approach, the difficulty lies in efficiently finding the *(a,c)*-combination which minimizes the free enthalpy at each temperature. An obvious solution to this problem is to fit two-dimensional functions to the non-equilibrium free enthalpies (or free energies at zero applied pressure) and to find the *(a,c)*-combination which minimizes this fit functions. Thus, in the QHAaniso approach, one attempts to fit two-dimensional functions to the data points at each temperature (and stress) and monitors how the equilibrium configurations *(a,c)* change as a function of temperature. This is exemplarily shown in Figure S 10(b) for using third-order polynomials in the variables *a* and *c* to fit the calculated non-equilibrium free enthalpies.

However, since, to the best of our knowledge, no equations of state such as the well-established one-dimensional equations of state mentioned above, exist for non-cubic materials, the choice of the fit functions is somewhat problematic. One possible choice would be a (two-dimensional) Taylor series in the lattice parameters *a* and *c* according to equation (S44), which shows the general expression of a polynomial of $N^{th}$ order relying on the fit parameters, $g_{ij}$. Depending on the chosen order, $N$, one obtains different results when applying the fitting and minimization procedure as shown for second to fifth order in Figure S 11 to Figure S 14.

$$G_{fit}^N(a,c) = \sum_{n=0}^{N} \sum_{i=0}^{n} \frac{1}{n!} \binom{n}{i} g_{n-i,i} a^{n-i} c^i \tag{S44}$$

Another choice for the fit function is shown in equation (S45). While such a fit function would again be a polynomial of $N^{th}$ order, the expansion is carried out in terms of Eulerian (subscript "E") or Lagrangian (subscript "L") finite strains. Such a model function would be the logical continuation of the Birch-Murnaghan EoS to more than one variable, relying on fit parameters, $p_{ij}$.

$$G_{fit,E/L}^N\left(\varepsilon_{a_{E/L}}, \varepsilon_{c_{E/L}}\right) = \sum_{n=0}^{N} \sum_{i=0}^{n} \frac{1}{n!} \binom{n}{i} p_{n-i,i} \varepsilon_{a_{E/L}}^{n-i} \varepsilon_{c_{E/L}}^{i} \tag{S45}$$

Also for fit functions according to equation (S45), one can observe quite a variation in the predicted thermal expansion tensors depending on whether one employs the Lagrangian definition of finite strain (see Figure S 15, Figure S 16, and Figure S 17 for the thermal expansion using equation (S45) as a fit function up to second, third, and fourth order, respectively) or the Eulerian definition of finite strain (see Figure S 18 and Figure S 19 for the thermal expansion using equation (S45) as a fit function up to second and third order, respectively). Note that in these plots, the corresponding results obtained with the PBE functional are shown in addition to the ones from the PBEsol functional. They are discussed separately in Section S5.2.

Many of the tested analytical model functions to fit the calculated non-equilibrium free enthalpies yield qualitatively correct estimates for the thermal expansion in the sense that $\alpha_{11}$ is negative, $\alpha_{33}$ is positive and $|\alpha_{11}| < |\alpha_{33}|$. The latter has, for example, not worked out for the QHAiso approach. Nevertheless, the strong dependence of the obtained result on the chosen model function is quite concerning.

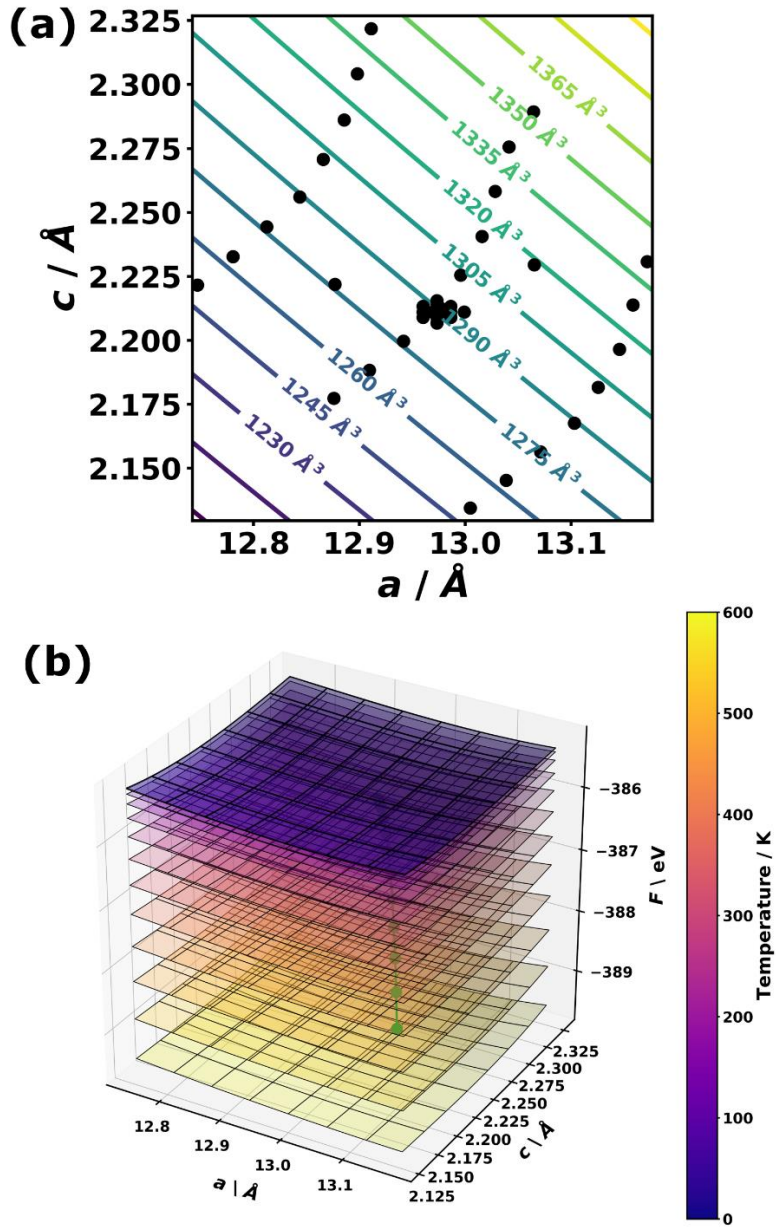

*Figure S 10: (a) Two-dimensional space of the lattice constants a and c, for which (non-equilibrium) free enthalpies were calculated from full phonon band structures. The black dots denote the 41 strained configurations with different lattice parameters, while the hyperbolic contour lines show the curves of constant unit-cell volume. (b) Fitted total free enthalpy at zero applied stress (=free energy) of MOF-74 as a function of the two lattice parameters (abscissa) and the temperature (color scale) obtained with the QHAaniso approach using a third-order polynomial to model the dependence on the lattice parameters, a and c. The green circles mark the minimum free energy and the pair of lattice constants, (a,c), that minimizes the free energy at*

*each temperature. The temperature dependence of those marked positions yields the thermal expansion tensor.*

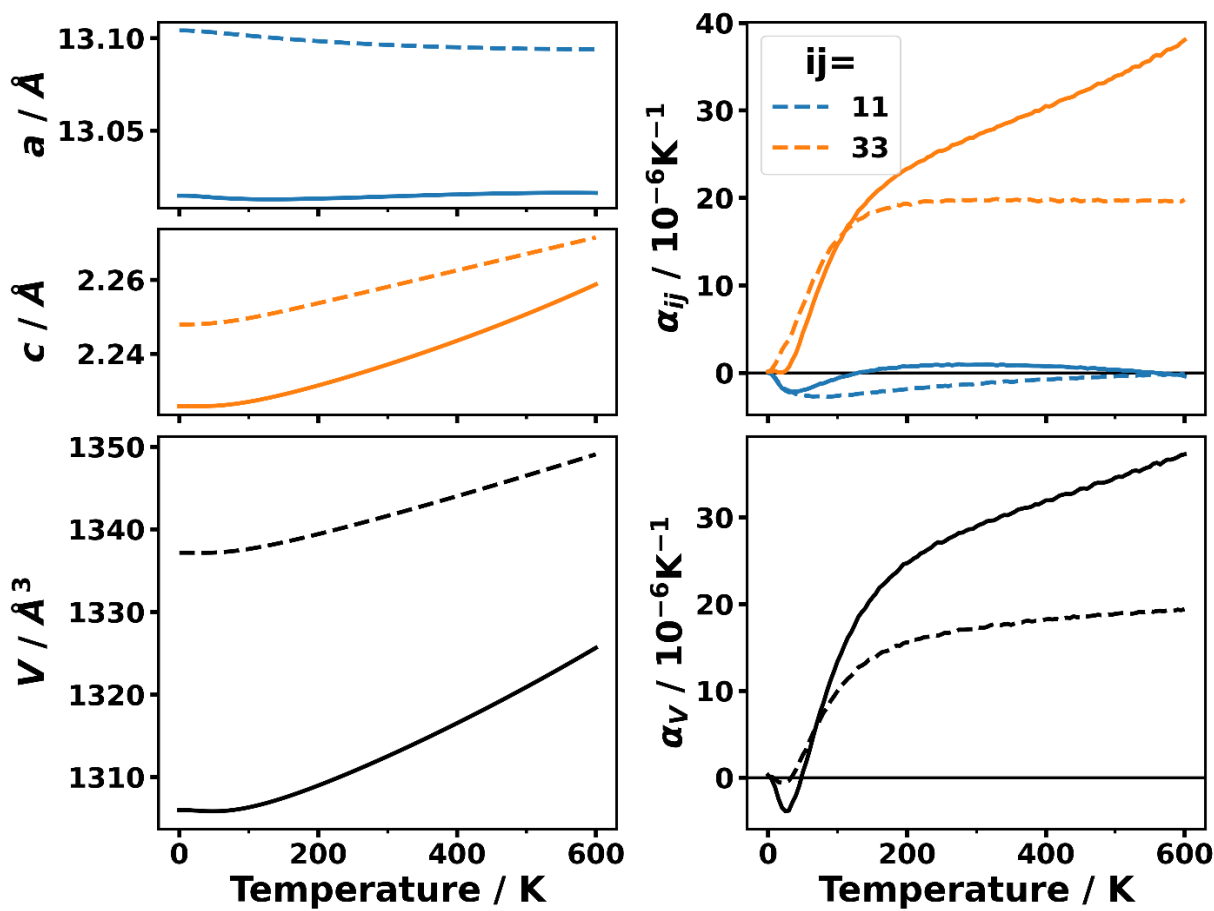

*Figure S 11: Temperature dependence of the two independent lattice parameters, a and c, and the unit-cell volume, V, of MOF-74 calculated with the QHA-procedure described in Section S3.5 (QHAaniso) based on a fit using a second-order polynomial to model the dependences of the total free enthalpy on the lattice parameters a and c. The right column shows the associated thermal expansion coefficients, $\alpha_{11}$, $\alpha_{33}$, and the volumetric thermal expansion coefficient $\alpha_V$.*

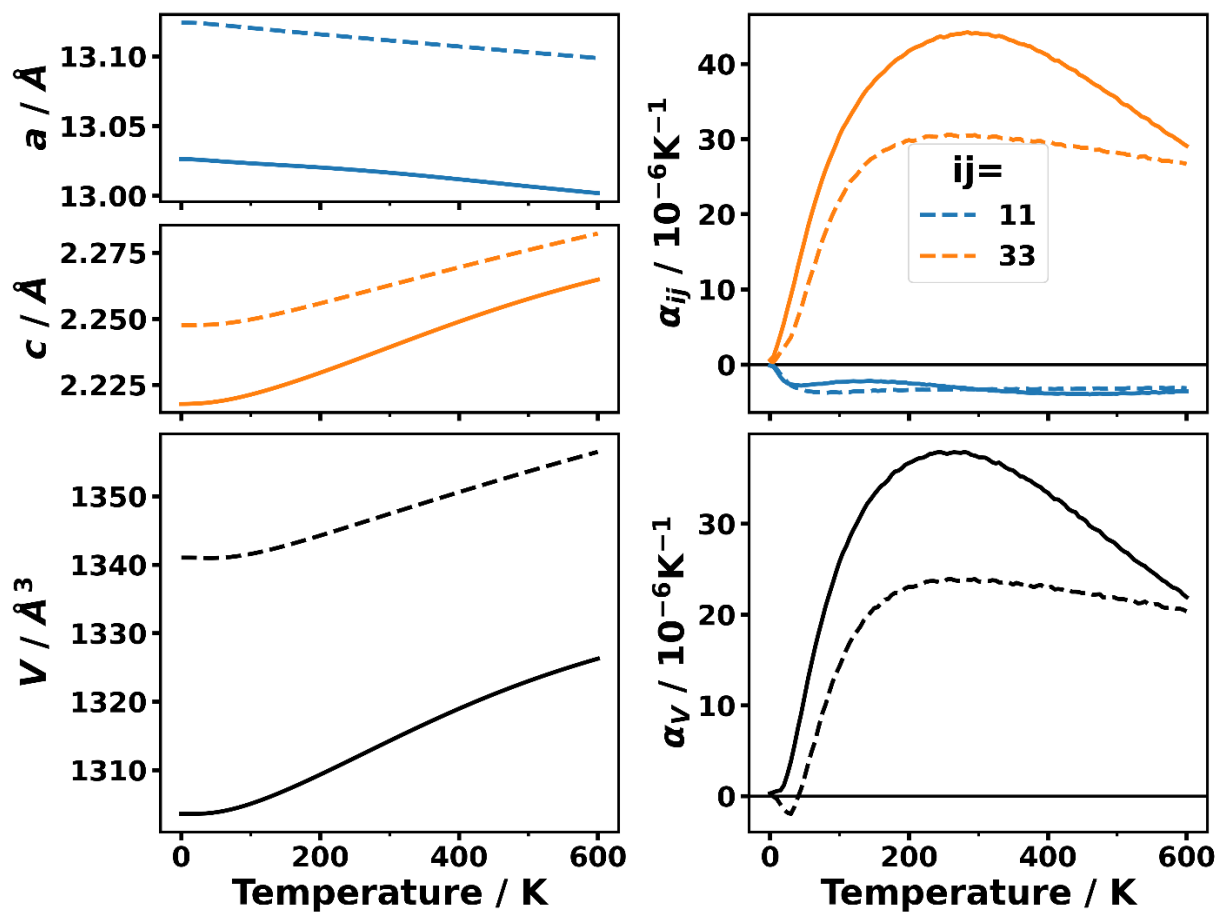

*Figure S 12: Temperature dependence of the two independent lattice parameters, a and c, and the unit-cell volume, V, of MOF-74 calculated with the QHA-procedure described in Section S3.5 (QHAaniso) based on a fit using a third-order polynomial to model the dependences of the total free enthalpy on the lattice parameters a and c. The right column shows the associated thermal expansion coefficients, $\alpha_{11}$, $\alpha_{33}$, and the volumetric thermal expansion coefficient $\alpha_V$.*

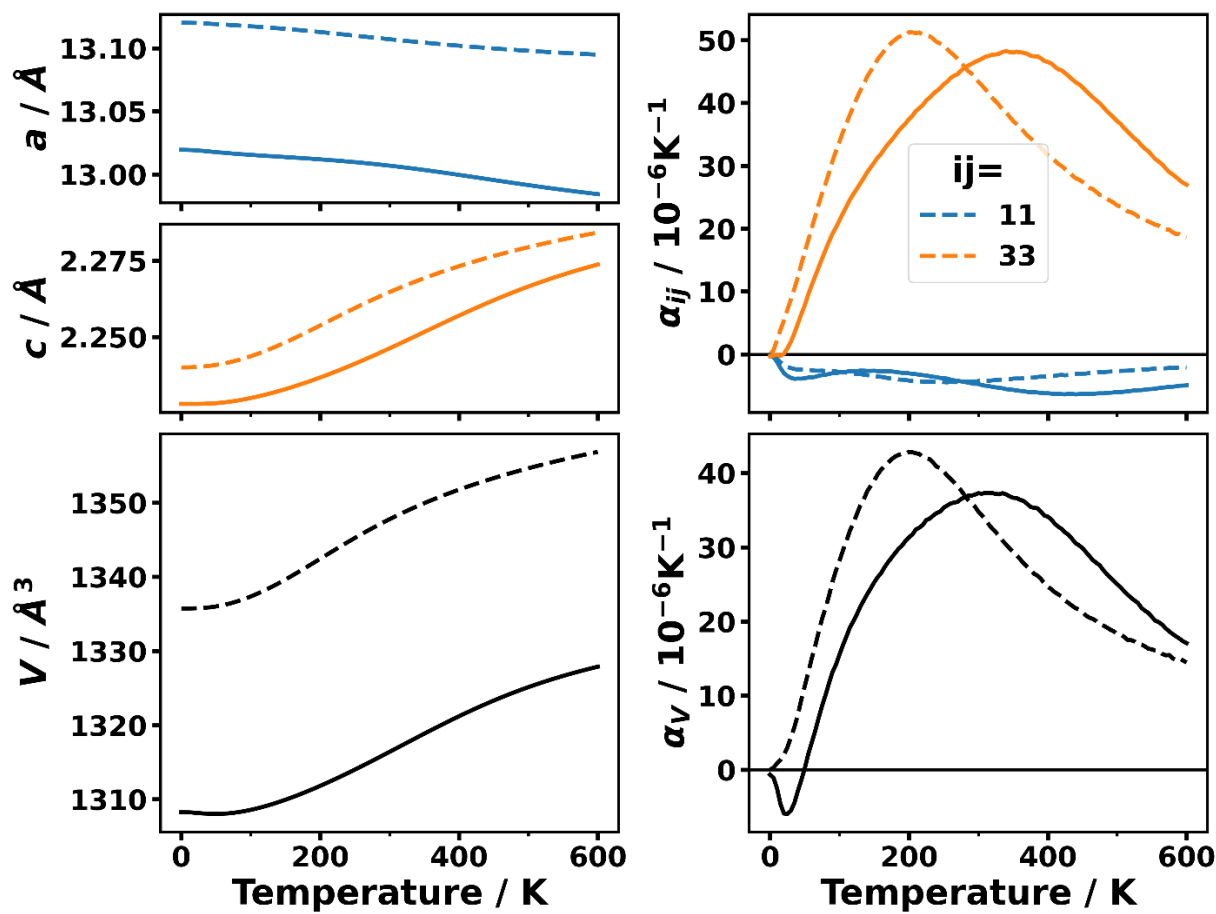

*Figure S 13: Temperature dependence of the two independent lattice parameters, a and c, and the unit-cell volume, V, of MOF-74 calculated with the QHA-procedure described in Section S3.5 (QHAaniso) based on a fit using a fourth-order polynomial to model the dependences of the total free enthalpy on the lattice parameters a and c. The right column shows the associated thermal expansion coefficients, $\alpha_{11}$, $\alpha_{33}$, and the volumetric thermal expansion coefficient $\alpha_V$.*

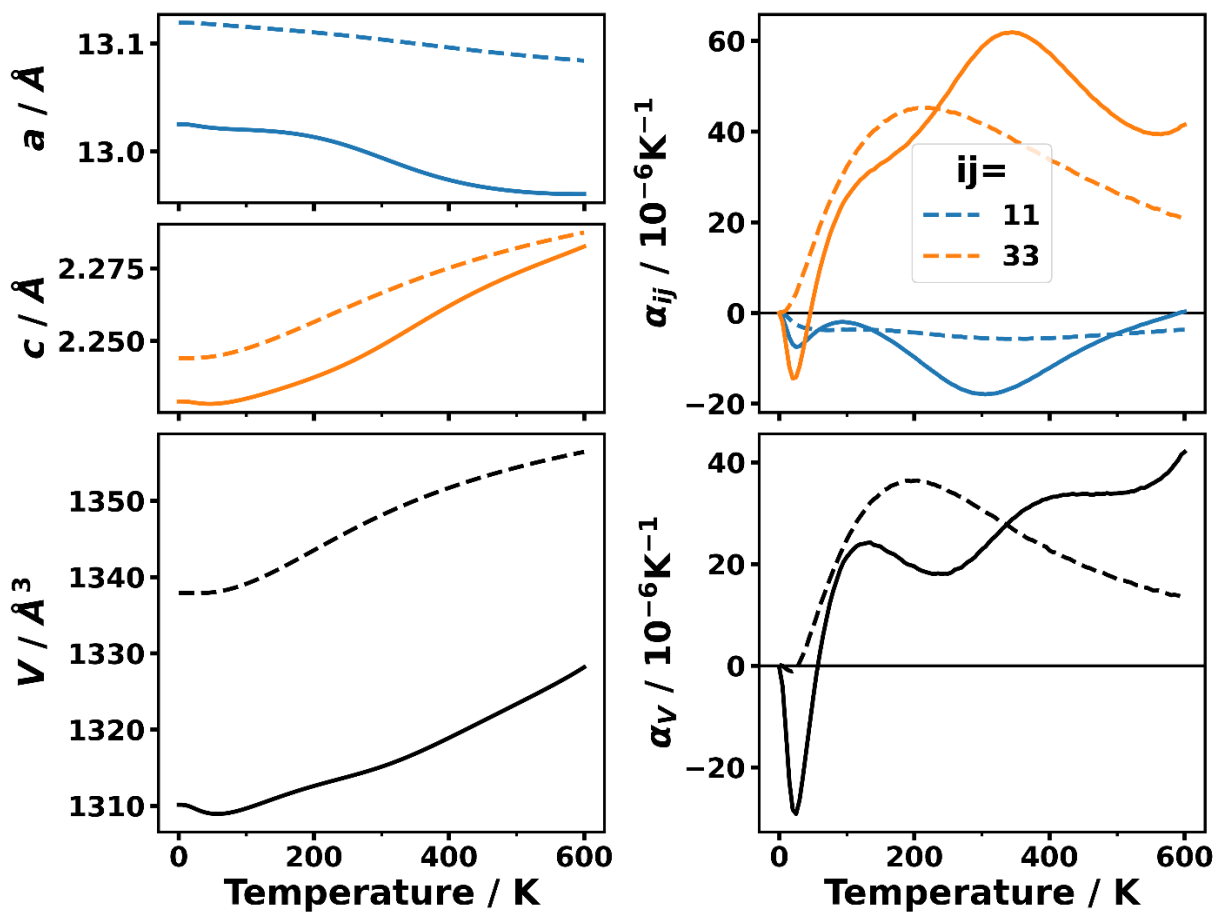

*Figure S 14: Temperature dependence of the two independent lattice parameters, a and c, and the unit-cell volume, V, of MOF-74 calculated with the QHA-procedure described in Section S3.5 (QHAaniso) based on a fit using a fifth-order polynomial to model the dependences of the total free enthalpy on the lattice parameters a and c. The right column shows the associated thermal expansion coefficients, $\alpha_{11}$, $\alpha_{33}$, and the volumetric thermal expansion coefficient $\alpha_V$. The solid lines show the results obtained with the PBEsol functional, while the dashed lines correspond to the results with the PBE functional.*

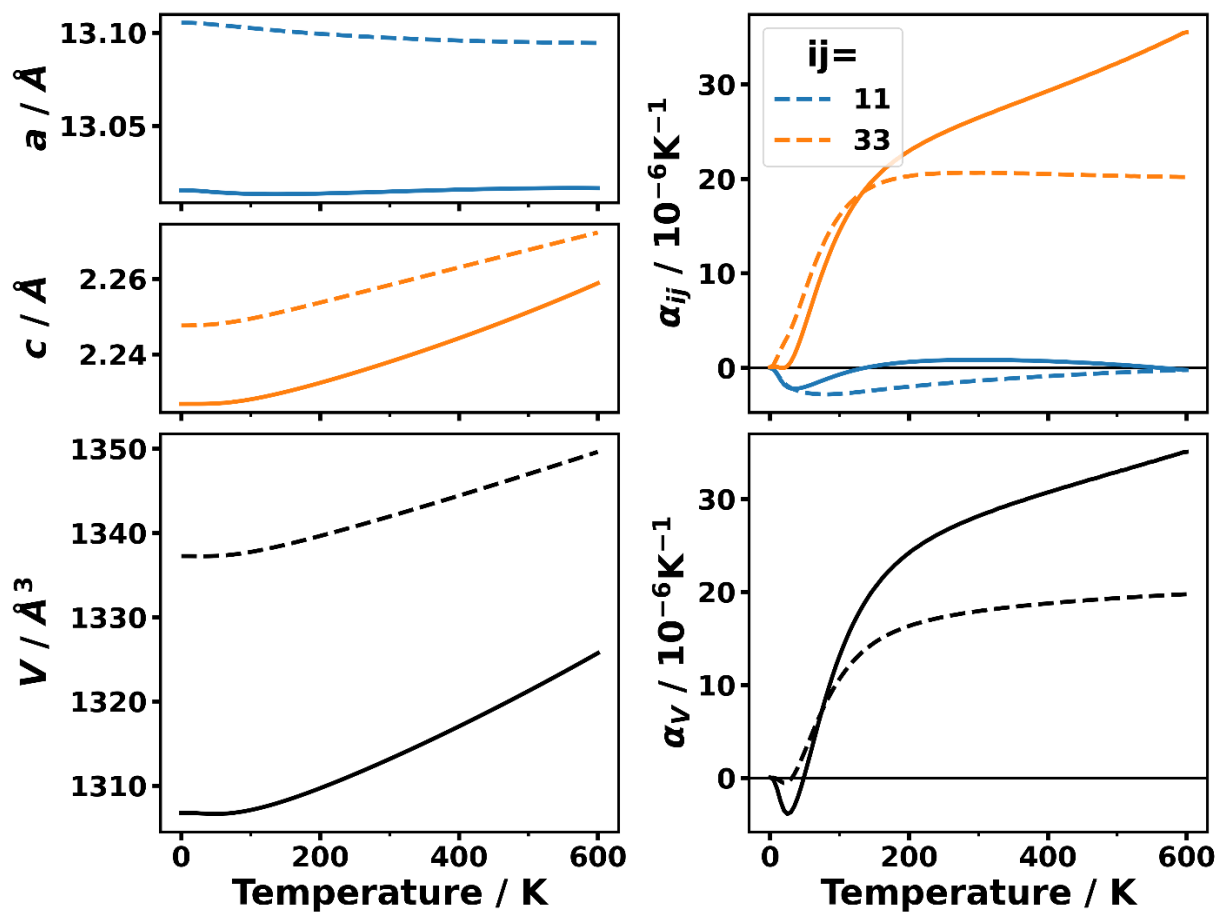

*Figure S 15: Temperature dependence of the two independent lattice parameters, a and c, and the unit-cell volume, V, of MOF-74 calculated with the QHA-procedure described in Section S3.5 (QHAaniso) based on a fit using a second-order polynomial to model the dependences of the total free enthalpy on the finite Lagrangian strains (Birch-Murnaghan-like EoS of second order). The right column shows the associated thermal expansion coefficients, $\alpha_{11}$, $\alpha_{33}$, and the volumetric thermal expansion coefficient $\alpha_V$. The solid lines show the results obtained with the PBEsol functional, while the dashed lines correspond to the results with the PBE functional.*

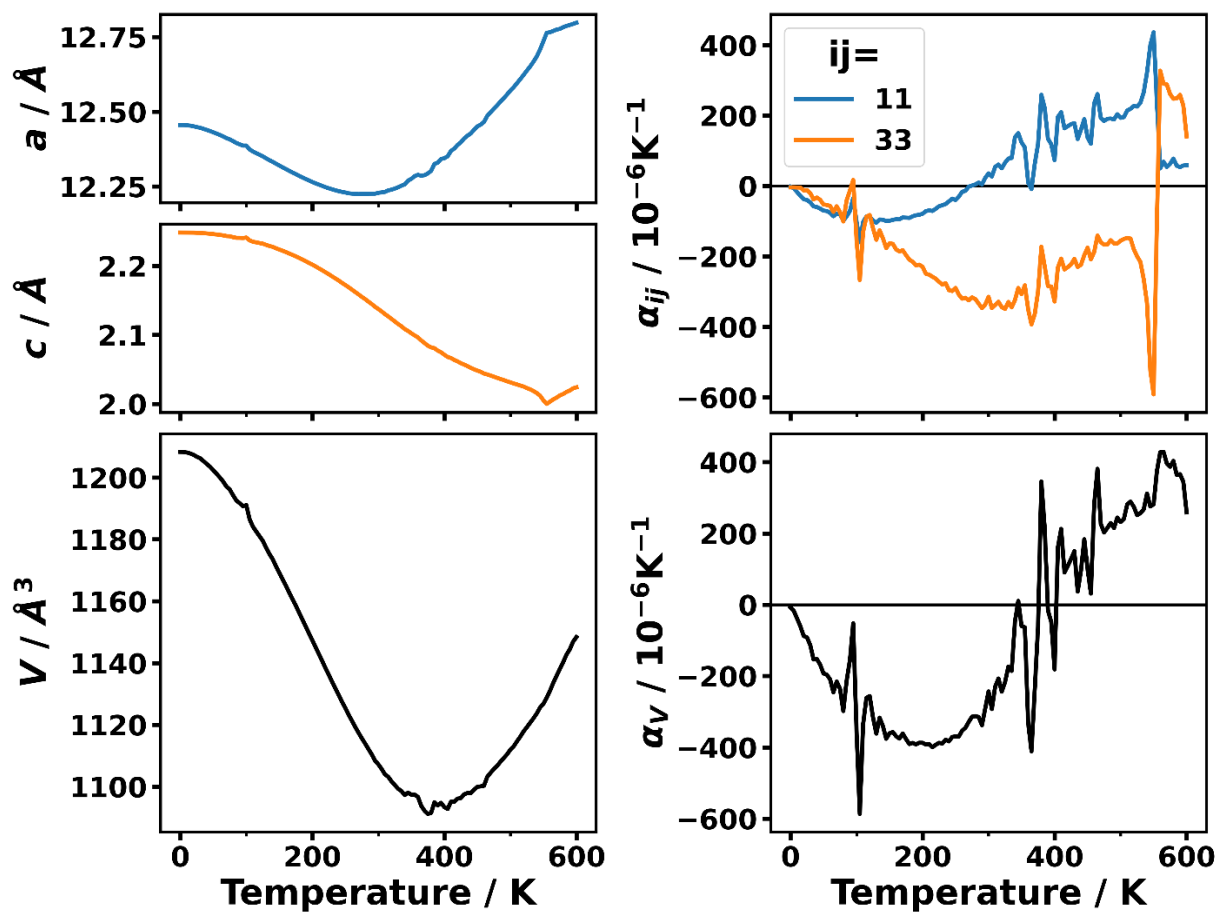

*Figure S 16: Temperature dependence of the two independent lattice parameters, a and c, and the unit-cell volume, V, of MOF-74 calculated with the QHA-procedure described in Section S3.5 (QHAaniso) based on a fit using a third-order polynomial to model the dependences of the total free enthalpy on the finite <u>Lagrangian</u> strains (Birch-Murnaghan-like EoS of third order). The right column shows the associated thermal expansion coefficients, $\alpha_{11}$, $\alpha_{33}$, and the volumetric thermal expansion coefficient $\alpha_V$. For reasons of visibility, the PBE results are omitted and are separately shown in Figure S 36.*

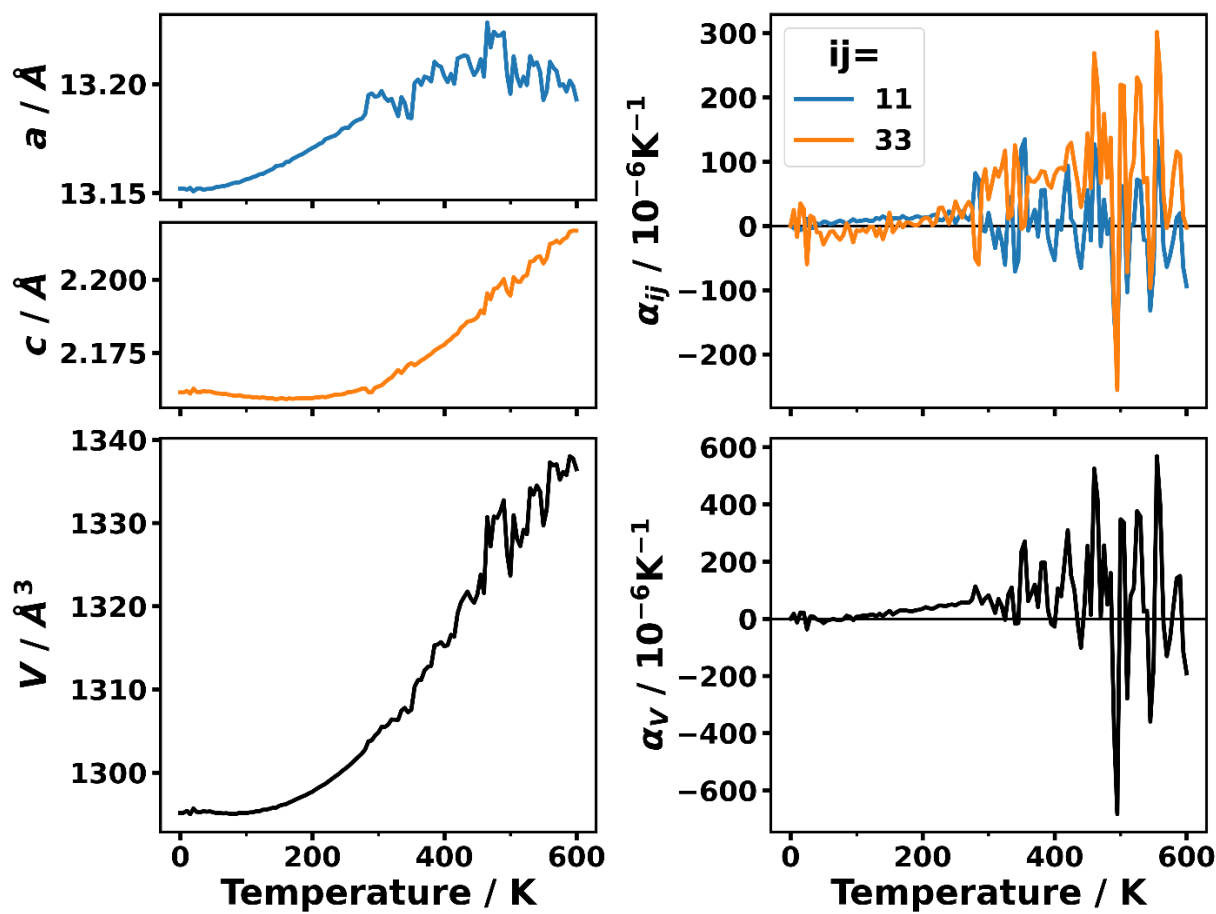

*Figure S 17: Temperature dependence of the two independent lattice parameters, a and c, and the unit-cell volume, V, of MOF-74 calculated with the QHA-procedure described in Section S3.5 (QHAaniso) based on a fit using a fourth-order polynomial to model the dependences of the total free enthalpy on the finite Lagrangian strains (Birch-Murnaghan-like EoS of fourth order). The right column shows the associated thermal expansion coefficients, $\alpha_{11}$, $\alpha_{33}$, and the volumetric thermal expansion coefficient $\alpha_V$. For reasons of visibility, the PBE results are omitted here and are separately shown in Figure S 37.*

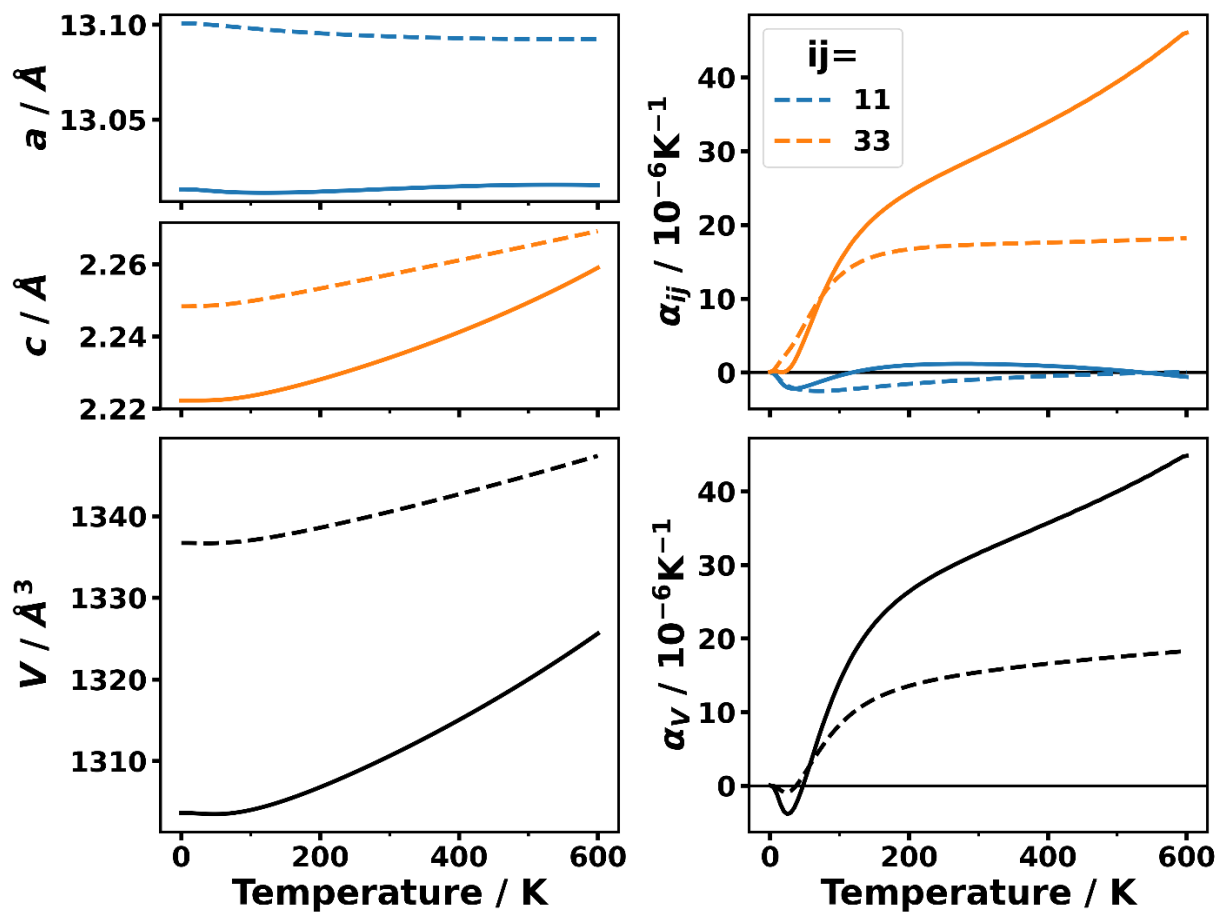

*Figure S 18: Temperature dependence of the two independent lattice parameters, a and c, and the unit-cell volume, V, of MOF-74 calculated with the QHA-procedure described in Section S3.5 (QHAaniso) based on a fit using a second-order polynomial to model the dependences of the total free enthalpy on the finite <u>Eulerian</u> strains (Birch-Murnaghan-like EoS of second order). The right column shows the associated thermal expansion coefficients, $\alpha_{11}$, $\alpha_{33}$, and the volumetric thermal expansion coefficient $\alpha_V$. The solid lines show the results obtained with the PBEsol functional, while the dashed lines correspond to the results with the PBE functional.*

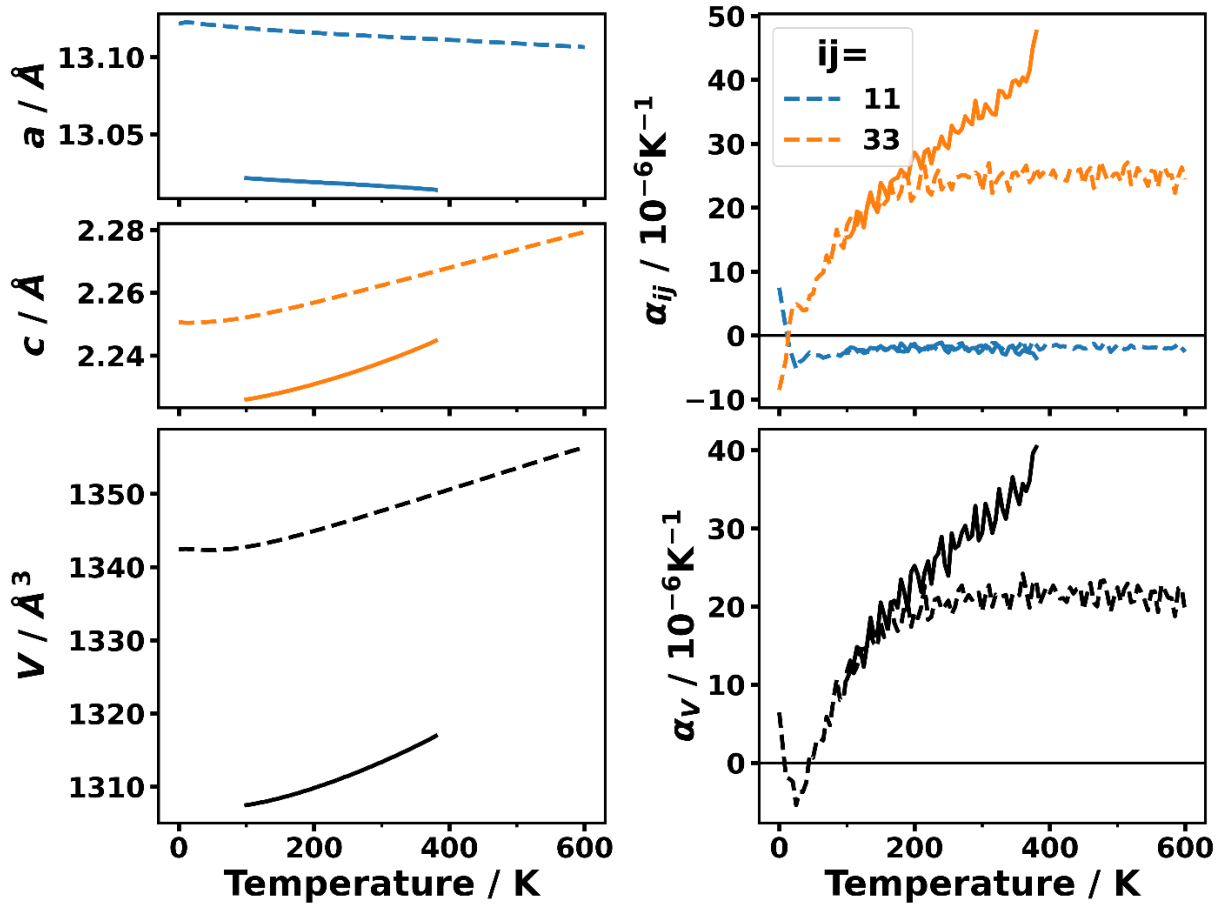

*Figure S 19: Temperature dependence of the two independent lattice parameters, a and c, and the unit-cell volume, V, of MOF-74 calculated with the QHA-procedure described in Section S3.5 (QHAaniso) based on a fit using a third-order polynomial to model the dependences of the total free enthalpy on the finite Eulerian strains (Birch-Murnaghan-like EoS of third order). The right column shows the associated thermal expansion coefficients, $\alpha_{11}$, $\alpha_{33}$, and the volumetric thermal expansion coefficient $\alpha_V$. The solid lines show the results obtained with the PBEsol functional, while the dashed lines correspond to the results with the PBE functional.*

In order to unambiguously test and compare the quality of the models, two indicators, the weighted root-mean-square error, *Rwp*, and the so-called goodness of fit, *S*, were calculated according to equations (S46) and (S47). These indicators are frequently used in XRD refinements to test how well the found crystal structure solution explains the experimental observations.

$Rwp$ is calculated as the sum of the squared difference between the actually calculated free energies, $F_i$ (at each temperature and zero stress), and the values the model function predicts at the same positions, $F_i^{model}$. This sum of quadratic deviations is then normalized by the sum of the (squared) calculated free energies, $F_i$. In principle, it would be possible to assign weights, $w_i$, to every data point separately. Here, all weights were set to unity.

$$Rwp = \sqrt{\frac{\sum_{i=1}^{N}(F_i - F_i^{model})^2 w_i}{\sum_{i=1}^{N} F_i^2 w_i}} \quad (S46)$$

The $Rwp$-value, which should be as small as possible for a good model, has, however, one disadvantage. It can become arbitrarily small if the model contains sufficiently many parameters. In the spirit of Occam's razor, the indicator $S$ (also referred to as "goodness of fit", GOF), divides the $Rwp$ value by the difference of the number of datapoints, $N$, and the number of parameters of the model, $M$. In that way, $S$ increases for models with large numbers of parameters.

$$S = \sqrt{\frac{Rwp}{N-M}} \quad (S47)$$

Figure S 20 shows the evolutions of the indicators, $Rwp$ and the $S$, as a function of the temperature for the polynomials according to equation (S44) up to fifth order ("poly2", "poly3", "poly4", and "poly5") and for Burch-Murnaghan-like model functions according to equation (S45) utilizing Eulerian finite strains of up to fourth order ("BM2_E", "BM3_E", and "BM4_E") or Lagrangian finite strains of up to fourth order ("BM2_E", "BM3_E", and "BM4_L"). The best goodness of fit (GOF, $S$) is obtained for the "BM4_L" model for low temperatures up to 300 K, with only slightly lower $S$-values than the "BM3_L" which becomes the best tested model between 300 K and ~450 K, before the "BM2_L" model shows the smallest $S$-values for temperatures >450 K. The $Rwp$-values of these three models do not change their relative order as a function of temperature such that the changing order of *S* must be a consequence of the fact that higher order models have more fit parameters, which increases *S*.

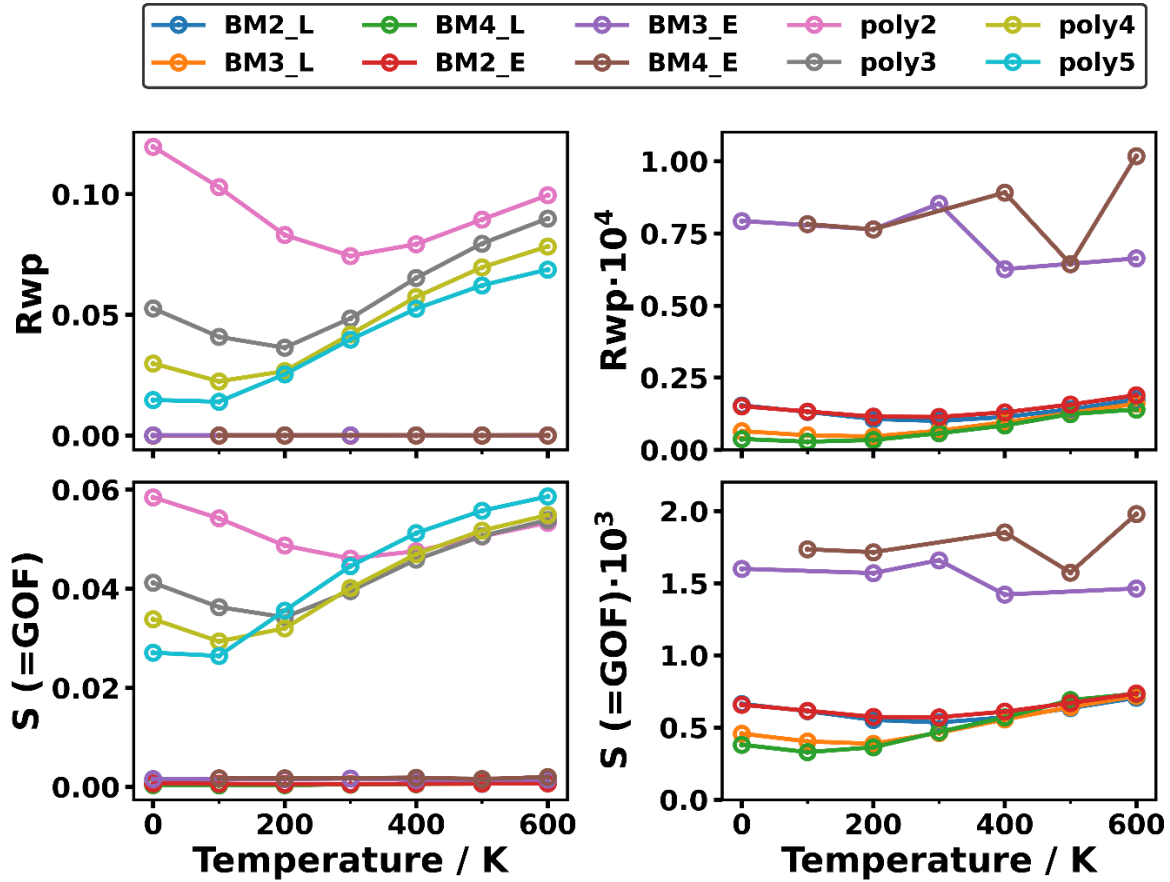

*Figure S 20: Quantitative comparison of the model functions used for the fits in terms of the resulting Rwp and S values (see equations (S46) and (S47), respectively). The considered model functions comprise polynomials in the variables a and c of second to fifth order ("poly2", "poly3", "poly4", and "poly5") and Birch-Murnaghan-like functions ("BM"), in which the free energy is expanded in (two-dimensional) Taylor series of either Eulerian ("_E") or Lagrangian ("_L") finite strains from second to fourth order. E.g., the model "BM2_L" corresponds to a Birch-Murnaghan-like EoS of second order with Lagrangian finite strains. The plots in the right column are zoomed-in views of the plots in the left column.*

Unfortunately, when one inspects the resulting thermal expansion of those three models using Lagrangian strains (see Figure S 15, Figure S 16, and Figure S 17), the temperature dependences of $a$ and $c$ are not very satisfactory. Moreover, especially for the "BM3_L" and "BM4_L" model functions, severe noise is superimposed on the $a(T)$ and $c(T)$ evolutions such that the temperature-derivatives are hardly perceivable in a reasonable way.

The Birch-Murnaghan-like model functions relying on Eulerian strains show only slightly worse $Rwp$ and $S$ values but result in more reasonable thermal expansion coefficients (see Figure S 18 and Figure S 19). In particular, the Birch-Murnaghan-like model of third order ("BM3_E") yields thermal expansion coefficients in good qualitative agreement with our expectations (in spite of being also relatively noisy).

As a final attempt to become less dependent on the choice of the model function for fitting the non-equilibrium free enthalpies, Gaussian progress regression (GPR) as implemented in the scikit-learn package [18] was employed. Here, the kernel consisted of a product of a constant kernel and a radial basis function kernel (RBF) with the respective hyperparameters (the constant and the RBF length scale) being optimized to maximize the logarithmic likelihood. The model vector was composed of the lattice constants, $a$ and $c$, and the associated total Gibbs free enthalpies (=free energies at zero pressure). Additionally, we assumed noise of 8 meV on the non-equilibrium free enthalpy data. The predicted thermal expansion of the two lattice parameters based on this method independent of a fixed analytical model function is shown in Figure S 21. This approach seems to be able to qualitatively capture the tendency of $a$ (primarily) decreasing with temperature and $c$ increasing with temperature.

However, as for all other models, especially the thermal expansion tensor element $\alpha_{33}$ is much larger than expected based on experimental observations and the results from Grüneisen theory. Notably, there is a fundamental disadvantage of the data used for all the QHA-based approaches presented so far: the strains used to vary the lattice parameters are much larger than the temperature-induced strains seen in the experiments. As a consequence, the changes in non-equilibrium free enthalpies are larger, which is favorable for fitting but might distort the picture. If one, however, uses only the free enthalpies in the close vicinity of the (electronically) optimized unit cell at 0 K, the variation of the non-equilibrium free enthalpies as a function of the *(a,c)*-configuration is too small to achieve a reasonable fit with any model function. A possible remedy for this problem is suggested in Section S3.6.

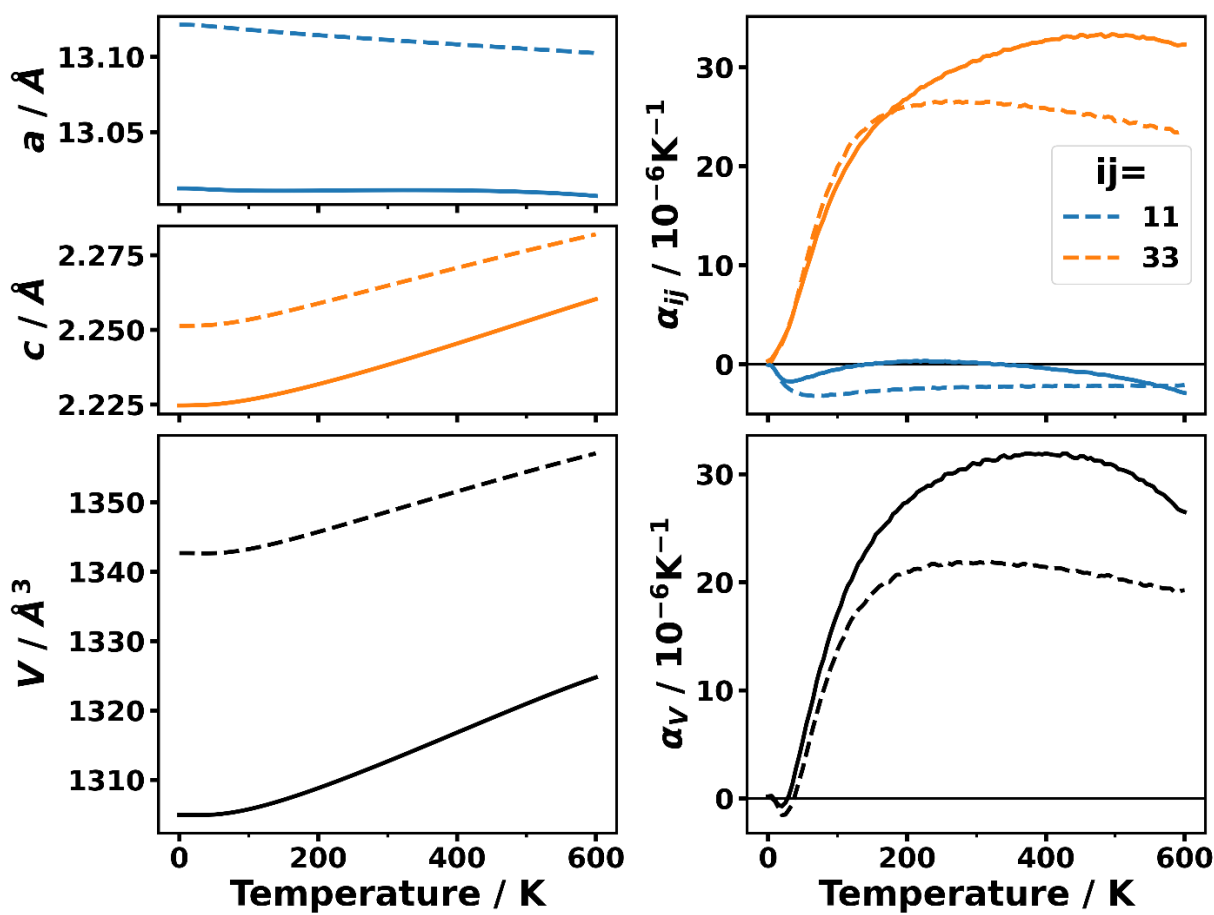

*Figure S 21: Temperature dependence of the two independent lattice parameters, a and c, and the unit-cell volume, V, of MOF-74 calculated with the QHA-procedure described in Section S3.5 (QHAaniso) based Gaussian process regression. The right column shows the associated thermal expansion coefficients, $\alpha_{11}$, $\alpha_{33}$, and the volumetric thermal expansion coefficient $\alpha_V$. The solid*

*lines show the results obtained with the PBEsol functional, while the dashed lines correspond to the results with the PBE functional.*

### S3.6 Quasi-harmonic approximation IV: numerically solving the analytical expression for Gibb's free enthalpy

The biggest challenge in the above-described QHAaniso approach is the choice of the model function for the non-equilibrium free enthalpies, which is used to find the strains which minimize this function at each temperature (and stress). Moreover, the fit turned out to be problematic for considering strains of a realistic order of magnitude.

As an alternative approach, we tried to solve this problem based on the analytic expression of the (non-equilibrium) Gibbs free enthalpy (see equation (S38)). This approach will be referred to as QHAanalytaniso. In the case of MOF-74, which does not appear to undergo a temperature-induced phase transition in experimentally considered temperature range, only two different (linearly independent) strain tensors are in agreement with the rhombohedral symmetry:

$$\boldsymbol{\varepsilon}' = \begin{pmatrix} \varepsilon_a & 0 & 0 \\ 0 & \varepsilon_a & 0 \\ 0 & 0 & 0 \end{pmatrix}, \quad \boldsymbol{\varepsilon}'' = \begin{pmatrix} 0 & 0 & 0 \\ 0 & 0 & 0 \\ 0 & 0 & \varepsilon_c \end{pmatrix} \tag{S48}$$

Using these strain tensors, $\boldsymbol{\varepsilon}'$ and $\boldsymbol{\varepsilon}''$, one can rewrite the non-equilibrium free enthalpies as a function of the scalar variables, $\varepsilon_a$ and $\varepsilon_c$:

$$G(T,\boldsymbol{\sigma}) = \min_{\varepsilon_a,\varepsilon_c}\{G'(T,\boldsymbol{\sigma},\varepsilon_a,\varepsilon_c)\}$$
$$= \min_{\varepsilon_a,\varepsilon_c}\{E_{elec}(\varepsilon_a,\varepsilon_c) + F_{ph}(\varepsilon_a,\varepsilon_c,T) - V[\varepsilon_a(\sigma_{11}+\sigma_{22}) + \varepsilon_c\sigma_{33}]\} \tag{S49}$$

As a next step, we use the total energy of the unstrained configuration, $E_0$, and the elements of the elastic tensor (in Voigt notation), $C_{ij}$, to express the electronic energy analytically as a Taylor series in terms of $\varepsilon_a$ and $\varepsilon_c$:

$$E_{elec}(\varepsilon_a,\varepsilon_c) = E_0 + 2(C_{11}+C_{12})V_0\frac{\varepsilon_a^2}{2} + C_{33}V_0\frac{\varepsilon_c^2}{2} + 2\,C_{13}\,V_0\,\varepsilon_a\,\varepsilon_c + + \mathcal{O}(\varepsilon_a^3,\varepsilon_c^3) \tag{S50}$$

Additionally, also the phonon free energy must be expressed in terms of the two scalar strain variables, $\varepsilon_a$ and $\varepsilon_c$. This is done *via* the expression of the free energy of the phonons (see equation (S51), with the index $\lambda$ containing the phonon band index and the wave vector) in which we assume an explicit dependence of the (angular) frequencies, $\omega_\lambda$, on the scalar strain variables according to equation (S52).

$$F_{ph}(\varepsilon_a, \varepsilon_c, T) = \frac{1}{N_q} \sum_\lambda \left\{ \frac{\hbar\omega_\lambda(\varepsilon_a, \varepsilon_c)}{2} + k_B T \ln\left[1 - \exp\left(-\frac{\hbar\omega_\lambda(\varepsilon_a, \varepsilon_c)}{k_B T}\right)\right] \right\} \quad (S51)$$

$$\omega_\lambda(\varepsilon_a, \varepsilon_c) = \omega_\lambda^0 + \left.\frac{\partial \omega_\lambda}{\partial \varepsilon_a}\right|_{\varepsilon_a=0} \varepsilon_a + \left.\frac{\partial \omega_\lambda}{\partial \varepsilon_c}\right|_{\varepsilon_c=0} \varepsilon_c + \mathcal{O}(\varepsilon_a^2, \varepsilon_c^2) \quad (S52)$$

The Taylor expansion in equation (S52) is only carried out up to first order because only the first order derivatives with respect to strain can be evaluated properly accounting for a potential reordering of band indices (see below). This can be achieved by expressing the derivative of phonon frequencies as derivatives of the dynamical matrix, $D$, and the (unstrained) phonon eigenvectors (polarization vectors), $e^0_\lambda$, with $X = \varepsilon_a, \varepsilon_c$ [6,19]:

$$\left.\frac{\partial \omega_\lambda}{\partial X}\right|_{X=0} = \frac{1}{2\omega_\lambda^0}\left.\frac{\partial \omega_\lambda^2}{\partial X}\right|_{X=0} = \frac{1}{2\omega_\lambda^0}(e_\lambda^0)^T \left.\frac{\partial D}{\partial X}\right|_{X=0} e_\lambda^0 \approx \frac{1}{2\omega_\lambda^0}(e_\lambda^0)^T \left.\frac{\Delta D}{\Delta X}\right|_{X=0} e_\lambda^0 \quad (S53)$$

In the last approximate equality in the equation above, the partial derivative is replaced by a suitable finite difference (FD) quotient. In this case, we used a second order scheme, with $D\{\pm\Delta X\}$ being the dynamical matrix of the configuration with positive or negative strain, $\Delta X$ with $X = \varepsilon_a$, $\varepsilon_c$ and $\Delta\varepsilon_{a/c} = 0.001$:

$$\left.\frac{\partial D}{\partial X}\right|_{X=0} \approx \left.\frac{D\{+\Delta X\} - D\{-\Delta X\}}{2\Delta X}\right|_{X=0} \quad with\ X = \varepsilon_a, \varepsilon_c \quad (S54)$$

Conversely, calculating the derivative of the $n^{th}$ phonon band directly without the dynamical matrix (*i.e.*, by applying a FD scheme directly to the phonon frequencies in a brute-force way) can be misleading because this band might be the $m^{th}$ band (with $m \neq n$) for a strained configuration.

Inserting equations (S52), (S51), and (S50) into equation (S49) yields an (involved) analytical function of the non-equilibrium Gibbs free enthalpies as a function of the scalar strain variables $\varepsilon_a$ and $\varepsilon_c$.

Figure S 22 shows the obtained dependence of the non-equilibrium free enthalpy on the scalar strain variables for a fixed temperature (300 K) and zero pressure with the QHAanalytaniso approach. From this analytical function, the minimum was determined numerically for each temperature (using the Nelder-Mead algorithm as implemented in the *minimize* routine of the scipy package[20,21]). The thermal expansion tensor as shown in Figure S 23 was obtained from the temperature evolution of the minima found in this way.

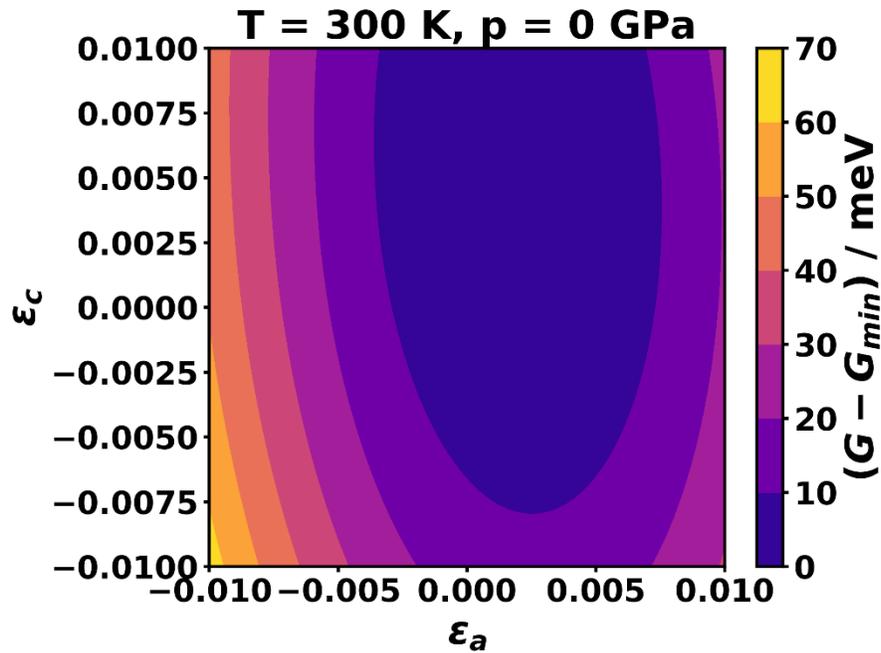

*Figure S 22: Contour plot of the analytical expression of the non-equilibrium free enthalpy, G, at zero applied stress and a temperature of 300 K of MOF-74 as a function of the two scalar strain variables, $\varepsilon_a$ and $\varepsilon_c$ (see equation (S48)) obtained with the QHAanalytaniso approach (see equations (S49), (S50), (S51), and (S52)).*

Figure S 23 shows that the obtained thermal expansion tensor elements are in good qualitative and also quantitative agreement with the expectations based on the experimentally observed trends and the results from the Grüneisen theory. Although the $\alpha_{33}$ element is smaller using this approach than in, *e.g.*, the QHAaniso approach, it is still slightly larger here than when obtained from the Grüneisen theory. This results also in somewhat increased volumetric thermal expansion. Compared to all the various QHA-based approaches presented above, this one, however, yields by far the most satisfying level of agreement with our expectations.

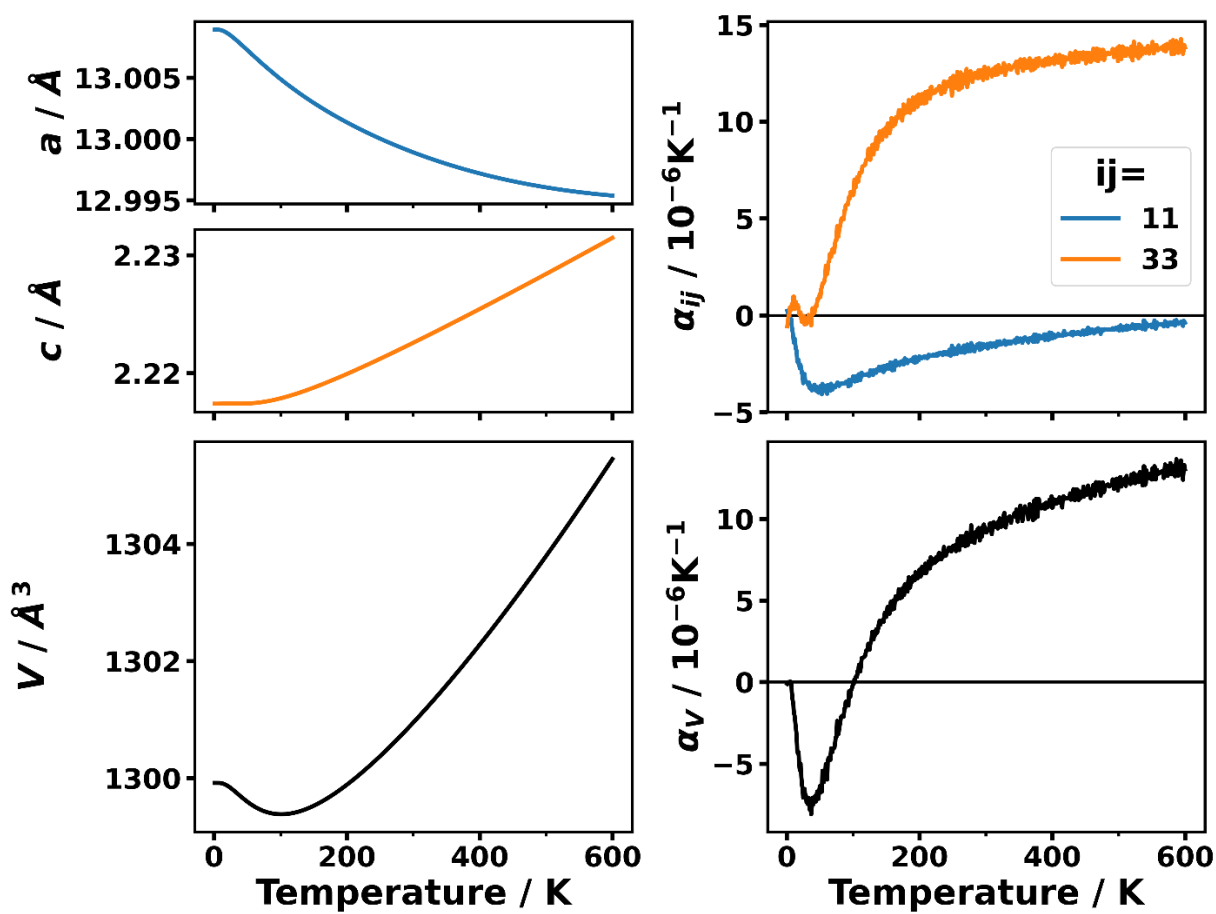

*Figure S 23: Temperature dependence of the two independent lattice parameters, a and c, and the unit-cell volume, V, of MOF-74 calculated with the QHA-procedure described in Section S3.6 (QHAanalytaniso) based on a second-order strain expansion of the electronic energy and a first-order strain expansion of phonon frequencies. The right column shows the associated thermal expansion coefficients, $\alpha_{11}$, $\alpha_{33}$, and the volumetric thermal expansion coefficient $\alpha_V$.*

## S3.7 Grüneisen theory of thermal expansion

In principle, the Grüneisen theory (GT) is very similar to the QHAanalytansio approach described above but is evaluated more directly. The following derivation of the thermal expansion tensor within GT is inspired by the isotropic/cubic case shown in Ref. [2]. The starting point for the GT is the general expression of the thermal expansion tensor (see equation (S20)). In GT, the temperature derivative at constant stress is evaluated directly by evaluating the total derivative of the strain tensor for the special case of constant stress ($d\sigma_{ij}=0$) in the (grand-) canonical ensemble (*i.e.,* in the thermodynamic variables temperature, $T$, and strain, $\varepsilon_{ij}$):

$$d\sigma_{ij} = \left.\frac{\partial \sigma_{ij}}{\partial T}\right|_{\varepsilon} dT + \left.\frac{\partial \sigma_{ij}}{\partial \varepsilon_{kl}}\right|_{T} d\varepsilon_{kl} = 0 \tag{S55}$$

Rearranging this equation so that the derivative of the strain tensor with respect to temperature (at constant stress) remains the only term on one side of the equation, yields:

$$\left.\frac{\partial \varepsilon_{ij}}{\partial T}\right|_{\sigma} (= \alpha_{ij}) = -\left.\frac{\partial \varepsilon_{ij}}{\partial \sigma_{kl}}\right|_{T} \left.\frac{\partial \sigma_{kl}}{\partial T}\right|_{\varepsilon} = -S_{ijkl} \left.\frac{\partial \sigma_{kl}}{\partial T}\right|_{\varepsilon} \tag{S56}$$

In the last equality of the equation above, the derivative of the strain tensor with respect to the stress tensor (at constant temperature) has been replaced by the (rank-4) compliance tensor, $S_{ijkl}$. To evaluate the above expression further, the temperature-derivative of the stress tensor is required. In the (grand-)canonical ensemble, the stress tensor is obtained from the strain-derivative of the thermodynamic potential (*i.e.*, the Helmholtz free energy, $F$):

$$\sigma_{kl} = \frac{1}{V}\frac{\partial F}{\partial \varepsilon_{kl}}\bigg|_T = \frac{1}{V}\frac{\partial}{\partial \varepsilon_{kl}}\bigg|_T \left(\frac{1}{N_q}\sum_\lambda \left\{\frac{\hbar\omega_\lambda(\boldsymbol{\varepsilon})}{2} + k_B T \ln\left[1 - \exp\left(-\frac{\hbar\omega_\lambda(\boldsymbol{\varepsilon})}{k_B T}\right)\right]\right\}\right)$$

$$= \frac{1}{V}\frac{1}{N_q}\sum_\lambda \left\{\hbar\left(\frac{1}{2} + \frac{1}{\exp\left(\frac{\hbar\omega_\lambda(\boldsymbol{\varepsilon})}{k_B T}\right) - 1}\right)\frac{\partial \omega_\lambda}{\partial \varepsilon_{kl}}\bigg|_T\right\} \quad (S57)$$

$$= \frac{1}{V}\frac{1}{N_q}\sum_\lambda \left\{\hbar\left(\frac{1}{2} + n_\lambda(T)\right)\frac{\partial \omega_\lambda}{\partial \varepsilon_{kl}}\bigg|_T\right\}$$

In the last step, one summand has been replaced by the Bose-Einstein distribution of a phonon with quantum number $\lambda$ at temperature $T$. This expression must be differentiated once more with respect to temperature (at constant strain) in order to obtain the factor necessary for evaluating equation (S56):

$$\frac{\partial \sigma_{kl}}{\partial T}\bigg|_\varepsilon = \frac{\partial}{\partial T}\bigg|_\varepsilon \left(\frac{1}{V}\frac{1}{N_q}\sum_\lambda \left\{\hbar\left(\frac{1}{2} + n_\lambda(T)\right)\frac{\partial \omega_\lambda}{\partial \varepsilon_{kl}}\bigg|_T\right\}\right)$$

$$= -\frac{1}{V}\frac{1}{N_q}\sum_\lambda \left\{\left(k_B\left(\frac{\hbar\omega_\lambda}{k_B T}\right)^2 \frac{\exp\left(\frac{\hbar\omega_\lambda}{k_B T}\right)}{\left[\exp\left(\frac{\hbar\omega_\lambda}{k_B T}\right) - 1\right]^2}\right)\left(-\frac{1}{\omega_\lambda}\frac{\partial \omega_\lambda}{\partial \varepsilon_{kl}}\right)\right\} \quad (S58)$$

$$= -\frac{1}{V}\frac{1}{N_q}\sum_\lambda \left\{c_v^\lambda(T)\,\gamma_{kl}^\lambda\right\}$$

In the second line of the equation above, the factors have been ordered such that the well-known mode contributions, $c_v^\lambda$, to the (isochoric) heat capacity, $C_V$, can be identified. The second factor in parenthesis in this line containing the strain-derivatives of the phonon frequencies, which defines the Grüneisen tensor, $\gamma_{kl}^\lambda$. Inserting equation (S58) into equation (S56), finally, yields the thermal expansion tensor (in accordance to the result in Ref. [4]):

$$\alpha_{ij} = \frac{\partial \varepsilon_{ij}}{\partial T}\bigg|_\sigma = \frac{1}{N_q}\sum_\lambda \frac{1}{V}S_{ijkl}\,\gamma_{kl}^\lambda\,c_v^\lambda(T) = \frac{1}{N_q}\sum_\lambda \alpha_{ij}^\lambda(T) \quad (S59)$$

The four factors in the summation (*i.e.*, the inverse volume, the compliance tensor, the Grüneisen tensor, and the mode contributions to the phonon heat capacity) determine the mode contributions to the thermal expansion tensor, $\alpha_{ij}^\lambda$, for each phonon with index $\lambda$.

The summation over phonon quantum numbers, $\lambda$, can be carried out to arrive at an expression independent of the phonon modes:

$$\alpha_{ij} = \left.\frac{\partial \varepsilon_{ij}}{\partial T}\right|_\sigma = \frac{1}{V}\frac{1}{N_q}\sum_\lambda S_{ijkl}\, \gamma_{kl}^\lambda\, c_v^\lambda(T) = \frac{1}{V} C_V S_{ijkl} \langle \gamma_{kl} \rangle \tag{S60}$$

Here, instead of the explicit summation over mode contributions in the thermal expansion tensor, quantities have been introduced which already incorporate this summation: the (phonon) heat capacity at constant volume, $C_V$, and the mean Grüneisen tensor, $\langle \gamma_{kl} \rangle$:

$$C_V(T) = \frac{1}{N_q}\sum_\lambda c_v^\lambda(T) \tag{S61}$$

$$\langle \gamma_{kl} \rangle(T) = \frac{\frac{1}{N_q}\sum_\lambda c_v^\lambda(T)\, \gamma_{kl}^\lambda}{\frac{1}{N_q}\sum_\lambda c_v^\lambda(T)} = \frac{\frac{1}{N_q}\sum_\lambda c_v^\lambda(T)\, \gamma_{kl}^\lambda}{C_V(T)} \tag{S62}$$

As it is of fundamental importance for the methodological route chosen in the main text, the practical steps necessary for the calculation of the thermal expansion tensor within the GT should be summarized in the following:

1. Calculating the complete (*i.e.*, in the entire first Brillouin zone) phonon band structure for the unstrained unit cell of the material.

2. Calculating the compliance tensor (or the elastic tensor) of the material. Within DFT, this is only possible at 0 K. That means that one, in principle, neglects the temperature dependence of the compliance tensor in the GT. For the present case, this is expected to have only a minor impact because the thermal expansion of MOF-74 is extra-ordinarily small such that the resulting changes in lattice parameters are likely to have essentially no impact on the calculated compliance tensor.

3. Calculating the necessary symmetry-inequivalent strain derivatives of the phonon frequencies to set up the mode Grüneisen tensors. Note that the Grüneisen theory explicitly considers the change in the phonon frequency with strain tensor to first order only. Consequently, one typically requires fewer strained configurations for which the (complete) phonon band structure must be calculated. In return, only small changes in the phonon frequencies with strain are considered because no higher-order derivatives are included in the GT. This might by a disadvantage compared to a suitable QHA-based approach for certain materials. For MOF-74, however, higher-order strain-derivatives are expected to have minor significance because of the extremely small thermal strains due to the thermal expansion. *I.e.,* in this case it is even preferable to recover a situation in which only the first-order contributions (at small strains) count.

4. Calculating the mean Grüneisen tensor and the heat capacity (using the mode heat capacity contributions from the unstrained cell).

5. Evaluating equation (S60).

As a last, not fully addressed aspect, the practical method for calculating the strain-derivatives of the phonon frequencies remains. It is, in fact, done relying on the same trick as shown above in equation (S53): instead of calculating the derivative of the phonon frequencies directly, this derivative is expressed as a derivative of the squared phonon frequency, which can, finally, be expressed as a derivative of the dynamical matrix, $\boldsymbol{D}$, [6,19] (making use of the phonon polarization vectors, $\boldsymbol{e}_\lambda$, of the unstrained cell):

$$\gamma_{ij}^\lambda = -\frac{1}{\omega_\lambda}\frac{\partial \omega_\lambda}{\partial \varepsilon_{ij}} = -\frac{1}{2\omega_\lambda^2}\frac{\partial \omega_\lambda^2}{\partial \varepsilon_{ij}} = -\frac{1}{2\omega_\lambda^2}(\boldsymbol{e}_\lambda)^T \frac{\partial \boldsymbol{D}}{\partial \varepsilon_{ij}} \boldsymbol{e}_\lambda \qquad (S63)$$

In practice, the derivative of the dynamical matrix is expressed *via* finite differences (FD). For reasons that will be explained below, we implemented a second-order and even a fourth-order FD scheme to increase the accuracy of this operation:

$$\frac{\partial \boldsymbol{D}}{\partial \varepsilon_{ij}} = \frac{\boldsymbol{D}(+\Delta\varepsilon_{ij}) - \boldsymbol{D}(-\Delta\varepsilon_{ij})}{2\Delta\varepsilon_{ij}} + \mathcal{O}\left((\Delta\varepsilon_{ij})^2\right) \tag{S64}$$

$$\frac{\partial \boldsymbol{D}}{\partial \varepsilon_{ij}} = \frac{-\boldsymbol{D}(+2\Delta\varepsilon_{ij}) + 8\boldsymbol{D}(+\Delta\varepsilon_{ij}) - 8\boldsymbol{D}(-\Delta\varepsilon_{ij}) + \boldsymbol{D}(-2\Delta\varepsilon_{ij})}{12\Delta\varepsilon_{ij}} + \mathcal{O}\left((\Delta\varepsilon_{ij})^4\right) \tag{S65}$$

Here, $\boldsymbol{D}(X)$ denotes the dynamical matrix of the system with a unit cell strained by the amount $X$.

Additionally, to reduce computational time, the symmetry of MOF-74 has been exploited. For rhombohedral crystals such as MOF-74, the mode Grüneisen tensor and the mean Grüneisen tensor only have two independent elements:

$$\boldsymbol{\gamma}^\lambda = \begin{pmatrix} \gamma_{11}^\lambda & 0 & 0 \\ 0 & \gamma_{11}^\lambda & 0 \\ 0 & 0 & \gamma_{33}^\lambda \end{pmatrix} ; \langle \boldsymbol{\gamma} \rangle = \begin{pmatrix} \langle \gamma_{11} \rangle & 0 & 0 \\ 0 & \langle \gamma_{11} \rangle & 0 \\ 0 & 0 & \langle \gamma_{33} \rangle \end{pmatrix} \tag{S66}$$

Instead of calculating the derivative of the phonon frequencies/dynamical matrix with respect to pure $\varepsilon_{11}$ or $\varepsilon_{22}$ strain, the derivative was calculated with respect to the following strain tensor – *i.e.*, applying $\varepsilon_{11}$ and $\varepsilon_{22}$ strain simultaneously:

$$\varepsilon = \begin{pmatrix} \epsilon & 0 & 0 \\ 0 & \epsilon & 0 \\ 0 & 0 & 0 \end{pmatrix} \tag{S67}$$

This has the advantage that, like for the case of applied $\varepsilon_{33}$ strain, the space group of the material is not changed, which would be the case if only $\varepsilon_{11}$ or $\varepsilon_{22}$ strain was applied separately. By simultaneously applying these strains (of the same magnitude) instead of straining the cell only with one of them, the required number of atomic displacements for a phonon calculation can be reduced from 162 to 54. Consequently, the needed computational resources are reduced by a factor of 3. However, at the end, one has to make sure that from such a calculation, still the correct tensor elements are extracted. This can be achieved by the following consideration exploiting that the derivatives of the frequencies with respect to $\varepsilon_{11}$ equal those with respect to $\varepsilon_{22}$ strain:

$$\frac{\partial \omega_\lambda}{\partial \epsilon} = \frac{\partial \omega_\lambda}{\partial \varepsilon_{11}}\frac{\partial \varepsilon_{11}}{\partial \epsilon} + \frac{\partial \omega_\lambda}{\partial \varepsilon_{22}}\frac{\partial \varepsilon_{22}}{\partial \epsilon} = \frac{\partial \omega_\lambda}{\partial \varepsilon_{11}} + \frac{\partial \omega_\lambda}{\partial \varepsilon_{22}} = 2\frac{\partial \omega_\lambda}{\partial \varepsilon_{11}} \tag{S68}$$

This means, the correct 11-component of the mode Grüneisen tensors are obtained in the following way:

$$\gamma_{11}^\lambda = \gamma_{22}^\lambda = -\frac{1}{\omega_\lambda}\frac{\partial \omega_\lambda}{\partial \varepsilon_{11}} = -\frac{1}{2\omega_\lambda}\frac{\partial \omega_\lambda}{\partial \epsilon} \tag{S69}$$

The parameters one can (freely) choose in practice are (i) the finite differences strain step size and (ii) the order of the FD scheme to approximate the derivative. The influence of (i) is shown in Figure S 24 using a step size of $10^{-3}$ and $2 \cdot 10^{-3}$ with a second-order FD scheme. Obviously, the larger step size is already too large to yield the same qualitative results as deduced from the experimental trends (negative $\alpha_{11}$). Prior to the discussion of further reducing the step size, a conclusion shall be drawn for the step size of $2 \cdot 10^{-3}$ already being too large. This is very likely the reason for the qualitative deviations in many of the QHA-based approaches: the probed changes in strains were (unrealistically) large, which, however, was necessary there to still be able to numerically resolve the small differences in free energies. This would explain why the QHA approaches typically tend to show larger thermal expansion values than the GT or the experimental trends. Thus, reducing the step size is supposed to increase the (mathematical) accuracy of approximating the derivative *via* the difference quotient.

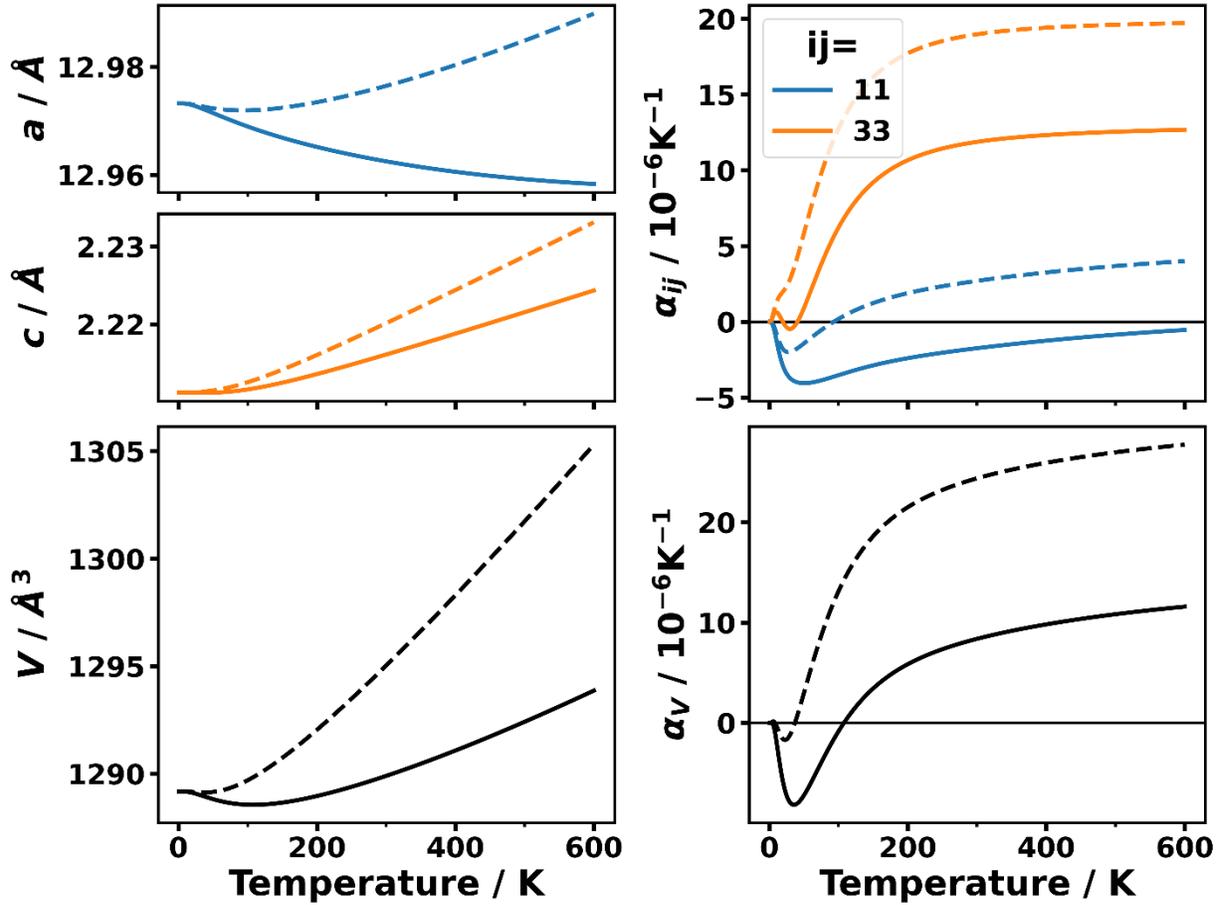

*Figure S 24: Temperature dependence of the two independent lattice parameters, a and c, and the unit-cell volume, V, of MOF-74 calculated with the Grüneisen theory described in Section S3.7. The right column shows the associated thermal expansion coefficients, $\alpha_{11}$, $\alpha_{33}$, and the volumetric thermal expansion coefficient $\alpha_V$. The solid lines show the results obtained with a finite differences scheme of second order using a strain step size of $10^{-3}$, while the dashed lines were obtained with a finite strain of twice that magnitude. The first Brillouin zone for the phonon calculations was sampled with a 20×20×20 mesh of wave vectors.*

The problem of trying to converge the results with respect to the finite strain step size is that convergence is unlikely to be observed for both, small and large step sizes: for the latter, the finite differences approximation becomes inaccurate, while for the former, numerical inaccuracies in the (imperfect) data overshadow the actual physical differences. In order to reduce the step size and, at the same time, improve the numeric robustness against fluctuations in the dynamical matrices for these small strains, a fourth-order finite differences scheme was implemented and applied. It is more precise by two orders of magnitude in the strain step size but considers the data from more finite strain steps (at the cost of doubling the computational effort to compute the necessary phonon band structures; see equation (S64) *vs.* (S65)). Thus, it is less prone to numerical noise. This means that to arrive at the same level of accuracy as the fourth-order scheme, the step size in the second-order scheme would have to be reduced by approximately two orders of magnitude. This could, however, only be done if the frequencies are infinitely accurately calculated. We note, in passing, that further reducing the order of magnitude of the finite strain step size would not be advisable as the changes become already difficult to resolve for the used strain magnitudes.

Figure S 25 shows a direct comparison between the thermal expansion behavior of MOF-74 predicted from the GT for a fourth-order (solid lines) and a second-order (dashed lines) FD scheme. It can be seen that the curves for the fourth-order scheme quantitatively but not qualitatively differ somewhat from the second-order results. This observation suggests that the fourth-order FD scheme yields results in the same direction as one would obtain when reducing the strain step size, yet without actually having to consider smaller step sizes for which the inherent numeric inaccuracies would increase. Thus, the fourth-order FD scheme presented here was finally employed to calculate the Grüneisen tensor and, subsequently, the thermal expansion tensor to achieve the desired level of accuracy.

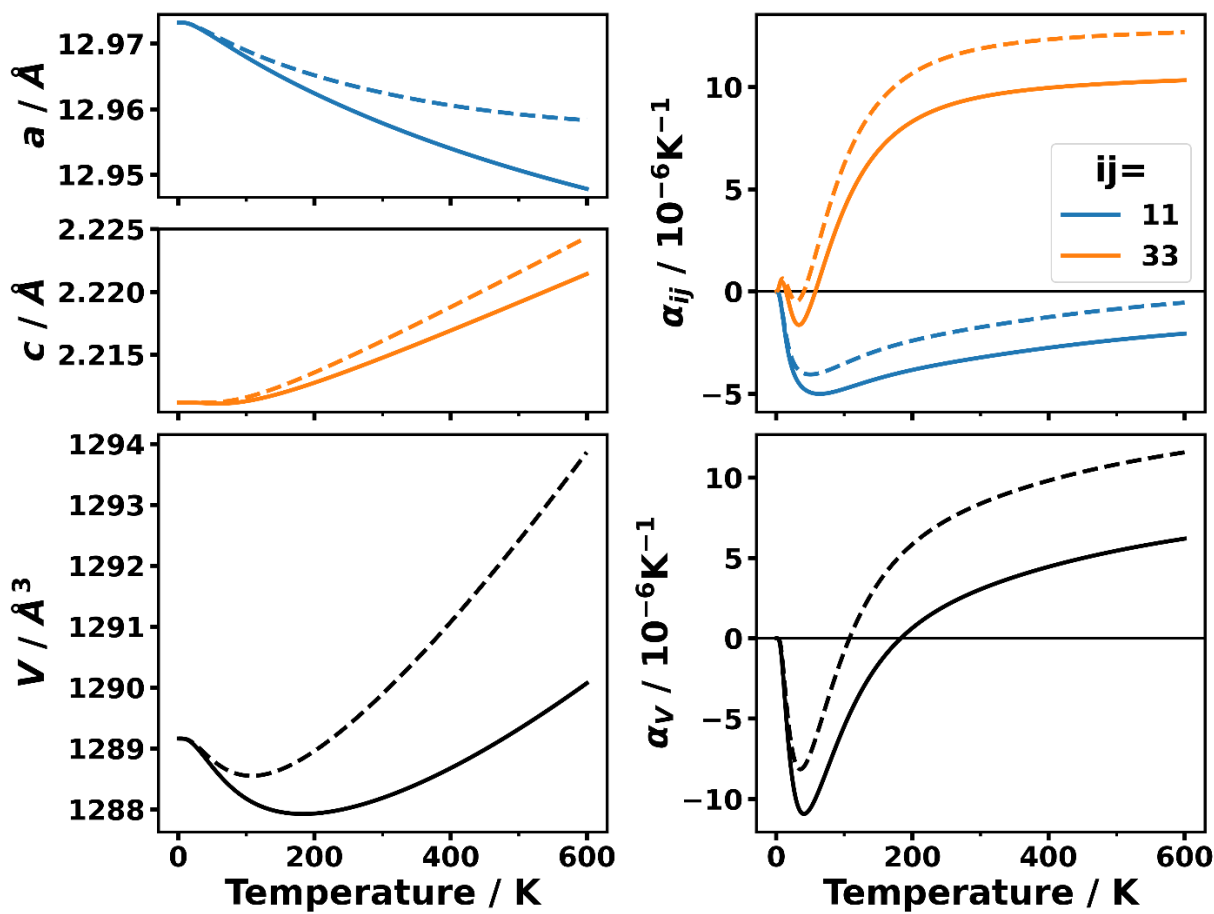

*Figure S 25: Temperature dependence of the two independent lattice parameters, a and c, and the unit-cell volume, V, of MOF-74 calculated with the Grüneisen theory described in Section S3.7. The right column shows the associated thermal expansion coefficients, $\alpha_{11}$, $\alpha_{33}$, and the volumetric thermal expansion coefficient $\alpha_V$. The solid lines show the results obtained with a finite differences scheme of fourth order, while the dashed lines correspond to a second-order scheme (both using a strain step size of $10^{-3}$). The first Brillouin zone for the phonon calculations was sampled with a 20×20×20 mesh of wave vectors.*

## S4. Computational details

### S4.1 Further settings for the density functional theory calculations

In addition to the general simulation parameters mentioned briefly in the main text, the following parameters for DFT calculations employing the *VASP* code (version 5.4.4) were chosen. The settings, which were held fixed for all calculations comprise (with *VASP* setting tags in parentheses; for details refer to the *VASP* manual[22]): the employed D3-BJ[23,24] *a posteriori* van der Waals correction (`IVDW = 12`), a convergence criterion for the self-consistent field calculation of $10^{-8}$ eV changes in the total energy (`EDIFF = 1E-8`), a Gaussian-type smearing of the electronic states with a width of 0.05 eV (`ISMEAR = 0; SIGMA = 0.05`), and high global precision (`PREC = Accurate`). Except for a few tests with the PBE functional [25,26] (see below), the PBEsol [27,28] (`GGA = PS`) was used to account for exchange-correlation contributions of the electrons in the system.

In order to find the unstrained unit-cell, the lattice parameters (together with the atomic positions) were optimized with a large plane wave energy cutoff of 1000 eV such that none of the residual forces exceeded 1 meV/Å. This was done directly with the conjugate gradient algorithm as implemented in *VASP* (`ISIF = 3`), yielding lattice parameters $a$ = 12.9732 Å and $c$ = 2.2117 Å.

The headers of the PAW-pseudopotentials that we used for the DFT calculations are listed in *Table S 2*.

*Table S 2: Headers of the used (standard) pseudopotentials for all atomic species.*

| Atomic species | Pseudopotential header |
|:---:|:---:|
| Zn | PAW_PBE Zn 06Sep2000 |
| O | PAW_PBE O 08Apr2002 |
| C | PAW_PBE C 08Apr2002 |
| H | PAW_PBE H 15Jun2001 |

## S4.2 Compliance Tensor

For the calculation of the compliance tensor, $S$, the elastic tensor, $C$, was calculated with a plane wave energy cutoff of 1000 eV and a 3×3×3 mesh of $k$-vectors as described in detail in Ref. [29]. Based on the detailed convergence tests in Ref. [29], these settings provide well-converged compliance tensor elements. Additionally, it should be stressed that the sum $(S_{11}+S_{12})$ appearing in the 11-component of the thermal expansion tensor in the Grüneisen theory, converges even much faster than the individual elements $S_{11}$ and $S_{12}$. The seven independent elements of the elastic tensor, $C$, and the compliance tensor, $S$, are listed in Table S 3. They can then be used to build the full tensors (in Voigt notation; superscript $V$; see Ref. [29] for more detail) according to the following equations.

$$C^V = \begin{pmatrix} C_{11} & C_{12} & C_{13} & C_{14} & C_{15} & 0 \\ C_{12} & C_{11} & C_{13} & -C_{14} & -C_{15} & 0 \\ C_{13} & S_{13} & C_{33} & 0 & 0 & 0 \\ C_{14} & -C_{14} & 0 & C_{44} & 0 & -S_{15} \\ C_{15} & -C_{15} & 0 & 0 & C_{44} & S_{14} \\ 0 & 0 & 0 & -C_{15} & C_{14} & \frac{1}{2}(C_{11}-C_{12}) \end{pmatrix} \quad (S70)$$

$$S^V = \begin{pmatrix} S_{11} & S_{12} & S_{13} & 2S_{14} & 2S_{15} & 0 \\ S_{12} & S_{11} & S_{13} & -2S_{14} & -2S_{15} & 0 \\ S_{13} & S_{13} & S_{33} & 0 & 0 & 0 \\ 2S_{14} & -2S_{14} & 0 & 4S_{44} & 0 & -4S_{15} \\ 2S_{15} & -2S_{15} & 0 & 0 & 4S_{44} & 4S_{14} \\ 0 & 0 & 0 & -4S_{15} & 4S_{14} & 2(S_{11}-S_{12}) \end{pmatrix} \quad (S71)$$

Table S 3: *The seven independent elements (in Voigt notation) of the elastic tensor, C (see equation (S70)), and the compliance tensor, S (see equation (S71)), of MOF-74 calculated withe the PBEsol functional with a plane-wave energy cutoff of 1000 eV. These shown tensor elements have been calculated within the study presented in Ref. [29].*

| ij | $C_{ij}$ / GPa | $S_{ij}$ / TPa$^{-1}$ |
|---|---|---|
| 11 | 22.79 | 180 |
| 33 | 14.87 | 68 |
| *12* | 18.31 | -155 |
| *13* | 1.75 | -3 |
| *44* | 17.13 | 22 |
| *14* | 3.50 | -34 |
| *15* | -0.73 | 7 |

### S4.3 Computational details and convergence tests concerning phonon-related quantities

For the phonon calculations, a supercell-based approach as provided by the *Phonopy* package [19] was employed. In order to reduce the computational cost at which the more demanding phonon band structure calculations would be carried out, the influence of the plane wave energy cutoff on the (more easily accessible) $\Gamma$-phonon frequencies (see Figure S 26) and anharmonic properties – *i.e.*, the Grüneisen tensor and the thermal expansion tensor at 300 K – (see Figure S 27) were studied. In order to save computational resources, only a second-order finite differences scheme was employed (with a step size of $2 \cdot 10^{-3}$) to calculate the Grüneisen tensors in these convergence tests.

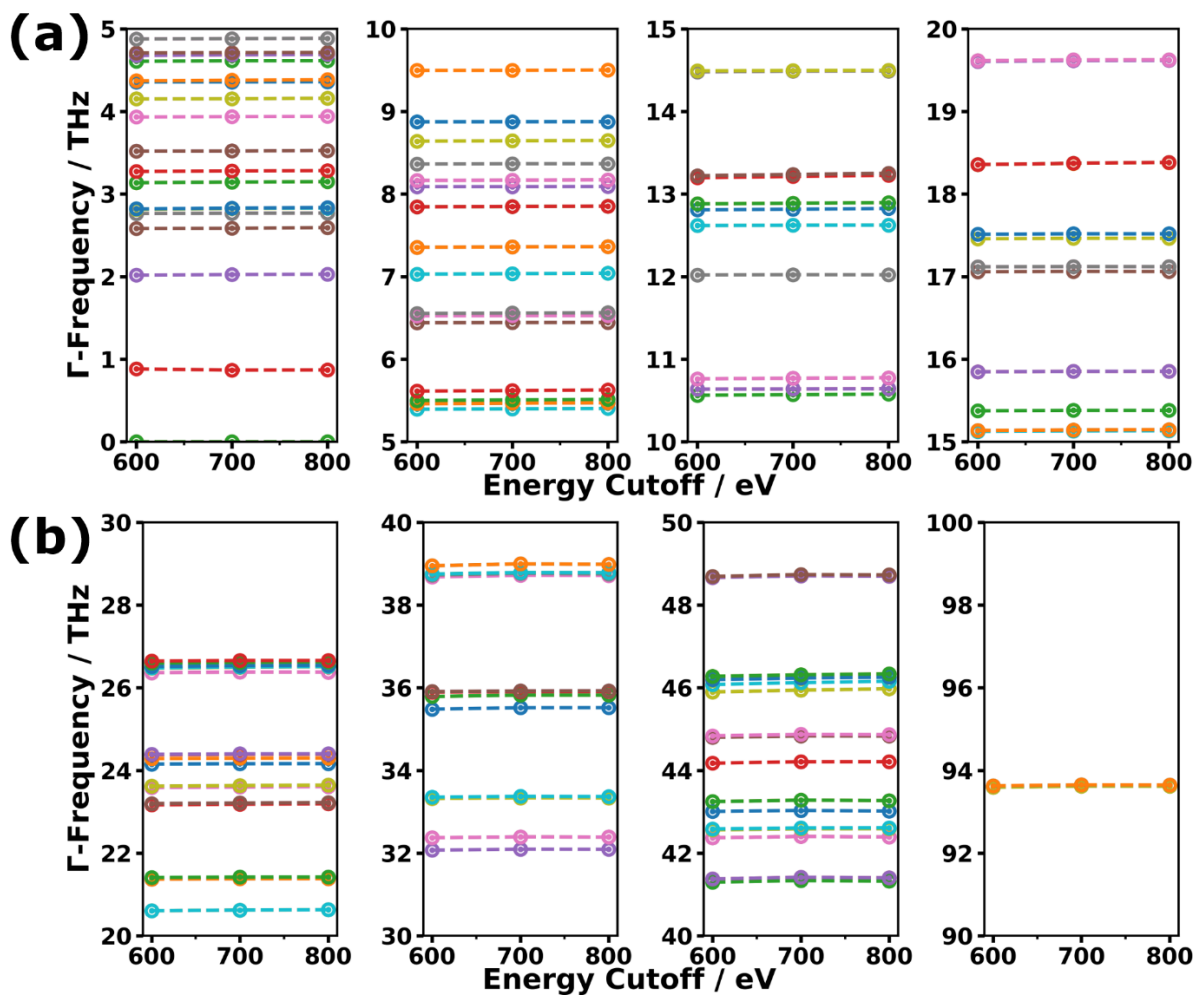

*Figure S 26: Γ-phonon frequencies as a function of the employed plane wave energy cutoff used for the phonon calculation with primitive unit cells of MOF-74 for (a) the low-frequency and (b) the high frequency range. For these tests, the **k**-meshes was fixed at a 3×3×3 mesh, and neither the atomic positions nor the lattice parameters were reoptimized with the respective cutoffs.*

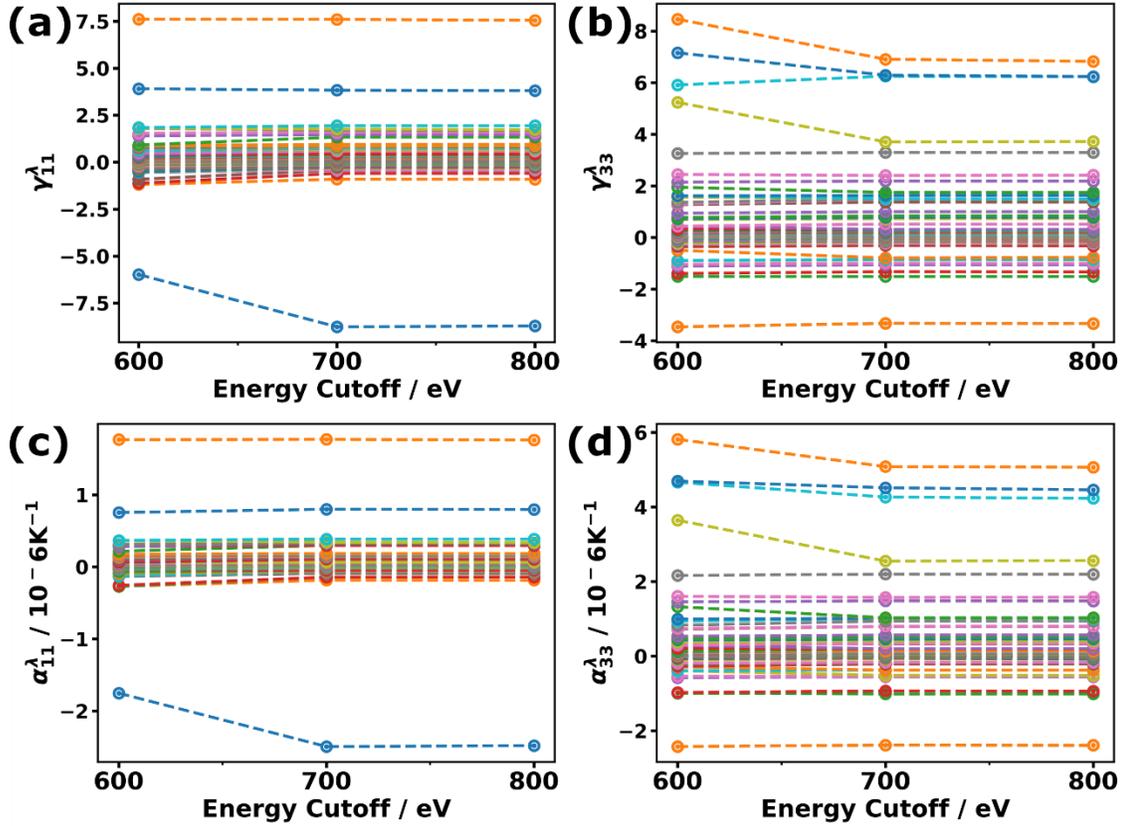

*Figure S 27: (a) 11-components of the mode Grüneisen tensor, $\gamma_{11}^{\lambda}$, (b) 33-components of the mode Grüneisen tensor, $\gamma_{33}^{\lambda}$, (c) 11-components of the mode contributions to the thermal expansion tensor, $\alpha_{11}^{\lambda}$, at 300 K and (d) 33-components of the mode contributions to the thermal expansion tensor, $\alpha_{33}^{\lambda}$, at 300 K at $\Gamma$ as a function of the employed plane wave energy cutoff used for the phonon calculation with primitive unit cells of MOF-74. For obtaining the mode Grüneisen tensor elements (and, subsequently, the mode thermal expansion tensors), a second-order finite differences scheme was applied with a strain step size of $2 \cdot 10^{-3}$ (see Section S3.7). Additionally, the **k**-meshes were fixed at 3×3×3, and neither the atomic positions nor the lattice parameters were reoptimized with the respective cutoffs.*

These convergence tests clearly show that not only the frequencies of the phonons at $\Gamma$ are converged at an excellent level as they do not show significant changes with the energy cutoff, but also the derived anharmonic properties show essentially no differences between the calculations carried out with energy cutoffs of 700 eV and 800 eV. Thus, we chose to reduce the energy cutoff to 800 eV for all phonon band structure (and Grüneisen tensor) calculations. In order to assure that the geometry is also in a minimum of the potential energy surface, the atomic positions (but not the lattice vectors) were reoptimized (to a maximum residual force of 1 meV/Å) with the cutoff of 800 eV prior to any phonon band structure calculations.

As another simulation parameter, we tested the convergence of Γ-phonon frequencies (see Figure S 28) and Grüneisen/thermal expansion tensor elements (see Figure S 29) with the density of electronic wave vectors, $k$. To be comparable, we calculated the $k$-density as the total number of $k$-points per reciprocal volume of the used unit cell, $V^* (= (2\pi)^{-1}V$; with $V$ being the volume in real space). Obviously, beyond a $k$-mesh that consists only of the (electronic) Γ-point, the zone-center phonon frequencies show no perceivable changes with increasing $k$-density, and densities >1640 Å$^3$ are expected to yield both well-converged frequencies and anharmonic properties. Moreover, the plots suggest that the $k$-density for the phonon band structure calculations employing 1×1×3 supercells of the conventional cell (see below), which were identified to be a good compromise between accuracy and computational cost (see below), is sufficiently dense to yield the desired level of convergence. The associated $k$-density (calculated as one $k$-point per a ninth of the reciprocal volume of the primitive unit cell) is shown in Figure S 28 and Figure S 29 as vertical dotted black lines.

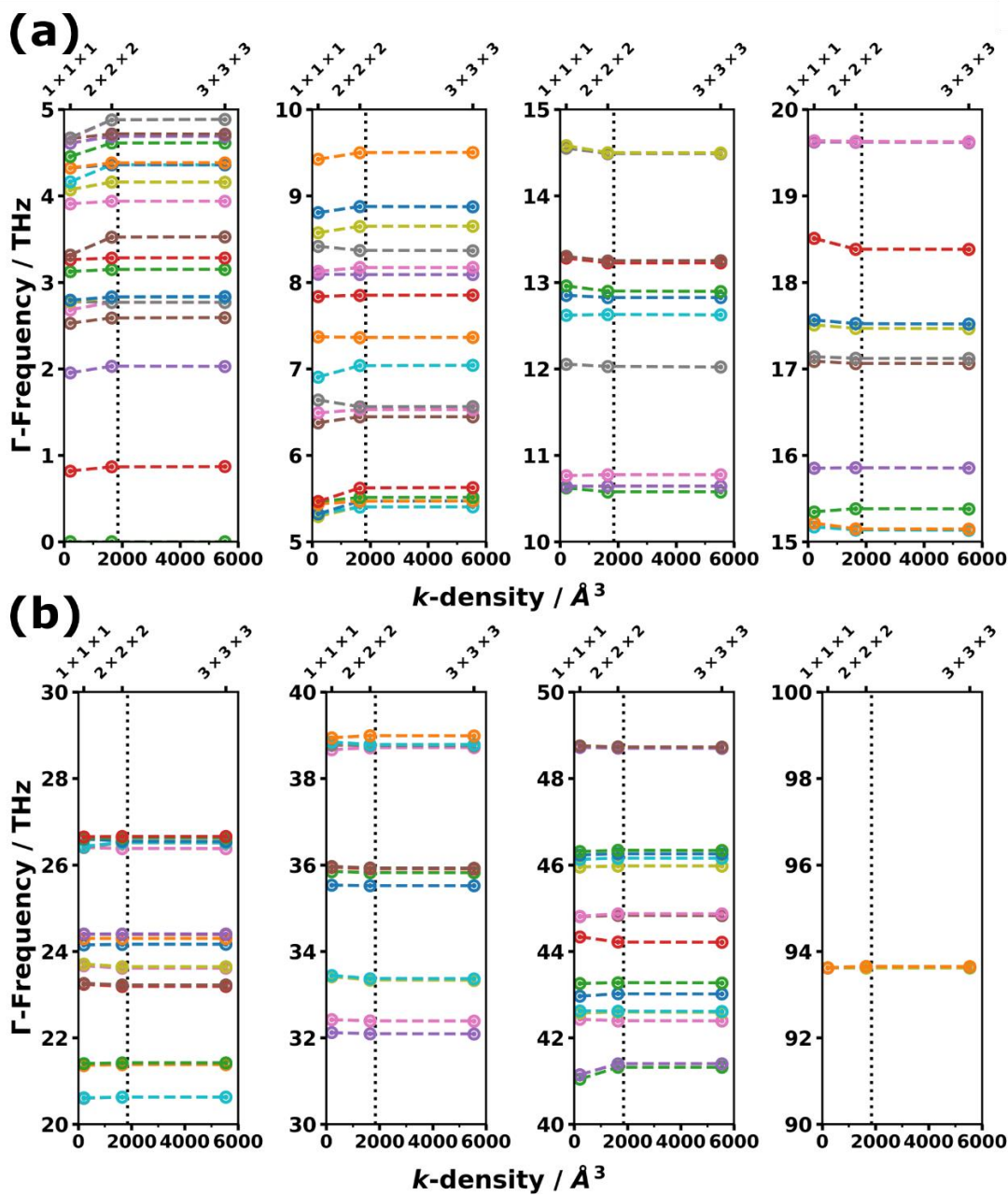

*Figure S 28: Γ-phonon frequencies as a function of the density of electronic wave vectors, **k**, used to sample the first Brillouin zone in primitive unit cells of MOF-74 for (a) the low-frequency and (b) the high frequency range. For these tests, the plane wave energy cutoff was fixed at 800 eV, and neither the atomic positions nor the lattice parameters were reoptimized with the respective **k**-meshes. The Γ-centered meshes can be seen in the horizontal axes on the top. The vertical dotted line corresponds to the wave vector density in reciprocal space for the supercell employed for the phonon band structure calculations (a 1×1×3 supercell of the conventional unit cell) sampled only with one **k**-point (Γ).*

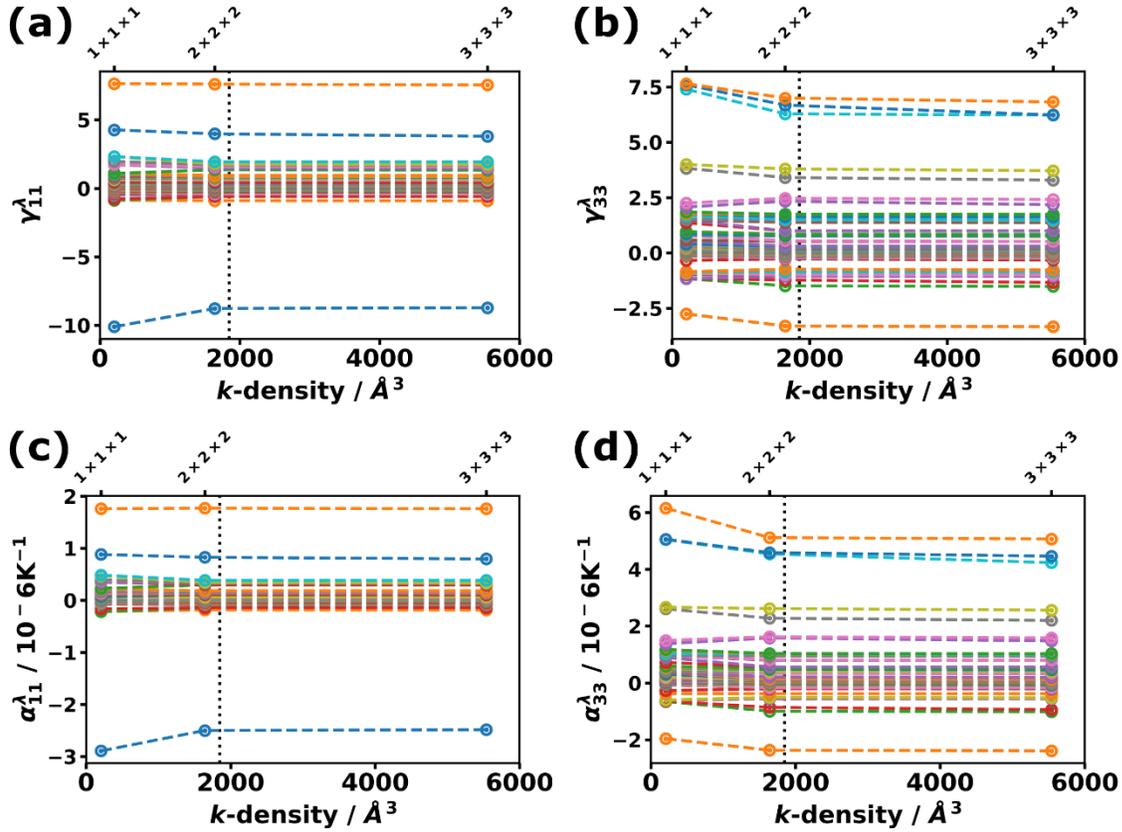

*Figure S 29: (a) 11-components of the mode Grüneisen tensor, $\gamma_{11}^\lambda$, (b) 33-components of the mode Grüneisen tensor, $\gamma_{33}^\lambda$, (c) 11-components of the mode contributions to the thermal expansion tensor, $\alpha_{11}^\lambda$, at 300 K and (d) 33-components of the mode contributions to the thermal expansion tensor, $\alpha_{33}^\lambda$, at 300 K at Γ as a function of the density of electronic wave vectors, **k**, used to sample the first Brillouin zone in primitive unit cells of MOF-74. The Γ-centered meshes can be seen in the horizontal axes on the top. The vertical dotted line corresponds to the wave vector density in reciprocal space for the supercell employed for the phonon band structure calculations (a 1×1×3 supercell of the conventional unit cell) sampled only with one **k**-point (Γ). For obtaining the mode Grüneisen tensor elements (and, subsequently, the mode thermal expansion tensors), a second-order finite differences scheme was applied with a strain step size of $2 \cdot 10^{-3}$ (see Section S3.7 ). Additionally, the plane wave energy cutoff was fixed at 800 eV, and neither the atomic positions nor the lattice parameters were reoptimized with the respective **k**-meshes.*

Finally, the convergence behavior of the calculated phonon-related quantities with respect to supercell size (in addition to the plane wave energy cutoff and the *k*-mesh) is an additional aspect that was studied. First, we would like to comment on the convergence behavior of the phonon band structure with respect to the supercell size. As pointed out in literature (see, *e.g.*, [30]), the probing of interatomic force constants in real space becomes very inefficient for unit cells with significantly anisotropic extents. Choosing a "more cubic" supercell instead is more beneficial, as it allows for a probing the force constants (approximately) isotropically. However, based on the primitive unit cell (see Figure S 30(a)), such a isotropic replication is impossible. Instead, one can rely on supercells of the conventional unit cell, which is displayed in Figure S 30(b). Note that the conventional unit cell contains three primitive unit cells replicated on a hexagonal Bravais lattice. The Cartesian lattice vectors of the conventional cell, $\vec{a}_n^{(c)}$, can be obtained from those of the primitive cell, $\vec{a}_m^{(p)}$, with the following transformation matrix, $M$:

$$\begin{pmatrix} 2a & 0 & 0 \\ -a & a\sqrt{3} & 0 \\ 0 & 0 & 3c \end{pmatrix} = \begin{pmatrix} \vec{a}_1^{\,c} \\ \vec{a}_2^{\,c} \\ \vec{a}_2^{\,c} \end{pmatrix} = M \begin{pmatrix} \vec{a}_1^{\,p} \\ \vec{a}_2^{\,p} \\ \vec{a}_2^{\,p} \end{pmatrix} = \begin{pmatrix} 1 & -1 & 0 \\ 0 & 1 & -1 \\ 1 & 1 & 1 \end{pmatrix} \begin{pmatrix} a & a/\sqrt{3} & c \\ -a & a/\sqrt{3} & c \\ 0 & -2a/\sqrt{3} & c \end{pmatrix} \quad (S72)$$

The conventional unit cell is, however, still much smaller in $\vec{a}_3$-direction ($3c = \sim 6.63$ Å; see Figure S 30(b)) than along the other two lattice vectors ($2a = \sim 25.95$ Å), which meet at an angle of 120° and are both perpendicular to $\vec{a}_3$ (see Figure S 30(a) and (b)). Therefore, a logical way to generate supercells, which allow probing the real space interatomic force constants more uniformly, would be to increase the size along the third lattice vector first.

As shown in Figure S 30(c), we calculated the phonon band structures using the primitive unit cell ("p1×1×1"), the conventional unit cell ("c1×1×1"), a 1×1×2 supercell of the conventional cell ("c1×1×2"), and a 1×1×3 supercell of the conventional cell ("c1×1×3"). Interestingly, already the primitive unit cell shows good accuracy of phonon band structures for wave vectors in those directions in which the first Brillouin zone is particularly small.

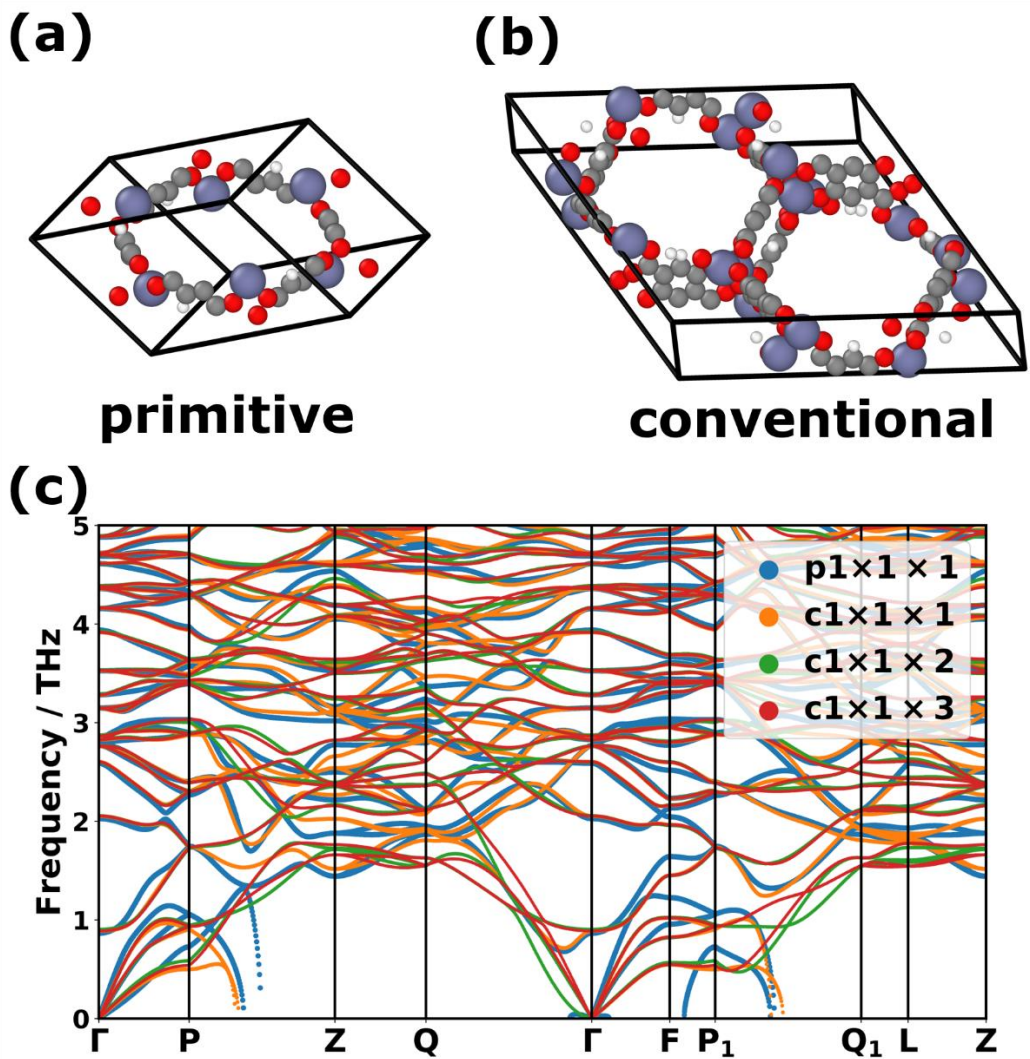

*Figure S 30: (a) Primitive and (c) conventional unit cells of MOF-74. Atomic color coding: Zn…purple, O…red, C…gray, H…white, plotted with Ovito (version 3.3.1) [31]. (c) Comparison of the PBEsol-calculated phonon band structures of MOF-74 as a function of the employed supercells based on either primitive ("p") or conventional ("c") supercells.*

One can see in Figure S 30(c) that the changes in the phonon band structures decrease as the extent of the supercell in the direction of the third dimension of the conventional cell is increased.

This is especially true for the "long" directions in reciprocal space such as ΓQ, PZ, or $P_1Q_1$. Somewhat surprisingly, the 1×1×2 supercell of the conventional cell ("c1×1×2") shows some deviations from the converged phonon band structure in regions in which even the conventional unit cell ("c1×1×1") seems to be converged at a better level (*e.g.*, along ΓP or $FP_1$). The reason for this is most probably the fact that for the 1×1×2 supercell of the conventional cell, the electronic structure calculations were carried out only with one electronic wave vector, *i.e.*, with a slightly reduced density of *k*-vectors than for all other supercells (~1231 Å$^3$ instead of ~ 5540 Å$^3$, ~1847 Å$^3$ and ~1847 Å$^3$ for the primitive unit cell and the 1×1×1, and 1×1×3 supercells of the conventional cell, respectively) and, thus, lies somewhat below the *k*-density that was identified above to yield well-converged results. In these tests, it was, however, technically impossible to hold the *k*-density constant for all employed supercells.

The same reason – *i.e.*, the slightly reduced density of *k*-vectors during the (electronic) calculation of interatomic forces – is also held responsible for the observation that the results obtained with this supercell slightly deviate from the trend observed for all the remaining supercell sizes regarding the convergence of mode Grüneisen and thermal expansion tensor elements (see Figure S 31). Note that here only Γ-phonons were tested explicitly because such tests would not be really sensible for supercells for which one observes negative phonon frequencies (such as the primitive and conventional unit cells).

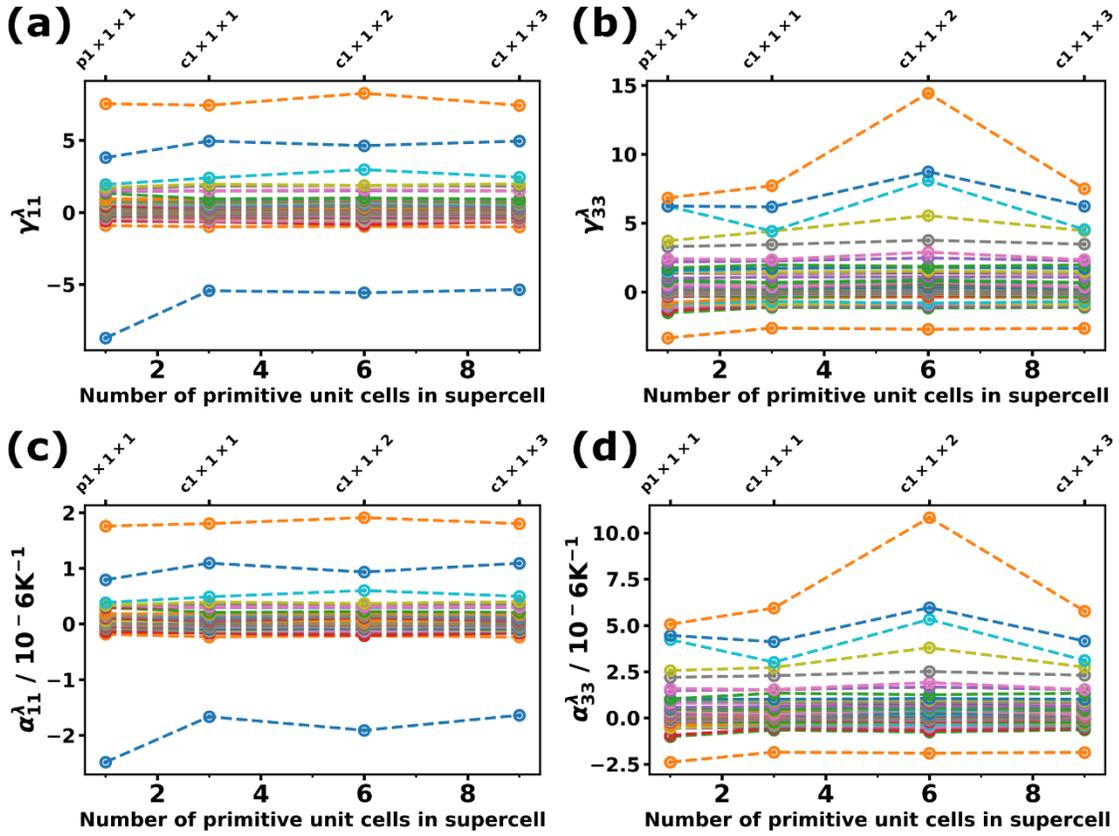

*Figure S 31: (a) 11-components of the mode Grüneisen tensor, $\gamma_{11}^\lambda$, (b) 33-components of the mode Grüneisen tensor, $\gamma_{33}^\lambda$, (c) 11-components of the mode contributions to the thermal expansion tensor, $\alpha_{11}^\lambda$, at 300 K and (d) 33-components of the mode contributions to the thermal expansion tensor, $\alpha_{33}^\lambda$, at 300 K at Γ as a function of the supercell size used for phonon calculations. The supercell sizes are determined by three integer multiplicators describing, how often the primitive (p) or the conventional (c) unit cell has been replicated. This is indicated at the top horizontal axes. The conventional 1×1×2 supercell was electronically sampled with a slightly reduced density of **k**-vectors compared to all other supercells (**k**-point density of ~1231 Å³ instead of ~5540 Å³, ~1847 Å³, and ~1847 Å³ for the primitive unit cell and the 1×1×1 and 1×1×3 conventional supercells, respectively). This is held responsible for the observed disruption in the convergence behavior for this supercell size – especially for $\gamma_{33}^\lambda$ and $\alpha_{33}^\lambda$. For obtaining the mode Grüneisen tensor elements (and, subsequently, the mode thermal expansion tensors), a second-order finite differences scheme was applied with a strain step size of $2 \cdot 10^{-3}$ (see Section S3.7 ).*

We note, in passing, that the Grüneisen tensors in Figure S 31 were only calculated based on a second-order finite differences scheme (strain step size $2 \cdot 10^{-3}$) to save computational time (by reducing the number of necessary phonon band structures by a factor of 2). Besides the outliers resulting from the 1×1×2 conventional supercell, the convergence behavior of the Grüneisen tensor and also the mode contributions to the thermal expansion tensor can be considered converged for supercells of at least the size of a 1×1×1 conventional cell (which, however, does not yield real-valued phonon frequencies in the entire first Brillouin zone such that larger supercells must be taken for obtaining non-imaginary phonon bands). As a good compromise between numerical accuracy and computational cost, we chose a 1×1×3 conventional supercell to calculate the phonon band structure and the mode Grüneisen tensor elements as the necessary ingredients for the thermal expansion tensor. As an additional supporting argument, this choice of supercell size is the same as Wang *et al.* [32] employed for their calculations of lattice thermal conductivity of MOF-74.

Finally, as a last simulation parameter, the impact of the used density of phonon wave vectors, $q$, on the calculated thermal expansion tensor was investigated. Figure S 32 shows that the thermal expansion tensor and the volumetric thermal expansion (calculated with a fourth-order finite differences scheme with a strain step size of $10^{-3}$, from phonon band structures based on 1×1×3 conventional supercells with an energy cutoff of 800 eV) rapidly converge with the $q$-mesh: while the results using only the Γ-phonons (1×1×1 $q$-mesh) shows notable deviations from the converged results with dense $q$-meshes, already a 5×5×5 $q$-mesh is extremely close to the results of the 20×20×20 $q$-mesh. As the choice of the $q$-mesh is not really time-critical, we consistently employed 20×20×20 $q$-meshes for calculating thermodynamic properties.

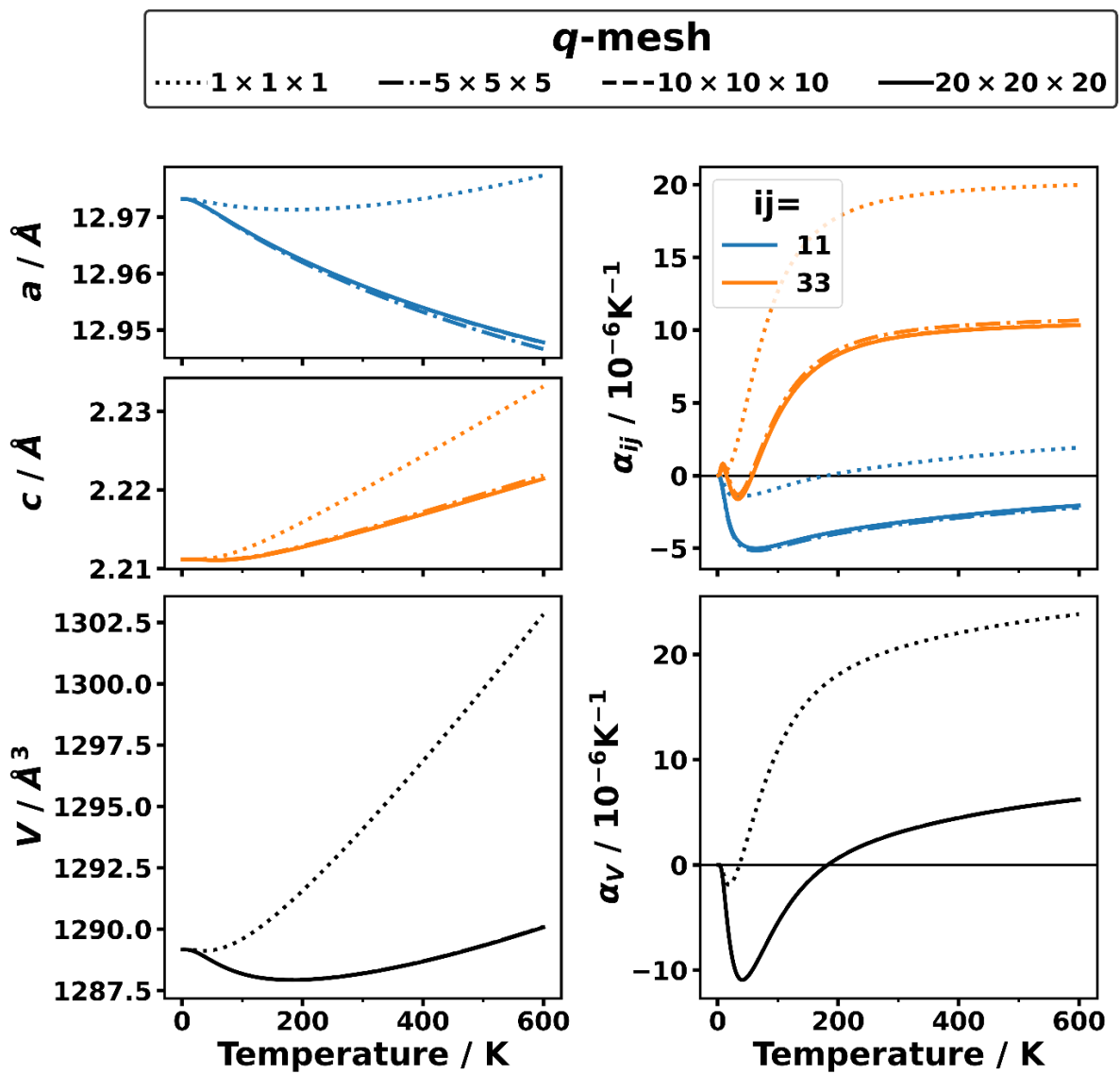

*Figure S 32: Temperature dependence of the two independent lattice parameters, a and c, and the unit-cell volume, V, of MOF-74 calculated with the Grüneisen theory described in Section S3.7. The right column shows the associated thermal expansion coefficients, $\alpha_{11}$, $\alpha_{33}$, and the volumetric thermal expansion coefficient $\alpha_V$. The different line styles (dotted, dash-dotted, dashed, and solid) denote different choices of meshes of phonon wave vectors, **q**, used to sample the first Brillouin zone for the phonon calculations. For obtaining the mode Grüneisen tensor elements (and, subsequently, the mode thermal expansion tensors), a fourth-order finite differences scheme was applied with a strain step size of $10^{-3}$ (see Section S3.7 ).*

## S5. Testing the choice of the functional: PBEsol vs. PBE

Finally, the thermal expansion tensor was also calculated with several of the approaches presented above using the PBE functional and the same settings as for the PBEsol-calculations. The only exception is the calculation of compliance tensor, $S$, and the elastic tensor, $C$, which were obtained at a plane wave energy cutoff of an available 1200 eV calculation [29] (compared to the cutoff of 1000 eV for PBEsol results). The values of the seven independent elements of the PBE-calculated compliance tensor amount to 165 $TPa^{-1}$, -142 $TPa^{-1}$, 77 $TPa^{-1}$, -5 $TPa^{-1}$, 22 $TPa^{-1}$, -31 $TPa^{-1}$, and 7 $TPa^{-1}$ for the $S_{11}$, $S_{12}$, $S_{33}$, $S_{13}$, $S_{44}$, $S_{14}$, and $S_{15}$ elements, from which the compliance tensor (in Voigt notation) can be constructed using equation (S71). Additionally, the choice of the functional slightly impacts the lattice parameters, $a$ and $c$, which, after a geometry optimization based on a Rose-Vinet equation of state as described in Ref. [29], amount to 13.065 Å and 2.230 Å, respectively.

### S5.1 QHAiso

As a first step, the QHAiso approach as described in Section S3.3 was applied to 13 unit cells of MOF-74 with the $c/a$-ratio, $\zeta$, optimized only electronically. Figure S 33 shows the non-equilibrium free energies as a function of the unit-cell volume (abscissa) and the temperature (labels and color scale). Also in this case, similar to the PBEsol results using the QHAiso approach, a distinct positive volumetric thermal expansion is observed (see Figure S 34). Additionally, both thermal expansion tensor elements are positive in almost the entire plotted temperature region.

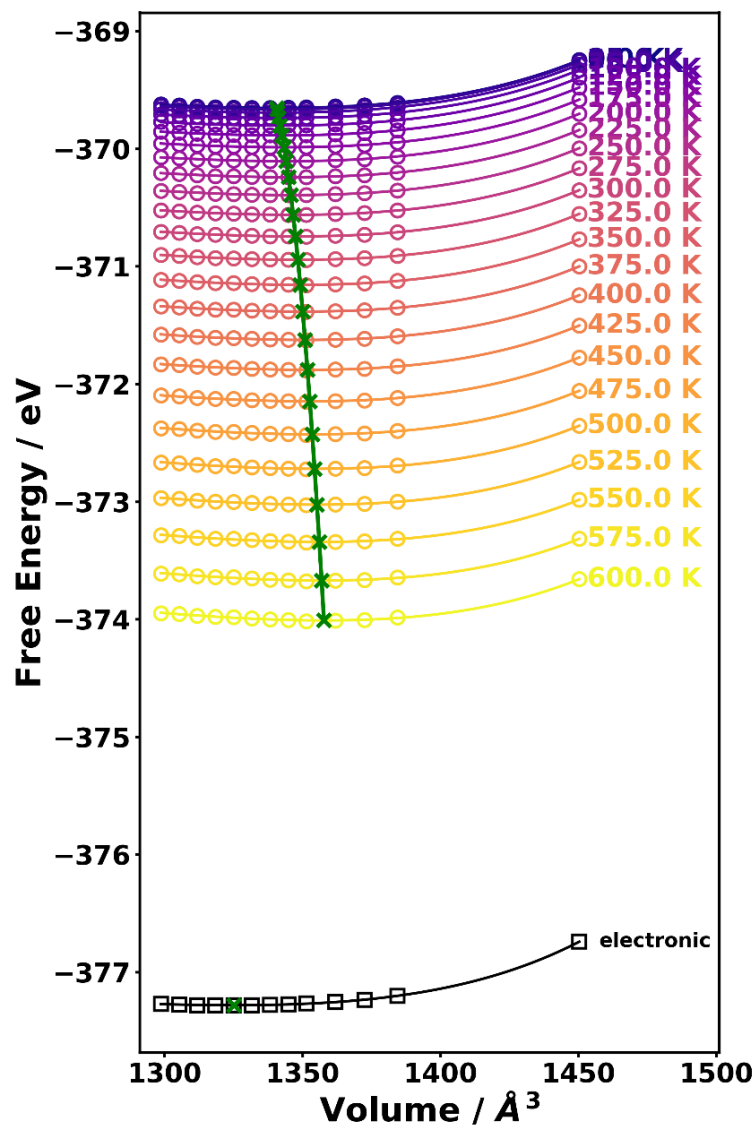

*Figure S 33: Total free enthalpy at zero applied stress (=free energy) of MOF-74 as a function of the unit-cell volume (abscissa) and the temperature (color scale and labels) obtained with the QHAiso approach using the PBE functional. The open symbols denote the total (i.e., electronic and phonon) free energies calculated as a function of the unit-cell volume with (electronically) optimized c/a-ratio, while the solid lines correspond to the fitted Rose-Vinet EoS at each temperature. The green crosses mark the minimum free energy and the volume that minimizes the free energy at each temperature. The temperature dependence of those marked volumes yields the volumetric thermal expansion. Additionally, the electronic energies (together with a fitted Rose-Vinet EoS) are shown.*

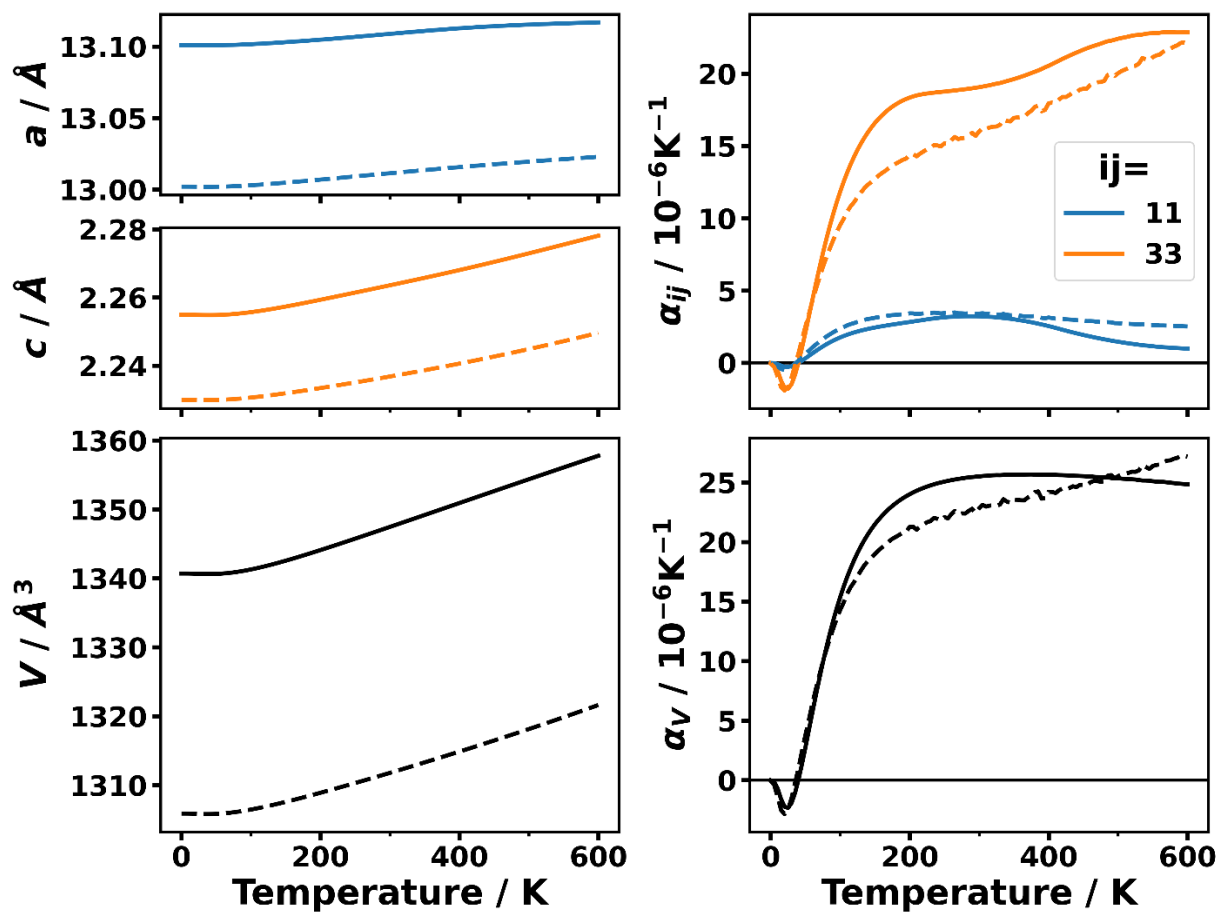

*Figure S 34: Temperature dependence of the two independent lattice parameters, a and c, and the unit-cell volume, V, of MOF-74 calculated with the QHA-procedure described in Section S3.3 (QHAiso) using the PBE functional. The right column shows the associated thermal expansion coefficients, $\alpha_{11}$, $\alpha_{33}$, and the volumetric thermal expansion coefficient $\alpha_V$. The solid lines show the results obtained with the PBE functional, while the dashed lines correspond to the results with the PBEsol functional.*

### S5.2 QHAaniso

Also the QHAaniso approach was applied to PBE-calculated free enthalpies. To this end, 36 configurations of pairs of lattice parameters, *a* and *c*, were considered (see Figure S 35(a)). For these configurations, the total (non-equilibrium) free enthalpy (or free energy at vanishing applied stress) was evaluated and fitted to various model functions as described in Section S3.5 for the PBEsol results. From those fits, the lattice parameters which minimize the (non-equilibrium) free enthalpies, can be identified as exemplarily shown in Figure S 35(b).

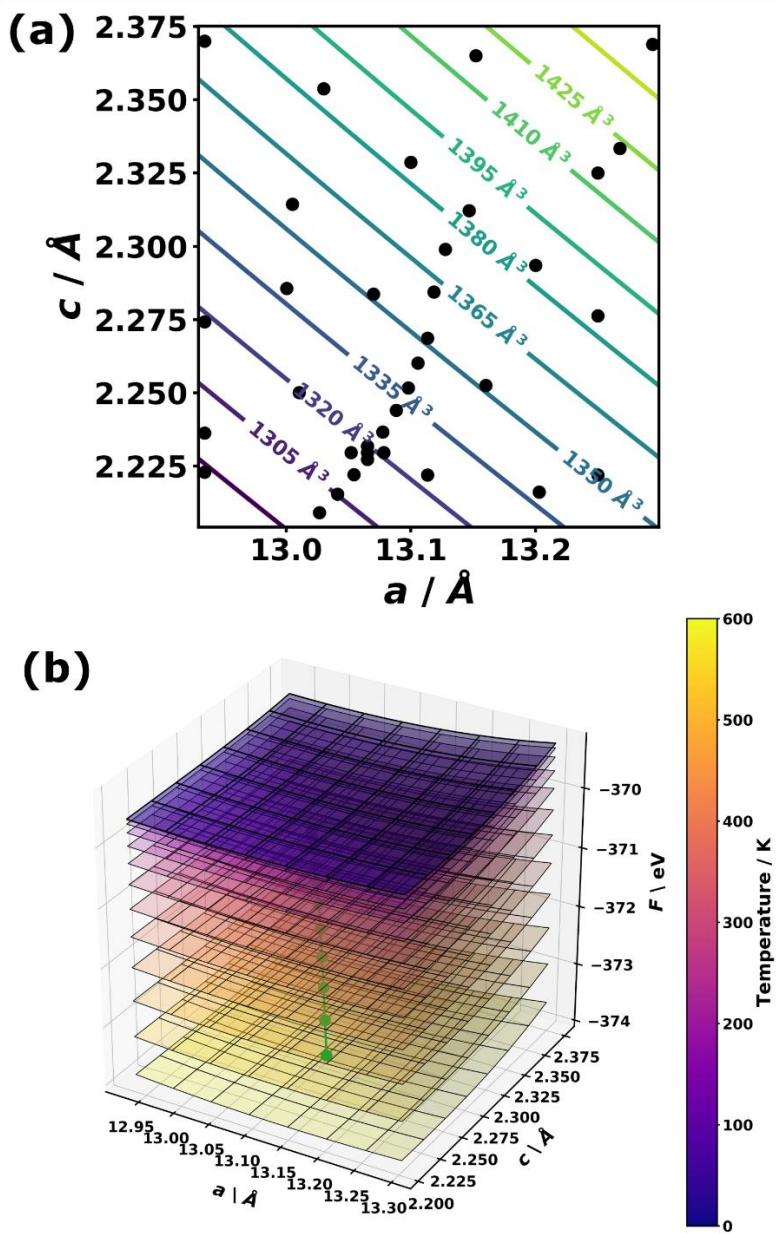

*Figure S 35: (a) Two-dimensional space of the lattice constants a and c, for which (non-equilibrium) free enthalpies were calculated from full (PBE-calculated) phonon band structures. The black dots denote the 36 strained configurations with different lattice parameters, while the hyperbolic contour lines show the curves of constant unit-cell volume. (b) Fitted total free enthalpy at zero applied stress (=free energy) of MOF-74 calculated using the PBE functional as a function of the two lattice parameters (abscissa) and the temperature (color scale) obtained with the*

*QHAaniso approach using a third-order polynomial in the space of the lattice parameters, a and c. The green circles mark the minimum free energy and the pair of lattice constants, (a,c), that minimizes the free energy at each temperature. The temperature dependence of those marked positions yields the thermal expansion tensor.*

The thermal expansion coefficients as obtained using various model fit functions as described in Section S3.5 are contained in Figure S 11 to Figure S 19 in a direct comparison with the PBEsol results. Note that in two cases (Birch-Murnaghan EoSs of third and fourth order based on Lagrangian strains) such a direct comparison was not possible for reasons of visibility. Thus, the results for these model functions are shown separately in Figure S 36 and Figure S 37. For the majority of used model functions qualitatively correct thermal expansion tensor components (*i.e.*, they have the expected signs) are obtained, although the thermal expansion of lattice parameter *c* is typically significantly overestimated resulting in notably positive volumetric thermal expansion. The best agreement with the expected thermal expansion tensor is exerted by the results obtained using a Birch-Murnaghan-like (BM) equation of state (EoS) of second or third order based on Eulerian strains (see Figure S 18 and Figure S 19), although $\alpha_V$ is much larger than expected in both cases.

Directly comparing the PBE and PBEsol results, it can be seen that the thermal expansion of lattice parameter *c* tends to be smaller with PBE than with PBEsol in almost all cases, while, in particular at low temperatures, the results with both functionals are in good agreement. The tendency to yield lower thermal expansion coefficients for the lattice parameter with the PBE functional might be a consequence of the notably increased (0 K) equilibrium value of *c* with PBE (~2.230 Å) compared to the PBEsol-calculated value (~2.212 Å).

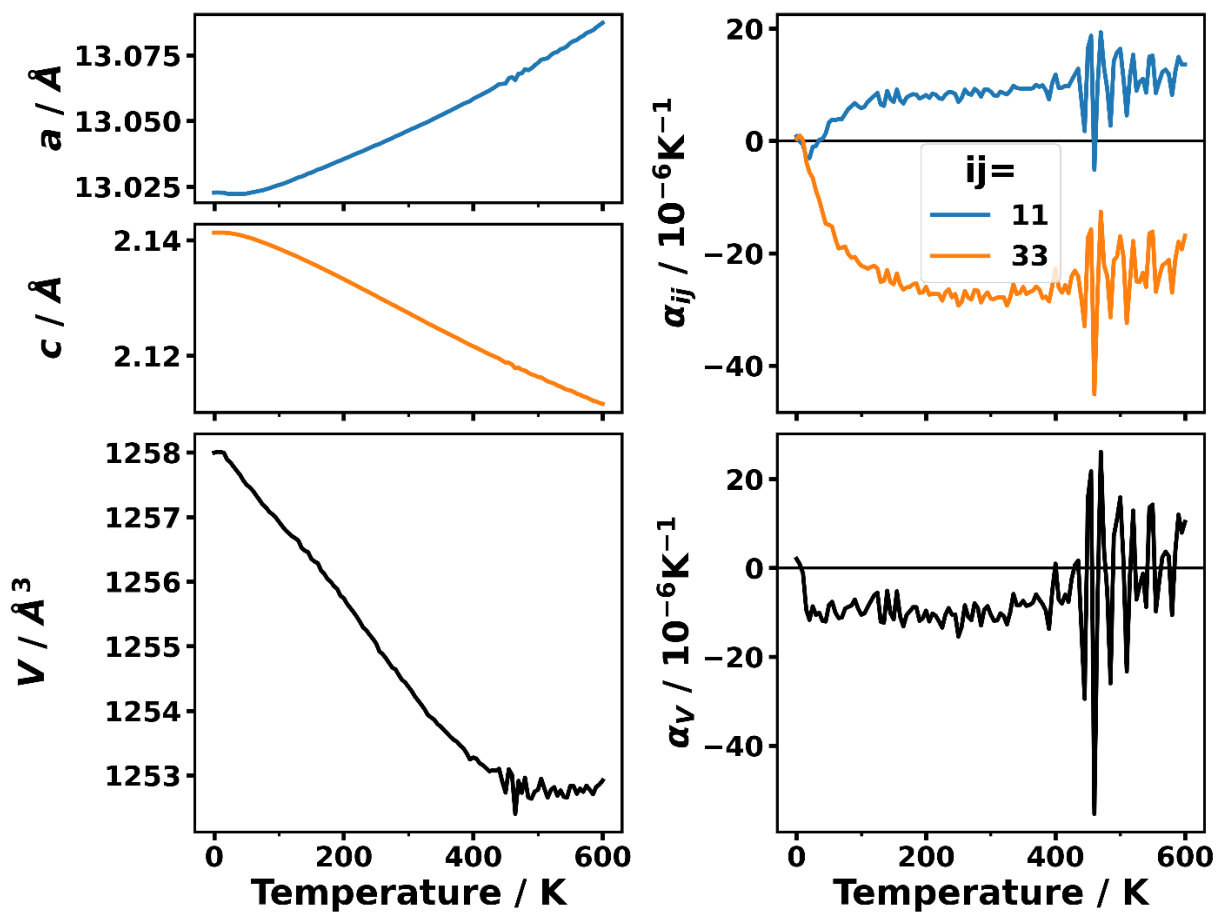

*Figure S 36: Temperature dependence of the two independent lattice parameters, a and c, and the unit-cell volume, V, of MOF-74 calculated with the QHA-procedure described in Section 0 (QHAaniso) using the PBE functional based on a fit using a third-order polynomial to model the dependences of the total free enthalpy on the finite Lagrangian strains (Birch-Murnaghan-like EoS of third order). The right column shows the associated thermal expansion coefficients, $\alpha_{11}$, $\alpha_{33}$, and the volumetric thermal expansion coefficient $\alpha_V$. For reasons of visibility, the corresponding PBEsol results from Figure S 16 are not shown.*

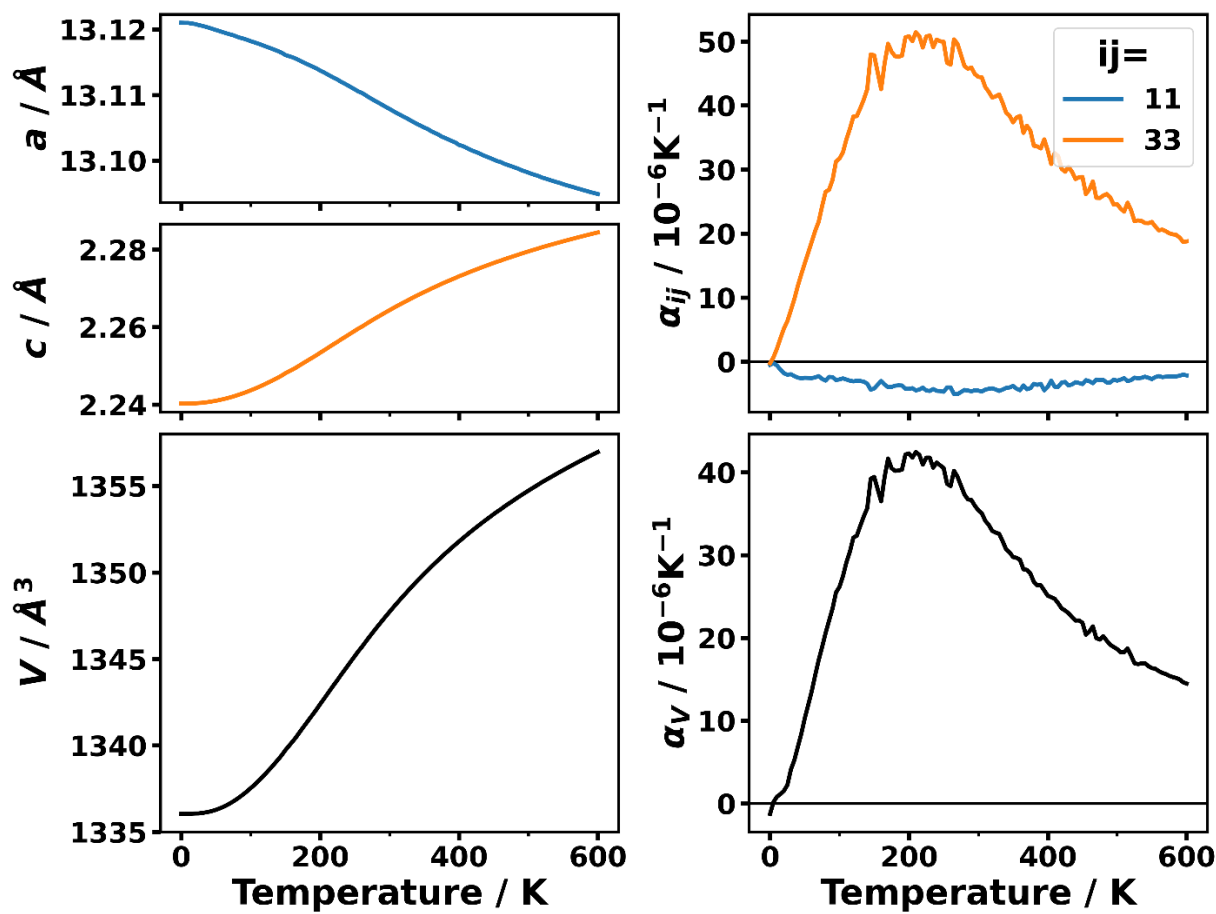

*Figure S 37: Temperature dependence of the two independent lattice parameters, a and c, and the unit-cell volume, V, of MOF-74 calculated with the QHA-procedure described in Section 0 (QHAaniso) using the PBE functional based on a fit using a fourth-order polynomial to model the dependences of the total free enthalpy on the finite <u>Lagrangian</u> strains (Birch-Murnaghan-like EoS of fourth order). The right column shows the associated thermal expansion coefficients, $\alpha_{11}$, $\alpha_{33}$, and the volumetric thermal expansion coefficient $\alpha_V$. For reasons of visibility, the corresponding PBEsol results from Figure S 17 are not shown.*

In order to unambiguously judge how well the various tested analytical model functions describe the actually calculated (non-equilibrium) free energies, the *Rwp* and *S* values according to equations (S46) and (S47) were evaluated based on the PBE-results. According to the "goodness of fit", *S*, Figure S 38 shows that the fourth- and third-order Birch-Murnaghan-like equations of state using Lagrangian strains are most suitable to describe the PBE-calculated free energies. Interestingly, in spite of the low *S*-value, the third order BM-like EoS based on Lagrangian strains produces the wrong sign in both independent thermal expansion coefficients. This emphasizes once more how delicate the fitting procedure is.

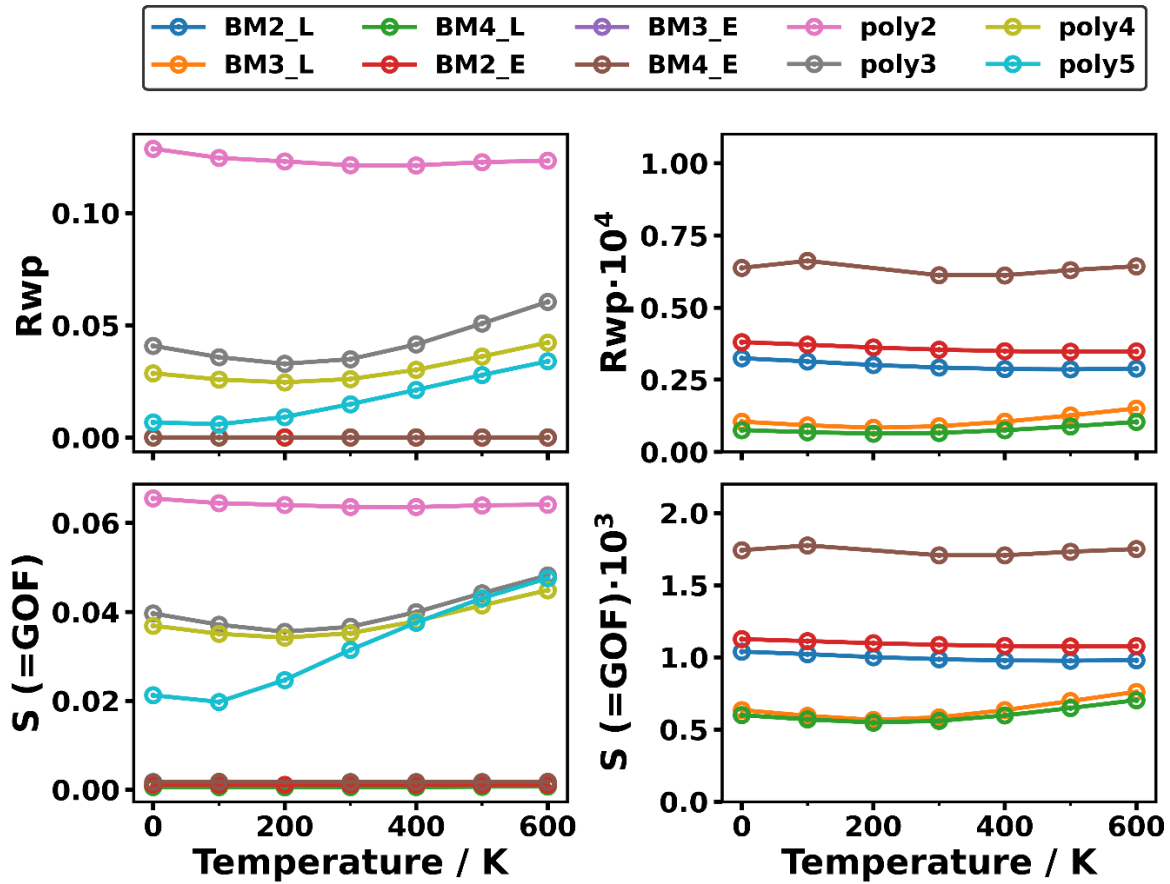

*Figure S 38: Quantitative comparison of the fitted model functions in terms of the Rwp and the S values (see equations (S46) and (S47), respectively) to fit the PBE-calculated free energies. Those comprise polynomials in the variables a and c of second to fifth order ("poly2", "poly3", "poly4", and "poly5") and Birch-Murnaghan-like functions ("BM"), in which the free energy is expanded in (two-dimensional) Taylor series of either Eulerian ("_E") or Lagrangian ("_L") finite strains to second to fourth order. E.g., the model "BM2_L" corresponds to a Birch-Murnaghan-like EoS of second order with Lagrangian finite strains. The right columns are zoomed-in views of the left columns.*

In contrast to the model-based fitting approaches, Gaussian Progress regression as described in Section S3.5 does not require a predefined model function. Using the same parameters as for the PBEsol data, qualitatively correct thermal expansion coefficients are obtained (see Figure S 21), with $\alpha_{33}$ being again notably overestimated compared to the PBEsol-based results of the Grüneisen theory and the experimental trends.

### S5.3 Grüneisen theory

As a final comparison between the PBE and PBEsol, the thermal expansion tensor was evaluated within the Grüneisen theory. To save computational time, only a second-order finite difference scheme was applied for obtaining the Grüneisen tensors with the PBE functional (strain step size $10^{-3}$). In this section, those results are compared to the PBEsol results based on the FD scheme of the same order to guarantee comparability (while the results presented in the main text are based on a fourth-order FD scheme). Figure S 39 shows the obtained thermal expansion tensor from the Grüneisen theory using the PBE (solid lines) and the PBEsol (dashed lines) functional as well as the PBEsol-calculated curves based on a fourth-order FD scheme as presented in the main text.

One can see in Figure S 39 that the thermal expansion tensor element $\alpha_{11}$, is in qualitative good agreement for both functionals, with slightly increased (absolute) magnitude for the PBE functional. However, with the PBE functional, also the thermal expansion tensor element $\alpha_{33}$ is distinctly negative, which yields a large negative volumetric thermal expansion.

The reason for that fundamental difference can be understood by analyzing the frequency-resolved mode Grüneisen tensors and the mode contributions to the thermal expansion tensor as shown in Figure S 40(a-d) (and Figure S 41(a-d) with linear frequency scales). Additionally, the PBEsol-calculated results (obtained with the same finite differences scheme as for PBE) are displayed in panels (e-h) in Figure S 40 and Figure S 41 to facilitate the comparison. While the 11-components appear to be relatively similar to the PBEsol results also quantitatively, the 33-components show fundamental qualitative differences: most of the mode Grüneisen tensor components, $\gamma_{33}^\lambda$, are negative as well, which essentially eliminates the prerequisite for reproducing the expected results of positive thermal expansion.

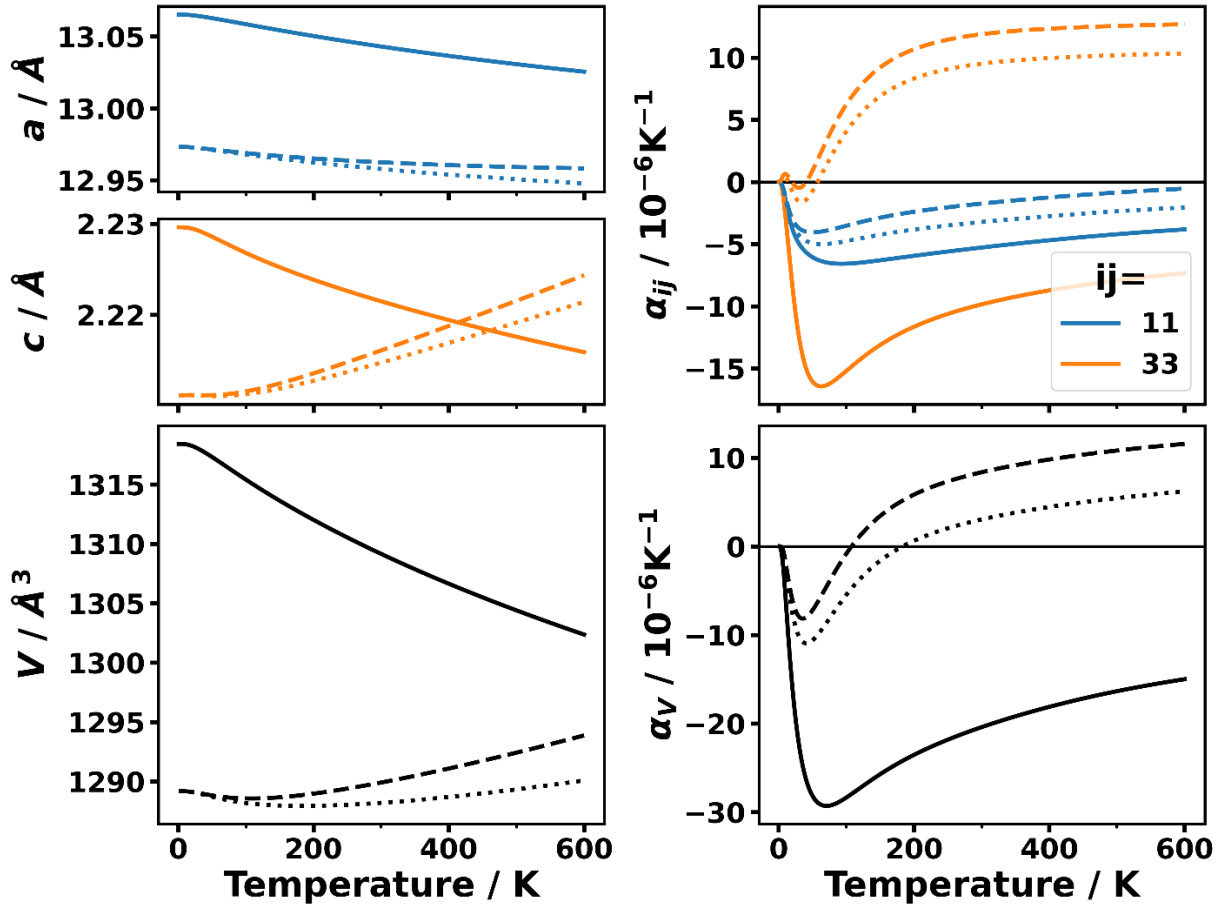

*Figure S 39: Temperature dependence of the two independent lattice parameters, a and c, and the unit-cell volume, V, of MOF-74 calculated with the Grüneisen theory described in Section S3.7. The right column shows the associated thermal expansion coefficients, $\alpha_{11}$, $\alpha_{33}$, and the volumetric thermal expansion coefficient $\alpha_V$. The solid lines show the results obtained with the PBE functional, while the dashed lines correspond to the results with the PBEsol functional (both using a second-order finite differences scheme with a strain step size of $10^{-3}$). Additionally, the PBEsol-calculated results presented in the main text (employing a fourth-order finite differences scheme*

*with the same step size) are shown as dotted lines. The first Brillouin zone for the phonon calculations was sampled with a 20×20×20 mesh of wave vectors.*

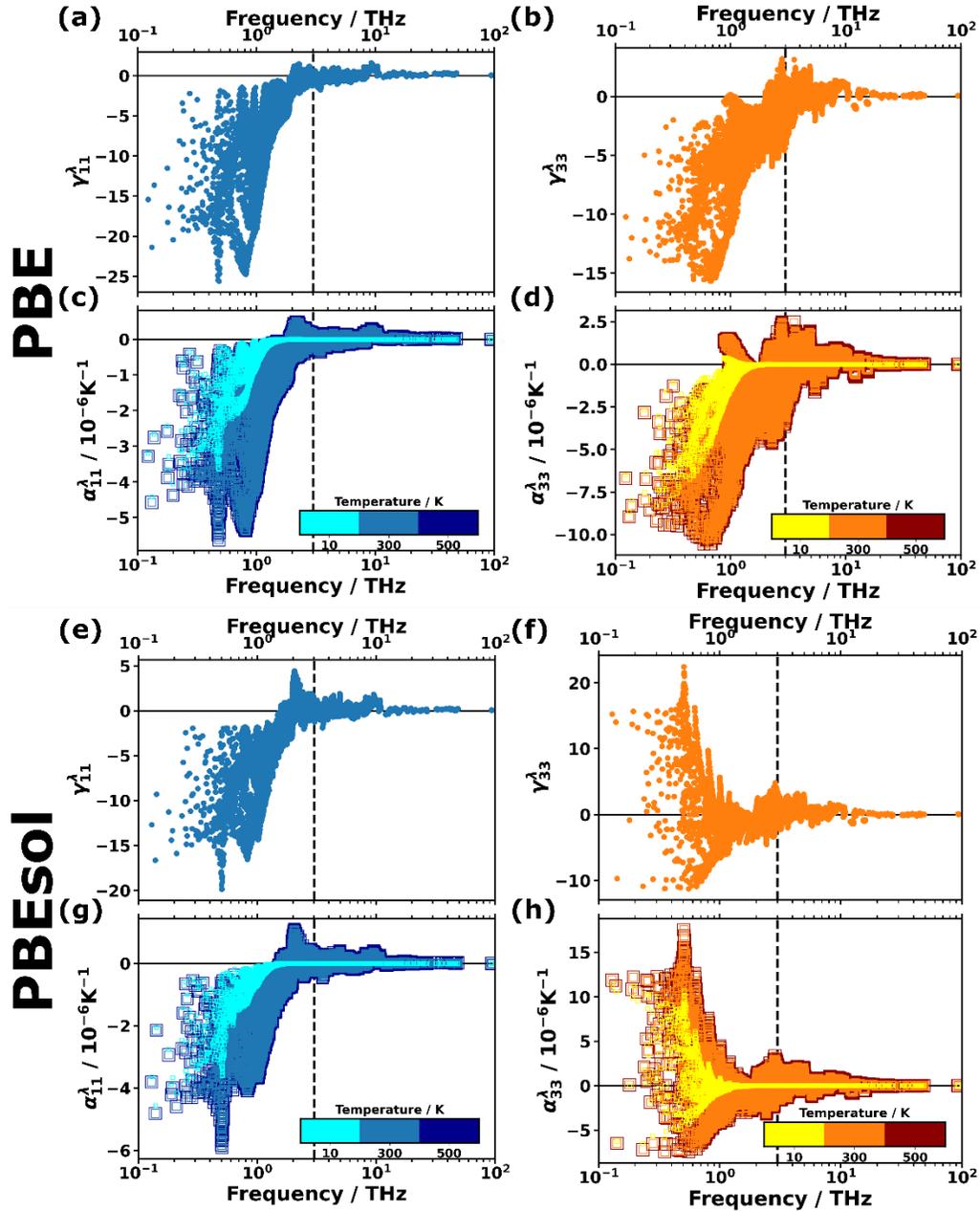

*Figure S 40: (a,b) PBE-calculated 11- and 33-components of the mode Grüneisen tensor and (c,d) mode contributions to the 11- and 33-components of the thermal expansion tensor at 10 K, 300 K, and 500 K of MOF-74. (e,f) PBEsol-calculated 11- and 33-components of the mode Grüneisen tensor and (g,h) mode contributions to the 11- and 33-components of the thermal expansion tensor*

at 10 K, 300 K, and 500 K of MOF-74. The colored data points correspond to the values for modes sampling the entire first Brillouin zone on a 20×20×20 mesh of wave vectors. For both functionals, a second-order finite differences scheme with a strain step size of $10^{-3}$ was employed to calculate the mode Grüneisen tensors. The symbols in panels (c), (d), (g) and (h) are plotted with increasing size for higher temperatures to facilitate the visual recognition. The vertical dashed lines in all panels are drawn at a frequency of 3 THz and serve as guide to the eye.

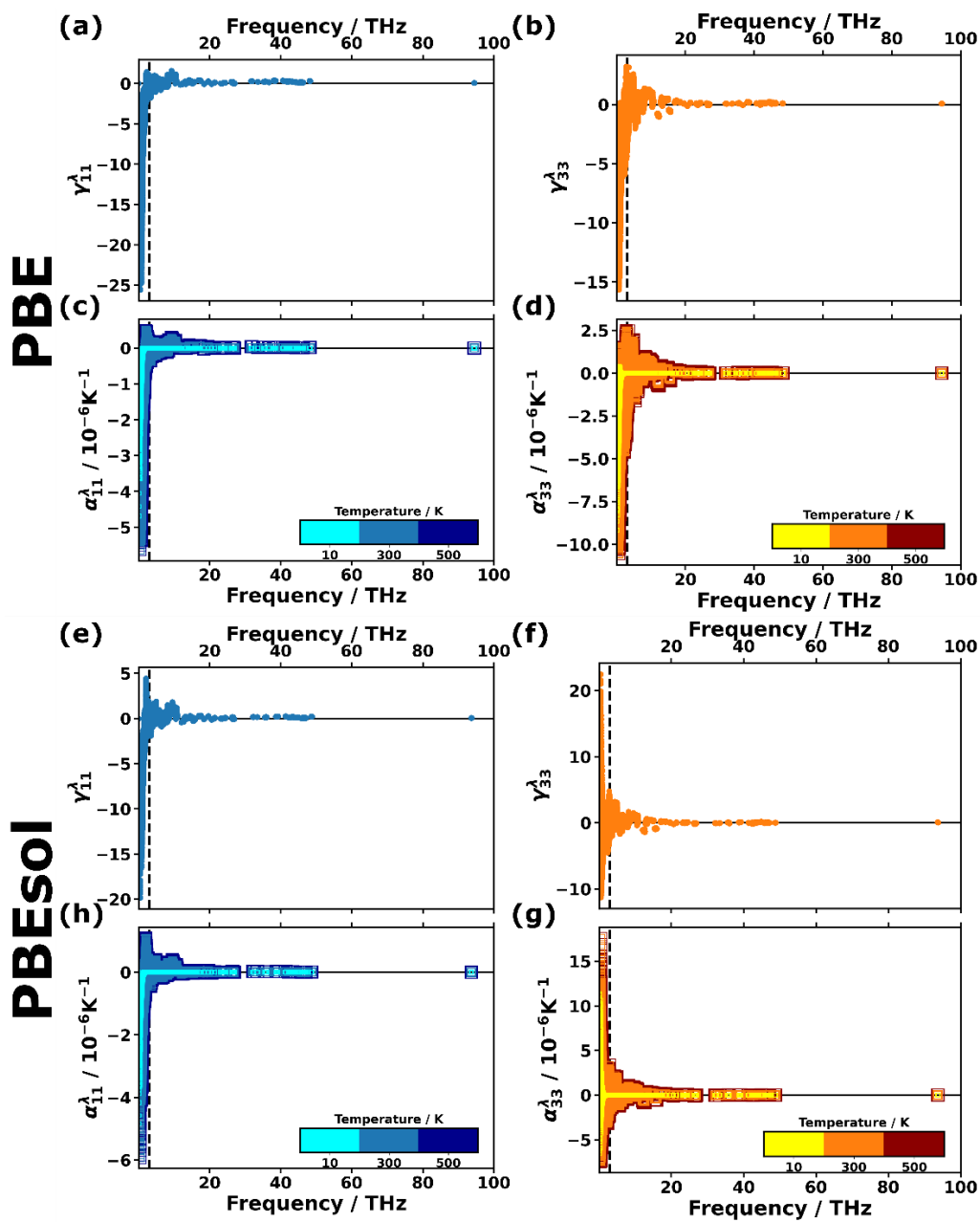

*Figure S 41: The same as Figure S 40 but with a linear frequency scale. (a,b) PBE-calculated 11- and 33-components of the mode Grüneisen tensor and (c,d) mode contributions to the 11- and 33-components of the thermal expansion tensor at 10 K, 300 K, and 500 K of MOF-74. (e,f) PBEsol-calculated 11- and 33-components of the mode Grüneisen tensor and (g,h) mode contributions to the 11- and 33-components of the thermal expansion tensor at 10 K, 300 K, and 500 K of MOF-74. The colored data points correspond to the values for modes sampling the entire first Brillouin zone on a 20×20×20 mesh of wave vectors. For both functionals, a second-order finite differences scheme with a strain step size of $10^{-3}$ was employed to calculate the mode Grüneisen tensors. The*

*symbols in panels (c), (d), (g) and (h) are plotted with increasing size for higher temperatures to facilitate the visual recognition. The vertical dashed lines in all panels are drawn at a frequency of 3 THz and serve as guide to the eye.*

These results are additionally supported by the low-frequency phonon band structure shown in Figure S 42 with the bands colored according to the mode Grüneisen tensor components and the mode contributions to the thermal expansion tensors (at 300 K). Although also for the PBE functional, the acoustic bands show the largest values of $\gamma_{ij}^{\lambda}$ and $\alpha_{ij}^{\lambda}$, they consistently exhibit negative values, in contrast to what is observed for the PBEsol functional. This qualitative difference can be rationalized by comparing the changes in the most relevant geometric descriptors as defined in Section S10.1 obtained with the PBEsol and the PBE functionals.

Table S 4 shows the signs of the changes in $L$, $A_\Delta$, and $\eta$ upon uniaxially compressing MOF-74 in *z*-direction and their net changes during a period of the lowest TA phonon mode at the F-point. One can see that, in the case of PBEsol, all geometric descriptors increase upon *z*-strain and also during this TA mode. Hence, following the observations stated in the main text, this clearly favors a positive mode Grüneisen tensor element. In contrast, with the PBE functional, $L$ changes upon *z*-compression in an opposite way as the (net) changes of the TA modes at F, while the minor (net changes) of $A_\Delta$ and $\eta$ occur in the same way as upon *z*-compression. The former supports the notion that with PBE one obtains negative $\gamma_{33}^{\lambda}$ elements and that this is not just related to a numerical artifact. The most probable reason for this qualitative difference in the change of the descriptors is that PBEsol is expected to improve geometric quantities (such as lattice parameters or bonding geometries) for solids [27,28,33]. Along the pore direction, the structure of MOF-74 more closely represents a (closely packed) solid. This might lead to subtle differences as those when using PBEsol rather than PBE.

*Table S 4: Comparison of the signs of the (net) changes of the geometric descriptors defined in Section S10.1 obtained with the PBEsol and the PBE functionals upon applying compressive strain in z-direction and as a consequence of the atomic displacements induced by the first transverse acoustic (TA) mode at the high-symmetry point F. For η, two signs are given as two neighboring nodes alternatingly expand/contract.*

| Geometric descriptor | Changes upon compressive uniaxial strain in z-direction | | Net changes during a period of the first TA phonon modes at F | |
|---|---|---|---|---|
| | PBEsol | PBE | PBEsol | PBE |
| $L$ | + | - | + | + |
| $A_\Delta$ | + | + | + | + |
| $\eta$ | +/+ | +/+ | -/+ | -/+ |

The first positive $\alpha_{33}^\lambda$-values occur for the first optical band close to the center of the first Brillouin zone, Γ, with much too small magnitude to cancel the overwhelming amount of (acoustic) phonons with negative $\alpha_{33}^\lambda$ below that frequency. The somewhat disturbing conclusion one has to draw is that with the PBE functional, the 33-components of the mode Grüneisen tensors do not show the expected results for MOF-74– *i.e.*, a positive 33- and a negative 11-component of the thermal expansion tensor as suggested by our experiments and by the experiments of Queen *et al.* conducted by means of neutron powder diffraction for magnesium-based MOF-74 [34]. Hence, this functional seems to fail in considering certain (potentially subtle) anharmonicities such that experimental trends cannot be reproduced properly for this particular system.

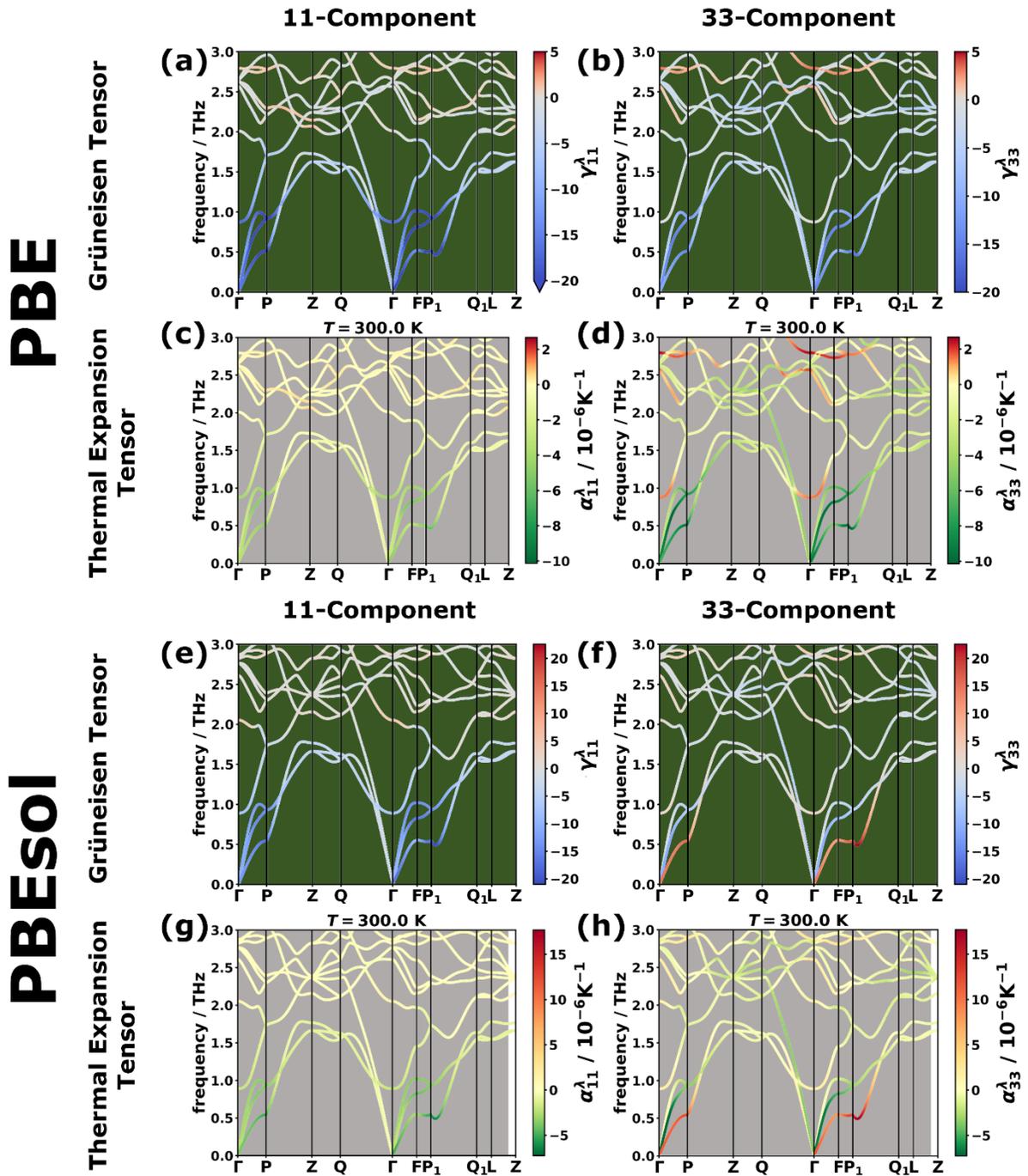

*Figure S 42: (a-d) PBE/D3-BJ-calculated phonon band structure of MOF-74 colored according to (a,b) the mode Grüneisen tensor components, $\gamma_{11}^\lambda$ and $\gamma_{33}^\lambda$, and (c,d) the mode contributions to the thermal expansion tensor components, $\alpha_{11}^\lambda$ and $\alpha_{33}^\lambda$. (e-h) PBEsol/D3-BJ-calculated phonon band structure of MOF-74 colored according to (a,f) the mode Grüneisen tensor components, $\gamma_{11}^\lambda$ and $\gamma_{33}^\lambda$, and (g,h) the mode contributions to the thermal expansion tensor components, $\alpha_{11}^\lambda$ and*

$\alpha_{33}^\lambda$. *For both functionals, a second-order finite differences scheme with a strain step size of $10^{-3}$ was employed to calculate the mode Grüneisen tensors.*

**S6. Density of states and density of states per logarithmic frequency interval**

In general, the density of states, *DOS*, as a function of the (continuous) phonon frequency, *f*, can be written as delta distributions centered at (sampled) discrete phonon frequencies, $f_\lambda$, summed over all phonon modes (with the index $\lambda$ containing the band index and the wave vector, *q*) and normalized by the number of wave vectors, $N_q$ [2,3]:

$$DOS(f) = \frac{1}{N_q} \sum_\lambda \delta(f - f_\lambda) \tag{S73}$$

The *DOS* shown in the main text was calculated from a 30×30×30 mesh of phonon wave vectors, *q*, even denser than other thermodynamic properties because, typically, convergence of the *DOS* is reached at somewhat higher *q*-densities.

Additionally, in practice, the delta distributions in the equation above are replaced with similar functions of finite widths. Here, Lorentzian functions with a width parameter, $\sigma$, were employed with $\sigma = 0.05$ THz. We note, in passing, that $2\sigma$ corresponds to the full width at half maximum of the finite-width peak.

$$\delta(f - f_\lambda) \to \frac{1}{\pi} \frac{\sigma}{(f - f_\lambda)^2 + \sigma^2} \tag{S74}$$

In order to calculate the density of states per logarithmic frequency interval, $DOS_\varphi$, as presented in the main text for the semi-logarithmic frequency plots, the following transformation was applied. As a first step, the relation between the phonon frequency, *f*, and the new variable, *i.e.*, the logarithmic frequency, $\varphi$, must be defined.

$$\varphi = \log_{10}\left(\frac{f}{1\text{ THz}}\right) \Leftrightarrow 10^\varphi \cdot 1\text{ THz} = f \tag{S75}$$

In general, a transformation of the density of states from frequency-space, *DOS*, to the space of any other variable, *x*, can be calculated as [2,3]:

$$DOS_x(x) = DOS(f(x)) \left|\frac{df}{dx}\right| \qquad (S76)$$

In the case of the "logarithmic frequency", $\varphi$, this transformation rule yields:

$$DOS_\varphi(\varphi) = DOS(f(\varphi)) \left|\frac{df}{d\varphi}\right| = DOS(f = 10^\varphi \cdot 1 \text{ THz}) \, 10^\varphi \ln 10 \qquad (S77)$$

This expression was directly used to compute the $DOS_\varphi$, with the expression $DOS(f = 10^\varphi \cdot 1 \text{ THz})$ having been evaluated by, first, interpolating the *DOS* in frequency space (using cubic splines) in order to have a mathematical function available, which could, in a second step, be used with the argument ($10^\varphi \cdot 1$ THz) instead of the original frequency arguments.

### S7. Animation of phonon modes

Animation of selected phonon modes are provided as separate file. More information such as the band indices, the frequencies, *etc.* of the animated phonons are found in Table S 5.

*Table S 5: Animations of phonon modes in MOF-74.*

| Frequency / THz | Band Index | Wave Vector (reduced coordinates) | File Name |
|---|---|---|---|
| 0.89 | 4 | [0, 0, 0] | mode004_00d89THz.mp4 |
| 2.05 | 5 | [0, 0, 0] | mode005_02d05THz.mp4 |
| 2.60 | 6 | [0, 0, 0] | mode006_02d60THz.mp4 |
| 2.77 | 7 | [0, 0, 0] | mode007_02d77THz.mp4 |
| 2.77 | 8 | [0, 0, 0] | mode008_02d77THz.mp4 |
| 2.83 | 9 | [0, 0, 0] | mode009_02d83THz.mp4 |
| 2.83 | 10 | [0, 0, 0] | mode010_02d83THz.mp4 |
| 2.87 | 11 | [0, 0, 0] | mode011_02d87THz.mp4 |
| 0.55 | 1 | [0.5, -0.5, 0] | Fmode_001_00d55THz.mp4 |

## S8. Mode contribution plots with linear frequency axes

In analogy to Figure 5 in the main text, which shows the mode contributions to the Grüneisen and the thermal expansion tensors with a logarithmic frequency scale, the equivalent plot with a linear frequency scale is shown in Figure S 43.

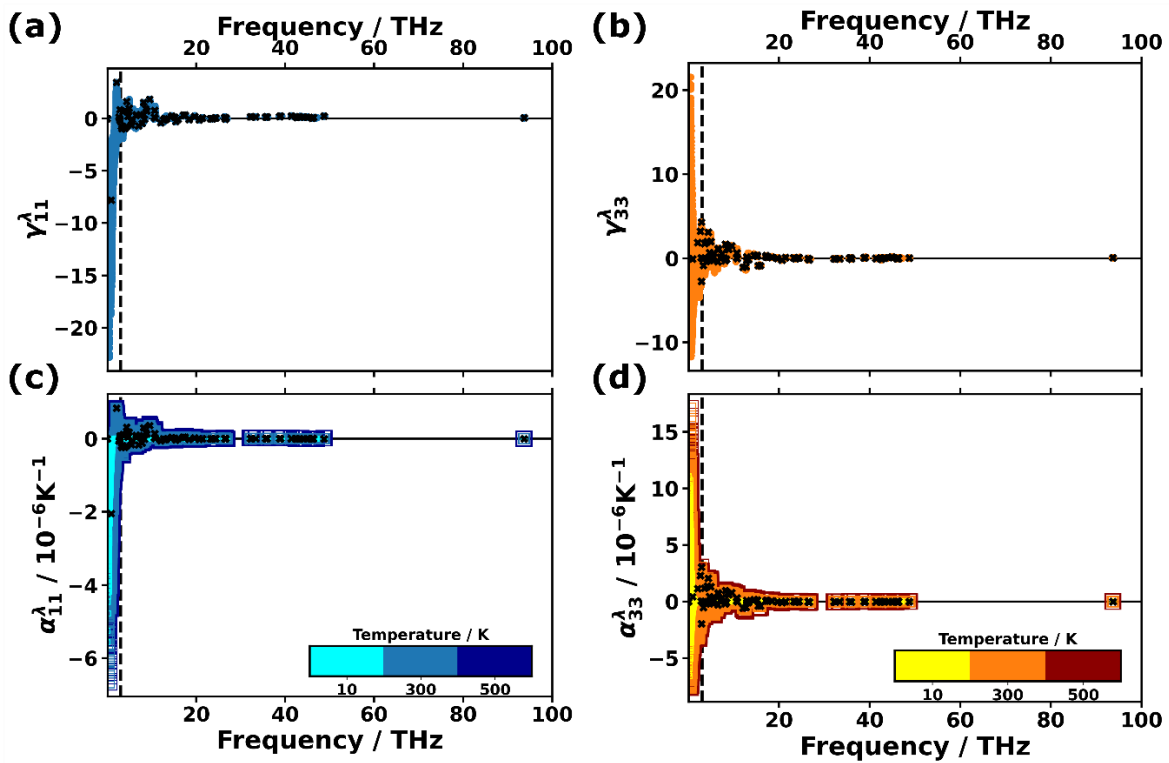

*Figure S 43: (a,b) 11- and 33-components of the mode Grüneisen tensor and (c,d) mode contributions to the 11- and 33-components of the thermal expansion tensor at 10 K, 300 K, and 500 K of MOF-74. The colored data points correspond to the values for modes sampling the entire first Brillouin zone on a 20×20×20 mesh of wave vectors. In contrast, the black crosses denote the values for phonons at the Γ-point. The symbols in panels (c) and (d) are plotted with increasing size for higher temperatures to facilitate the visual recognition. The vertical dashed lines in all panels are drawn at a frequency of 3 THz and serve as guide to the eye.*

## S9. Further analysis of the x-ray diffraction experiments

### S9.1 Alternative plots of the power x-ray diffraction data

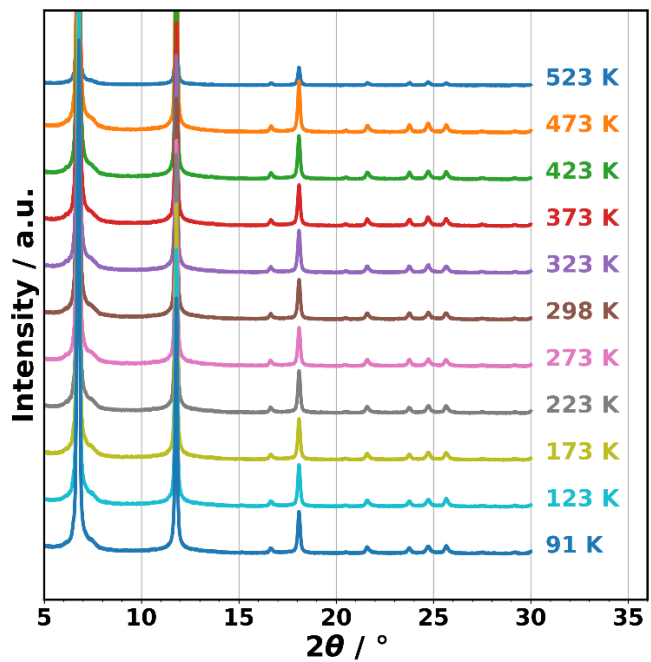

*Figure S 44: Powder XRD patterns of the activated MOF-74 sample vertically shifted for various temperatures with a linear intensity scale.*

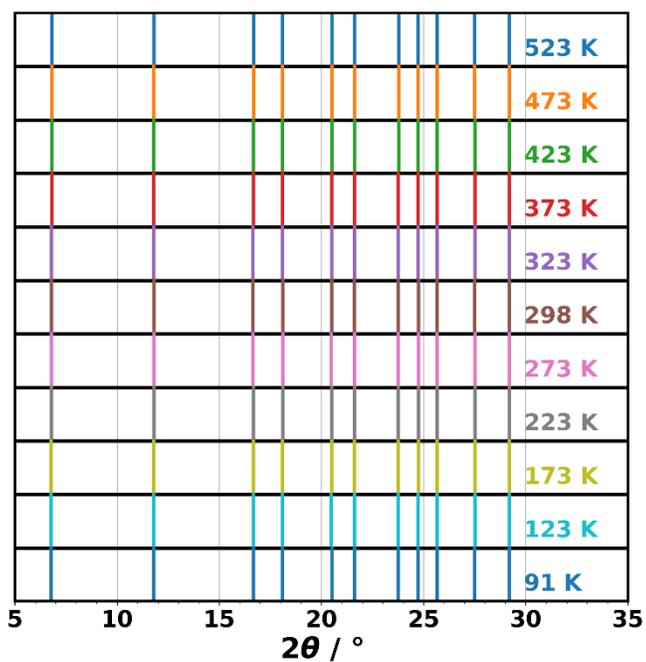

*Figure S 45: Fitted peak positions of the experimental powder XRD patterns of the activated MOF-74 sample for various temperatures.*

## S10. Geometric descriptors used to characterize the MOF-74 structure
### S10.1 Definition of the descriptors

To provide more insight into the deformation mechanisms induced by phonon displacements and by the applied strains, a similar strategy as in Ref. [29] was employed: we defined a number of geometrical descriptors in order to quantify the distortions from the equilibrium (*i.e.*, 0-K optimized structure) as a result of the occupation of a phonon mode or a non-vanishing strain. The used geometrical descriptors are shown in Figure S 46.

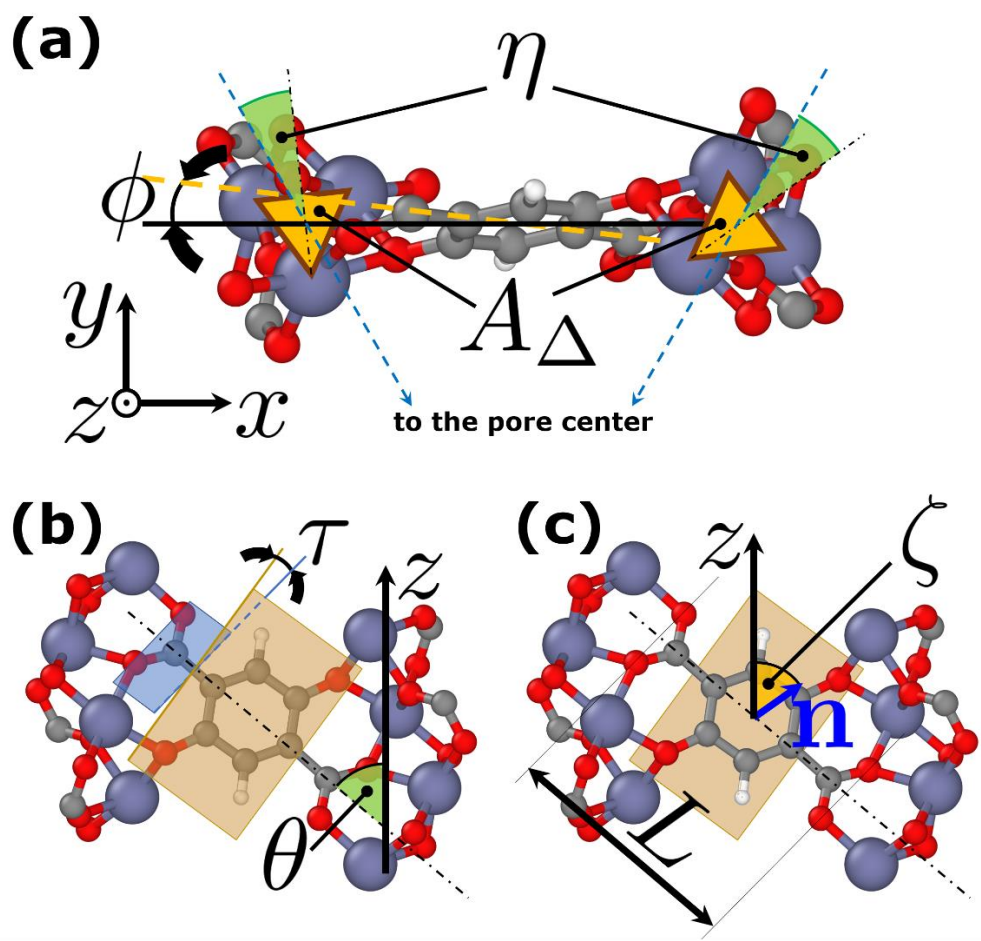

*Figure S 46: Schematic definition of the geometrical descriptors used to characterize the framework geometry of MOF-74 in (a) top view and (b,c) front view. The descriptors comprise: the cross-sectional area, $A_\Delta$, the node rotation angle, $\eta$, the linker inclination, $\phi$, the linker tilt, $\theta$, the dihedral angle between the COO$^-$ groups and the phenylene ring, $\tau$, the linker length, $L$, and*

*the polar angle, $\zeta$, between the z-axis (i.e., the channel direction) and the normal vector, **n**, of the phenylene plane.*

The quantities used to describe the structural distortions comprise:

i. the cross-sectional area, $A_\Delta$, defined as the area of the triangle formed by the projection of the centers of the three inequivalent metal atoms per node onto the *xy*-plane,

ii. the node rotation angle, $\eta$, measured between the lines connecting the center of the nodes with the center of the pore and the line connecting the inner corner of the projected triangle with the node axis,

iii. the linker inclination angle, $\phi$, measured between the projection of the linker axis onto the *xy*-plane and the straight line connecting the axes of two neighboring nodes,

iv. the dihedral angle between the plane of the COO⁻ group and the plane of the phenylene ring of the linker,

v. the polar linker tilt angle, $\theta$, measured between the long molecular axis of the linker backbone and the node axis (=z-axis),

vi. the polar angle, $\zeta$, between the normal vector of the plane of the phenylene ring and the node axis (=z-axis), and

vii. the total length of the linker, $L$.

### S10.2 Evolution under strain

In order to quantify the strain-induced changes to the internal structure of MOF-74, the evolution of the descriptors defined in Section S10.1 as a function of the magnitude of strain was monitored in terms of scalar strain variables, $\varepsilon_a$ (*i.e.*, straining the lattice parameter *a*; *i.e.*, isotropic stress in the *xy*-plane) and $\varepsilon_c$ (*i.e.*, straining the lattice parameter *c*). The changes are shown in Figure S 47. Note that – with one exception – all descriptors show monotonic and often even (close to) linear behavior as a function of the strain. The exception is the linker length, $L$, which, due to the particularly small changes, shows additional noise superimposed to the still perceivable trends.

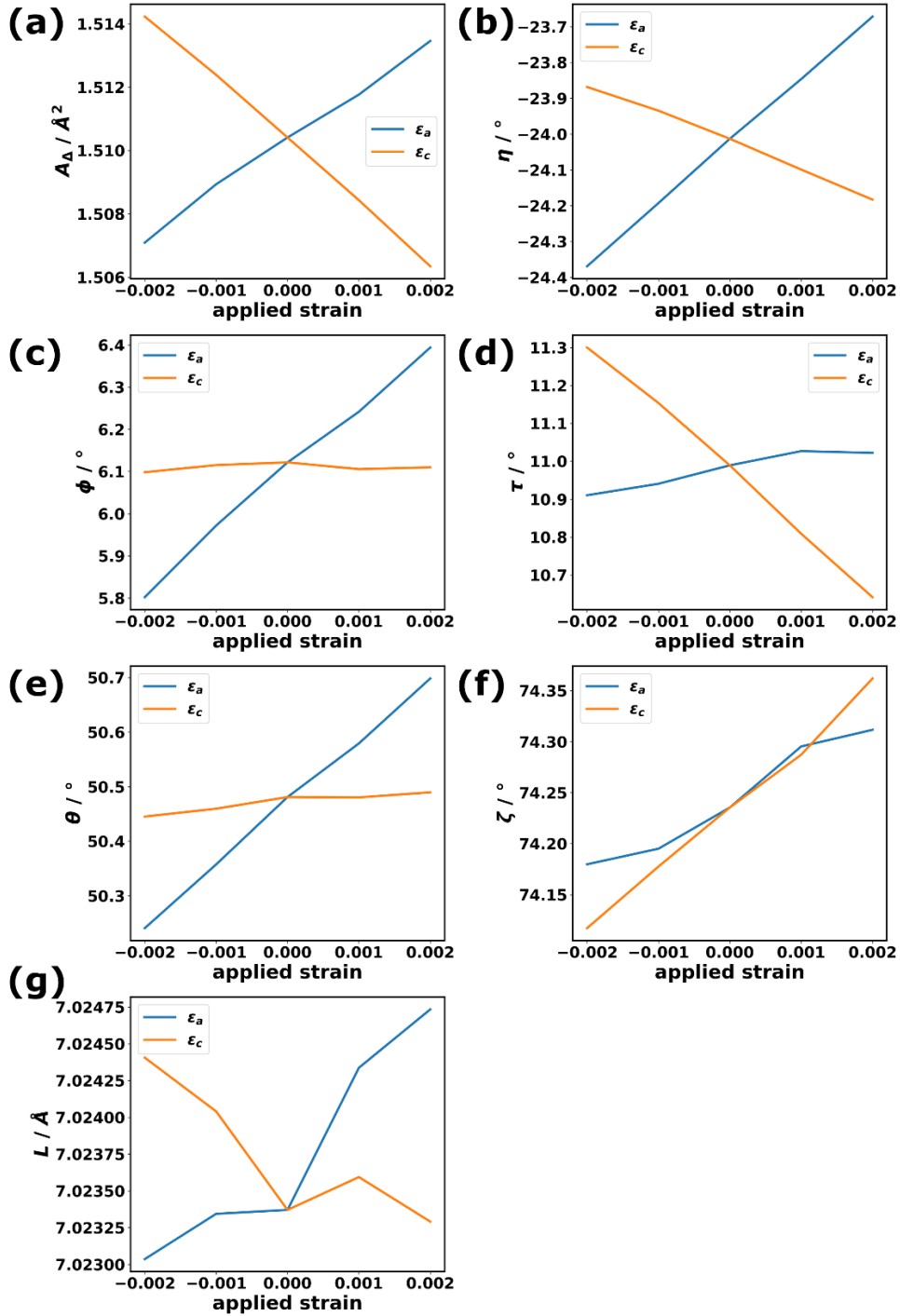

Figure S 47: Evolution of the geometric descriptors defined in Section S10.1 as a function of the scaler strain variables $\varepsilon_a$ (i.e., straining the lattice parameter a) and $\varepsilon_c$ (i.e., straining the lattice parameter c). For the definition of the associated strain tensors see equation (S48).

Except for the linker length, *L*, and the linker inclination, *ϕ*, all descriptors show the same qualitative behavior upon strain in *x*- and *z*-direction as shown in the Supporting Information of Ref. [29] (although there, uniaxial strain has also been applied in *x*-direction, while here isotropic strain in the *xy*-plane and uniaxial strain in *z*-direction were chosen to account for the more natural situation of MOF-74 which would, in the absence of external stresses, thermally expand equally in the *xy*-plane). The qualitative differences in the evolutions of *L* and *ϕ* with applied strain concern, however, only the case of uniaxial strain in *z*-direction. With the here employed magnitudes of compressive *z*-strain (up to -0.002), *ϕ* shows hardly any perceivable changes, while in Ref. [29] (using a much larger *z*-strain of -0.03) *ϕ* somewhat increases. At the here employed rather small strain levels (additionally preventing the lateral expansion which would result from applied stress rather than from strain), the combination of the laterally growing and rotating nodes and this rather unchanged inclination of the linkers lead to the tendency of the linkers length, *L*, to slightly increase upon (small) uniaxial compressions in *z*-directions. For the comparably large strain levels employed in Ref. [29], the notable change in the linker inclination leads to a small decrease in *L* instead. This is supported by the observed non-linearity of *L* with respect to the applied stress magnitude in *z*-direction in Ref. [29].

### S10.3   Evolution during atomic motions associated with Γ-phonons

In order to analyze the evolution of the geometric descriptors defined above, the structure of MOF-74 was displaced along the Γ-modes, $\lambda$, according to the well-known relations (see, *e.g.*, Ref. [19]) for the displacement, *u*, of atom $\alpha$ in the Cartesian direction *i*:

$$u_{i,\alpha}^\lambda(t) = Q(t)\, Re\left\{\frac{\xi_{i,\alpha}^\lambda}{||\xi^\lambda||}\right\} \quad (S78)$$

$$\text{with} \quad \xi_{i,\alpha}^{\lambda} = \frac{e_{i,\alpha}^{\lambda}}{\sqrt{m_\alpha}} \quad \text{(S79)}$$

$$\text{and} \quad ||\xi^{\lambda}|| = \sqrt{\sum_{i,\alpha} |\xi_{i,\alpha}^{\lambda}|^2} \quad \text{(S80)}$$

Here, $m_\alpha$ is the mass of atom $\alpha$, and $e_{i,\alpha}^{\lambda}$ denotes the component of the eigenvector of phonon mode $\lambda$ belonging to atom $\alpha$ in the Cartesian direction $i$. The time dependence of the atomic displacements is covered by the time dependence of the normal mode coordinate, $Q(t)$, for which we employed the functional form

$$Q(t) = A \sin(\Omega t) \quad \text{(S81)}$$

with an amplitude $A = 1$ Å for the numeric analysis, and $A = 5$ Å for the (exaggerated) animations. It is important to note that although every single atom strictly follows a sinusoidal atomic displacement over time, the geometric descriptors defined in Section S10.1 are (non-linear) functions of these positions and, thus, do not show sinusoidal time dependences in general. This is emphasized by the example of the time evolution of the node cross-sectional areas, $A_\Delta$, of two nodes shown in Figure S 48.

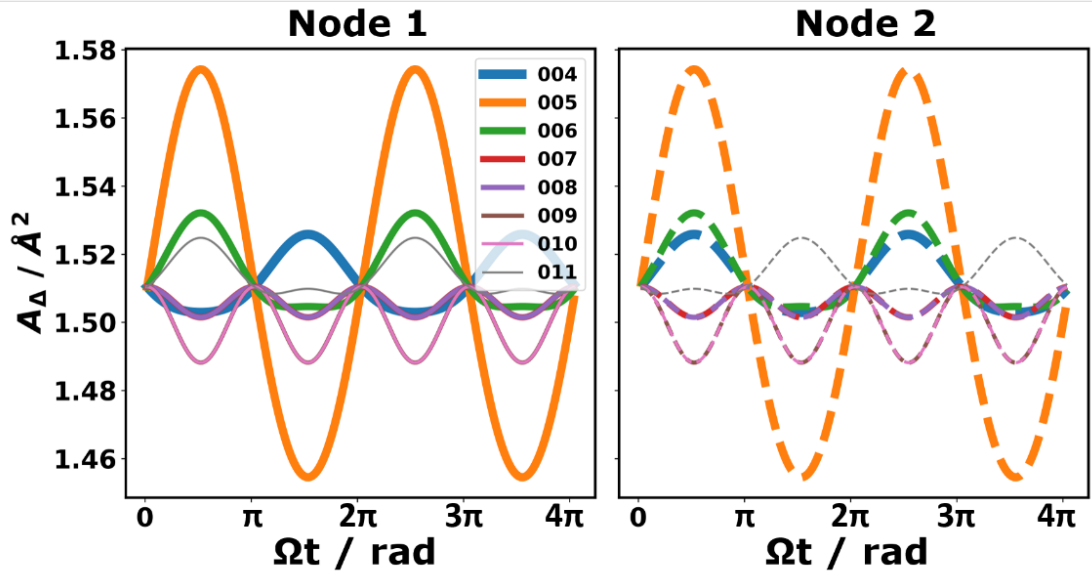

*Figure S 48: Evolution of the node cross-section area, $A_\Delta$, of two neighboring nodes ("node 1" and "node 2") with time, $t$, multiplied with the angular frequency, $\Omega$, of the sinusoidal time dependence of the normal mode coordinate, Q (see equation (S81)), using an amplitude of A = 1*

Å. *The lines in different colors show the time dependence of $A_\Delta$ calculated for the $\Gamma$-phonons with band indices 4 to 11 (according to the legend).*

The non-sinusoidal shape of the time dependence of the above defined geometric descriptors has the interesting consequence that, averaged over an integer number of periods, the average values do not vanish. *I.e.*, on average, a phonon mode produces a net change, $\overline{\Delta X}$, of a descriptor, $X$, and, more obviously, a root mean square (RMS) change, $\left(\overline{\Delta(X)^2}\right)^{1/2}$. These two quantities were evaluated for the first eight optical $\Gamma$-phonons and are shown in the following figures. As only the most important changes are used to rationalize the signs of the Grüneisen tensor elements in the main text, we leave Figure S 49 to Figure S 55 uncommented at this point.

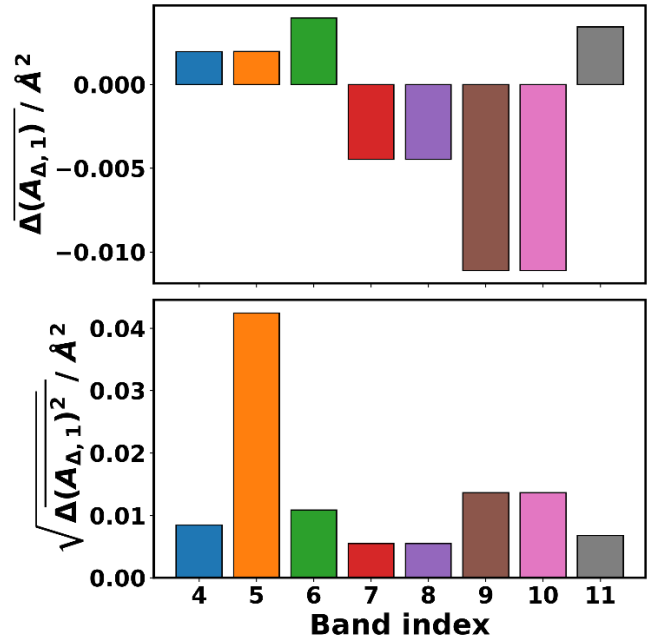

*Figure S 49: Average (top) and root mean square (bottom) change in the cross-sectional area, $A_\Delta$, (node 1) for the first eight optical phonon modes at $\Gamma$.*

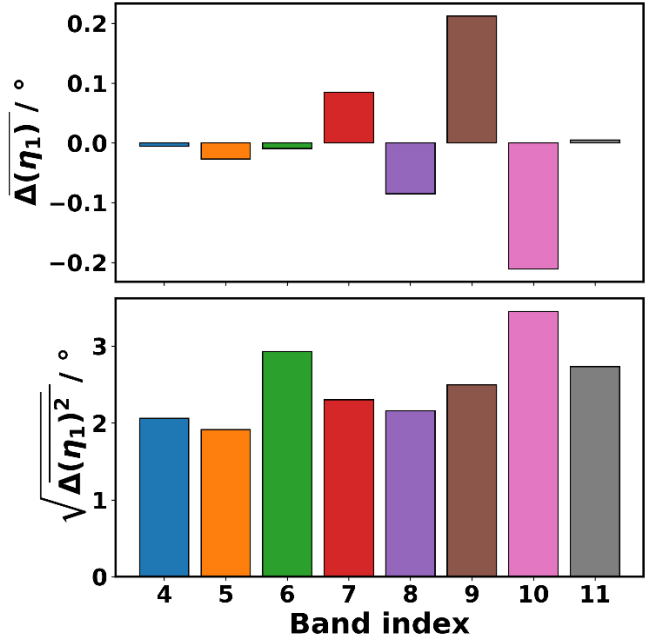

*Figure S 50: Average (top) and root mean square (bottom) change in the node rotation angle, η, (node 1) for the first eight optical phonon modes at Γ.*

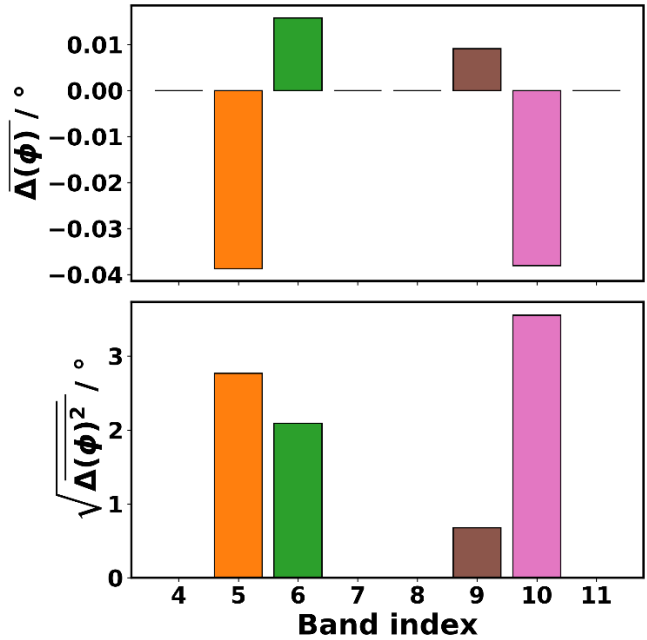

*Figure S 51: Average (top) and root mean square (bottom) change in the linker inclination angle, ϕ, for the first eight optical phonon modes at Γ.*

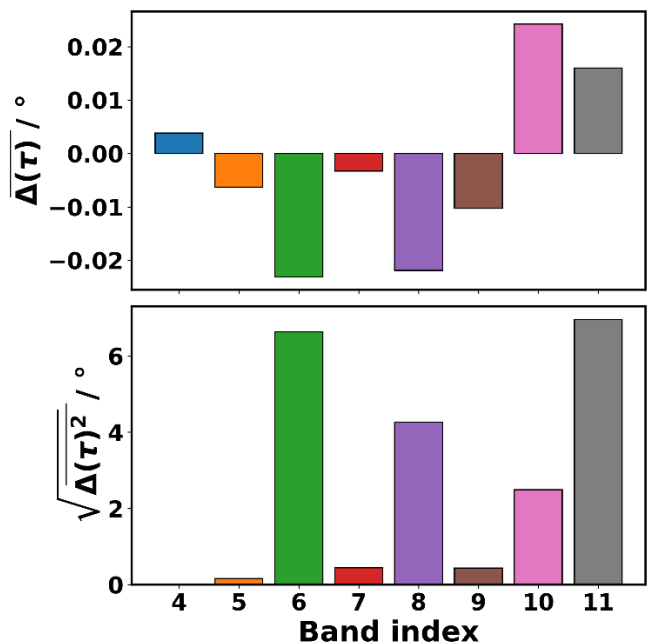

*Figure S 52: Average (top) and root mean square (bottom) change in the linker dihedral angle, τ, for the first eight optical phonon modes at Γ.*

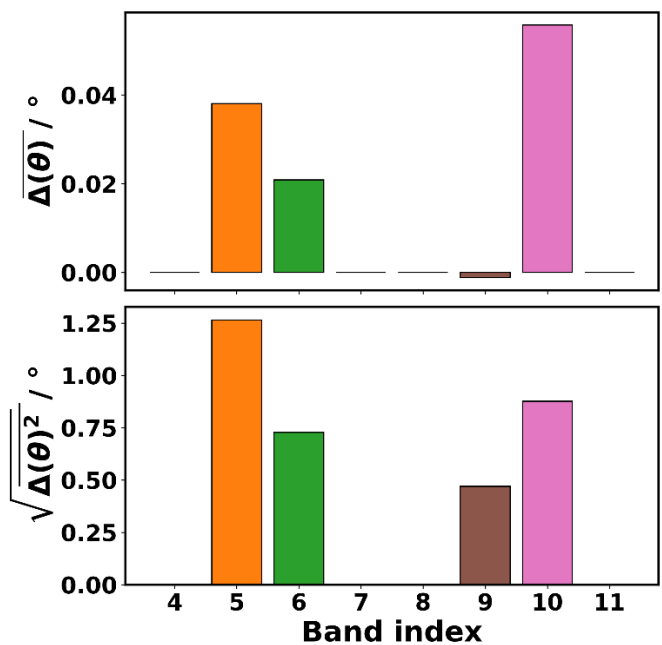

*Figure S 53: Average (top) and root mean square (bottom) change in the linker tilt angle, θ, for the first eight optical phonon modes at Γ.*

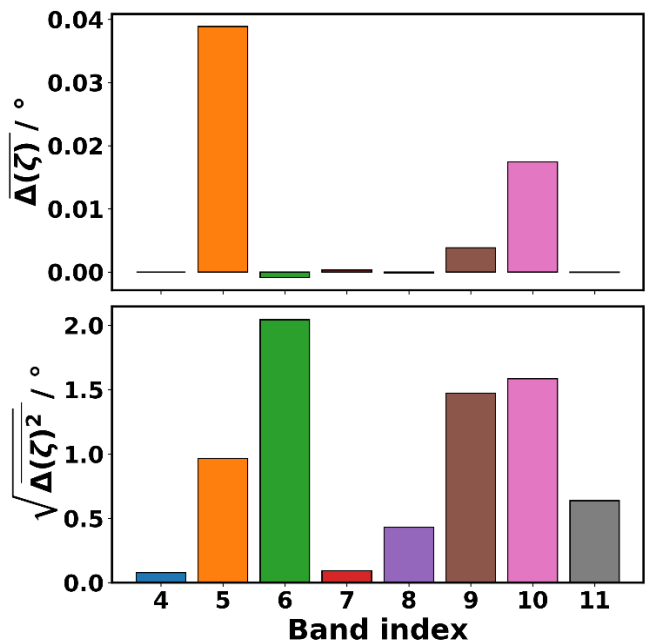

*Figure S 54: Average (top) and root mean square (bottom) change in the polar angle of the normal vector of the phenylene plane, ζ, for the first eight optical phonon modes at Γ.*

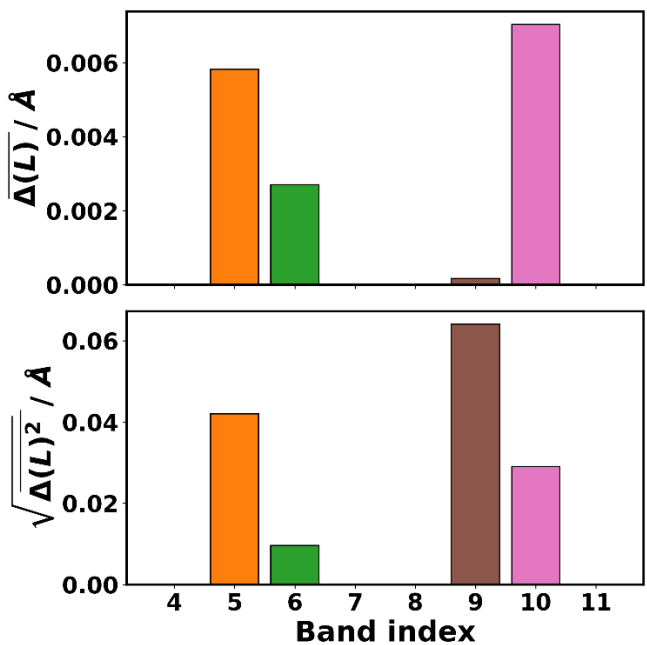

*Figure S 55: Average (top) and root mean square (bottom) change in the linker length, L, for the first eight optical phonon modes at Γ.*